\documentclass[12pt]{article}
\usepackage{multirow}
\usepackage{float}
\usepackage{pdflscape}
\usepackage{adjustbox}
\usepackage{amssymb}
\usepackage[utf8]{inputenc}
\usepackage{mathtools}
\usepackage{amsmath,amsfonts,graphicx}
\usepackage{bbm}
\usepackage{algorithm}
\usepackage{algpseudocode}
\usepackage{url}
\usepackage{afterpage}

\usepackage[colorlinks, citecolor={blue}]{hyperref}

\newtheorem{theorem}{Theorem}[section]
\newtheorem{lemma}[theorem]{Lemma}

\usepackage{natbib}

\newcommand{\blind}{1}

\begin{document}

\def\spacingset#1{\renewcommand{\baselinestretch}%
{#1}\small\normalsize} \spacingset{1}


\if1\blind
{
  \title{\bf Covariate Adjusted Functional Mixed Membership Models}
  \author{Nicholas Marco\thanks{Corresponding author. \href{nicholas.marco@duke.edu}{nicholas.marco@duke.edu}}\\
    Department of Statistical Science, Duke University, USA.\\
    and \\
    Damla \c{S}ent\"{u}rk \\
    Department of Biostatistics, University of California,
Los Angeles, USA. \\
    and \\
    Shafali Jeste \\
    Division of Neurology and Neurological Institute,\\ Children’s Hospital Los Angeles, Los Angeles, USA.\\
    and \\
    Charlotte DiStefano \\
    Division of Psychiatry, Children’s Hospital Los Angeles, Los Angeles, USA.\\
    and \\
    Abigail Dickinson \\
    Department of Psychiatry and Biobehavioral Sciences,\\ University of California, Los Angeles, USA.\\
    and \\
    Donatello Telesca \hspace{.2cm}\\
    Department of Biostatistics, University of California,
Los Angeles, USA.\\
    }
  \maketitle
} \fi

\if0\blind
{
  \bigskip
  \bigskip
  \bigskip
  \begin{center}
    {\LARGE\bf Functional Mixed Membership Models}
\end{center}
  \medskip
} \fi

\bigskip

\newpage
\begin{abstract}
    Mixed membership models are a flexible class of probabilistic data representations used for unsupervised and semi-supervised learning, allowing each observation to partially belong to multiple clusters or features. In this manuscript, we extend the framework of functional mixed membership models to allow for covariate-dependent adjustments. The proposed model utilizes a multivariate Karhunen-Loève decomposition, which allows for a scalable and flexible model. Within this framework, we establish a set of sufficient conditions ensuring the identifiability of the mean, covariance, and allocation structure up to a permutation of the labels. This manuscript is primarily motivated by studies on functional brain imaging through electroencephalography (EEG) of children with autism spectrum disorder (ASD). Specifically, we are interested in characterizing the heterogeneity of alpha oscillations for typically developing (TD) children and children with ASD. Since alpha oscillations are known to change as children develop, we aim to characterize the heterogeneity of alpha oscillations conditionally on the age of the child. Using the proposed framework, we were able to gain novel information on the developmental trajectories of alpha oscillations for children with ASD and how the developmental trajectories differ between TD children and children with ASD.
\end{abstract}
\noindent%
\vfill

\newpage
\spacingset{1.5} 

\section{Introduction}
Cluster analysis is an unsupervised exploratory task that aims to group similar observations into ``clusters'' to explain the heterogeneity found in a particular dataset. 
When an outcome of interest is indexed by covariate information, it is often informative to adjust clustering procedures to account for covariate-dependent patterns of variation.
Covariate-dependent clustering has 
found applications in post hock analyses of clinical trial data \citep{muller2011product}, where conditioning on factors such as dose response, tumor grade, and other clinical factors is often of interest when defining patients response subgroups. In addition to clinical trial settings, covariate-dependent clustering has also become popular in fields such as genetics \citep{qin2006clustering}, flow cytometry \citep{hyun2023modeling}, neuroimaging \citep{guha2022bayesian,binkiewicz2017covariate}, and spatial statistics \citep{page2016spatial}. 

In statistics and machine learning, covariate-dependent clustering procedures have known a number of related nomenclatures, including \textit{finite mixture of regressions}, \textit{mixture of experts}, and \textit{covariate adjusted product partition models}, which tend to refer to specific ways in which covariate information is used in data grouping decisions. The term finite mixture of regressions \citep{mclachlan2019finite, faria2010fitting, grun2007applications, khalili2007variable, hyun2023modeling, devijver2015finite} refers to the fitting of a mixture model, where the mean structure depends on the covariates of interest through a regression framework. The mixture of experts model \citep{jordan1994hierarchical, bishop2002bayesian} is similar to a finite mixture of regressions model in that it assumes that the likelihood is a weighted combination of probability distribution functions. However, in the mixture of experts model, the weights are dependent on the covariates of interest, providing an additional layer of flexibility beyond that of traditional finite mixture of regressions models. Although the mixture of experts model was originally designed for supervised learning settings, recent developments have expanded its application to unsupervised settings \citep{kopf2021mixture,tsai2020mice}.  Lastly, covariate adjusted product partition models \citep{muller2011product, park2010bayesian} are a Bayesian non-parametric version of covariate adjusted clustering. Analogous to mixture of experts models, covariate adjusted product partition models allow the cluster partitions to depend on the covariates of interest. 

In this manuscript, we extend the class of functional mixed membership models proposed by \citet{marcoFunctional} to allow latent features to depend on covariate information. Mixed membership models \citep{Erosheva04,heller2008statistical,Gruhl2014,marcoFunctional, marcoMultivariate}, sometimes referred to as partial membership models, can be thought of as a generalization of traditional clustering, where each observation is allowed to partially belong to multiple clusters or features. Although there have been many advancements in covariate adjusted clustering models for multivariate data, to the best of our knowledge, little work has been done on incorporating covariate information in functional mixed membership models or functional clustering models. Two important exceptions are \citet{yao2011functional}, who specified a finite mixture of regressions model where the covariates are functional data and the clustered data are scalar, and \citet{gaffney1999trajectory} who proposed a functional mixture of regressions model where the function is modeled by a deterministic linear function of the covariates. Here, we consider the case where we have a covariate vector in $\mathbb{R}^p$ and the outcome we cluster is functional.

 Our work is primarily motivated by functional brain imaging studies on children with autism spectrum disorder (ASD) through electroencephalography (EEG). Specifically, \citet{marcoFunctional} analyzed how alpha oscillations differ between children with ASD and typically developing (TD) children. Since alpha oscillations are known to change as children develop, the need for an age-dependent mixed membership model is crucial to ensure that shifts in the power spectrum of alpha oscillations do not confound measurements of alpha power \citep{haegens2014inter}. Unlike mixture of experts models and covariate adjusted product partition models, we aim to specify a mixed membership model in which the allocation structure does not depend on the covariates of interest. Although previous studies have shown that the alpha peak shifts to a higher frequency and becomes more attenuated \citep{scheffler2019covariate, Rodriguez2017}, the results can be confounded if this effect is not observed in all individuals. In our model, since we assume that each individual's allocation parameters do not change with age, we can infer how alpha oscillations change as children age at an individual level, making a covariate adjusted mixed membership model ideal for this case study.  

 This manuscript starts with a brief review of functional clustering and covariate-dependent clustering frameworks, such as the mixture of experts model and the finite mixture of regressions model. Using these previous frameworks as a reference, we derive the general form of our covariate adjusted mixed membership model. In Section \ref{sec: MKL} we review the multivariate Karhunen-Loève (KL) theorem, which allows us to have a concise representation of the $K$ latent functional features. In Section \ref{sec: ModelSpecification}, we leverage the KL decomposition to fully specify the covariate adjusted functional mixed membership model. A review of the identifiability issues that occur in mixed membership models, as well as a set of sufficient conditions to ensure identifiability in a covariate adjusted mixed membership model can be found in Section \ref{sec: ModelIdentifiability}. Section \ref{sec: simulation_study} covers two simulation studies which explore the empirical convergence properties of the proposed model and evaluate the performance of various information criteria in choosing the number of features in a covariate adjusted functional mixed membership model. Section \ref{sec: ASD} illustrates the usefulness of the covariate adjusted functional mixed membership model by analyzing EEG data from children with ASD and TD children. Lastly, we conclude this manuscript with discussion on some of the challenges of fitting these models, as well as possible theoretical challenges when working with covariate adjusted mixed membership models.


\section{Covariate Adjusted Functional Mixed Membership Model}
\label{sec: CAFMMM}

Functional data analysis (FDA) focuses on the statistical analysis of sample paths from a continuous stochastic process $f: \mathcal{T} \rightarrow \mathbb{R}$, where $\mathcal{T}$ is a compact subset of $\mathbb{R}^d$ \citep{wang2016functional}. In FDA, one commonly assumes that the random functions are elements of a Hilbert space, or more specifically that the random functions are square-integrable functions $\left( f \in L^2 \text{ or } \int_{\mathcal{T}} \mid f(t)\mid ^2 \text{d}t < \infty \right)$. In this manuscript, we assume that the observed sample paths are 
generated from a Gaussian process (GP), 
fully characterized by a mean function, 
$\mu(t) = \mathbb{E}\left(f(t)\right)$, and a covariance function, $C(s,t) = \text{Cov}\left(f(s), f(t)\right)$, for $t,s \in \mathcal{T}$. Since mixed membership models can be considered a generalization of finite mixture models, we will show how the finite mixture of regressions and mixture of experts models relate to our proposed mixed membership models. We relate our contribution to covariate adjusted product partition models only superficially, due to the significant differences in both theory and implementation for this model class 
\citep{muller2011product, park2010bayesian}. For the theoretical developments discussed in this section, we will assume that the number of clusters or features, $K$, is known \textit{a-priori}. Although the number of clusters or features is often unknown, Section \ref{sec: sim1} shows that the use of information criteria or simple heuristic methods such as the ``elbow'' method can be informative in choosing the number of features.

\subsection{Mixture of Regressions, Mixture of Experts and Covariate Adjusted Mixed Membership}
\label{sec: CovClust}

Functional clustering generally assumes that each sample path is drawn from one of $K$ underlying cluster-specific sub-processes \citep{james2003clustering, chiou2007functional, jacques2014model}. In the GP framework, assuming that $f^{(1)}, \dots, f^{(K)}$ are the $K$ underlying cluster-specific sub-processes with corresponding mean  $\left(\mu^{(1)}, \dots \mu^{(K)}\right)$ and covariance functions $\left(C^{(1)}, \dots, C^{(K)}\right)$, the general form of a GP finite mixture model is typically expressed as:
\begin{equation}
    \label{eq: FMM}
    P\left(f_i \mid \rho^{(1:K)}, \mu^{(1:K)}, C^{(1:K)}\right) = \sum_{k=1}^K \rho^{(k)}\;\mathcal{GP}\left(f_i\mid \mu^{(k)},\, C^{(k)}\right),
\end{equation}
where $\rho^{(k)}$ ($\sum_{k=1}^K \rho^{(k)} =1$) are the mixing proportions and $f_i$ are the sample paths for $i = 1, \dots, N$. 
Let $\mathbf{x}_i = \left(X_{i1} \dots X_{iR}\right)^\prime \in \mathbb{R}^R:= \mathcal{X}$ be the covariates of interest associated with the $i^{th}$ observation. 
Extending the multivariate finite mixture of regressions model \citep{mclachlan2019finite, faria2010fitting, grun2007applications, khalili2007variable, hyun2023modeling, devijver2015finite} to a functional setting, the covariate adjustments would be encoded through the mean by defining a set of cluster-specific regression functions $\mu^{(k)}(\mathbf{x}_i): \mathcal{X}\times \mathcal{T} \rightarrow \mathbb{R}$. The mixture of GPs in (\ref{eq: FMM}) would then be generalized to include covariate information as follows:
\begin{equation}
    \label{eq: CAGPMOR}
    P\left(f_i \mid \mathbf{x}_i, \rho^{(1:K)}, \mu^{(1:K)}, C^{(1:K)} \right) =  \sum_{k=1}^K \rho^{(k)}\mathcal{GP}\left(f_i\mid \mu^{(k)}(\mathbf{x}_i),\,C^{(k)}\right),
\end{equation}
where cluster specific GPs are now assumed to depend on mean functions $\mu^{(k)}(\mathbf{x})$, which are allowed to vary over the covariate space.
Alternative representations of covariate adjusted mixtures have been proposed to allow the mixing proportions to depend on covariate information through generalized regression functions $\rho^{(k)}(\mathbf{x}_i)$ \citep{jordan1994hierarchical} or implicitly defining covariate adjusted partitions through a cohesion prior \citep{park2010bayesian}.  

In both the finite mixture of regressions and mixture of experts settings, covariate adjustment does not change the overall interpretation of the finite mixture modeling framework, and the underlying assumption is that a function $f_i$ exhibits uncertain membership to \emph{only one of $K$} well-defined subprocesses $f^{(1:K)}$. Following \citet{marcoFunctional}, we contrast this representation of functional data with one where a sample path $f_i$ is allowed mixed membership to \emph{one or more} well-defined sub-processes $f^{(1:K)}$. In the mixed membership framework, the underlying subprocesses $f^{(1:K)}$ are best interpreted as latent functional features, which may partially manifest in a specific path $f_i$.


Our representation of a functional mixed membership model hinges on the introduction of a new set of latent variables $\mathbf{z}_i = (Z_{i1},\ldots, Z_{iK})'$, where $Z_{ik} \in [0,1]$ and $\sum_{k=1}^K Z_{ik} = 1$. Given the latent mixing variables $\mathbf{z}_i$, each sample path is assumed to follow the following sampling structure:
\begin{equation}
    f_i \mid \mathbf{Z}_1,\dots, \mathbf{Z}_N =_d \sum_{k=1}^K Z_{ik} f^{(k)}.
    \label{eq: FMM_f}
\end{equation}
Thus we can see that under the functional mixed membership model, each sample path is assumed to come from a convex combination of the underlying GPs, $f^{(k)}$. To avoid unwarranted restrictions on the implied covariance structure, the functional mixed membership model does not assume that the underlying GPs are mutually independent. Thus, we will let $C^{(k,j)}$ represent the cross-covariance function between the $k^{th}$ GP and the $j^{th}$ GP, for $1 \le k \ne j \le K$. Letting $\boldsymbol{\mathcal{C}}$ be the collection of covariance and cross-covariance functions, we can specify the sampling model of the functional mixed membership model as
\begin{equation}
    \label{eq: GPMMM}
    f_i \mid \mathbf{z}_i, \mu^{(1:K)}, \boldsymbol{\mathcal{C}} \sim \mathcal{GP}\left(\sum_{k=1}^K Z_{ik}\mu^{(k)}, \sum_{k=1}^K Z_{ik}^2 C^{(k)} + \sum_{k=1}^K \sum_{k'\ne k} Z_{ik}Z_{ik'} C^{(k,k')}\right).
\end{equation}
Leveraging the mixture of regressions framework in (\ref{eq: CAGPMOR}),  the proposed covariate adjusted 
functional mixed membership model becomes:
\begin{equation}
    \label{eq: CAGPMMM}
    f_i \mid \mathbf{X}, \mathbf{z}_1,\dots, \mathbf{z}_N, \mu^{(1:K)}(\mathbf{x}_i), \boldsymbol{\mathcal{C}} \sim \mathcal{GP}\left(\sum_{k=1}^K Z_{ik}\mu^{(k)}(\mathbf{x}_i), \sum_{k=1}^K Z_{ik}^2 C^{(k)} + \sum_{k=1}^K \sum_{k'\ne k} Z_{ik}Z_{ik'} C^{(k,k')}\right).
\end{equation}
In Section \ref{sec: MKL}, we review the multivariate Karhunen-Loève (KL) construction \citep{ramsay2005principal, happ2018multivariate}, which allows us to have a joint approximation of the covariance structure of the $K$ underlying GPs. Using the joint approximation, we are able to concisely represent the $K$ GPs, facilitating inference on higher-dimensional functions such as surfaces. Using the KL decomposition, we are able to fully specify our proposed covariate adjusted functional mixed membership model (Equation \ref{eq: CAGPMMM}) in Section \ref{sec: ModelSpecification}. 

\subsection{Multivariate Karhunen-Loève Characterization}
\label{sec: MKL}
The proposed modeling framework in Equation (\ref{eq: CAGPMMM}) 
depends on a set of mean and covariance functions, which are naturally  
represented infinite-dimensional parameters. Thus, in order to 
work in finite dimensions, we assume that the underlying GPs are smooth and lie in the $P$-dimensional subspace, $\boldsymbol{\mathcal{S}} \subset L^2(\mathcal{T})$, spanned by a set of linearly independent square-integrable basis functions, $\{b_1, \dots, b_P\}$ ($b_p: \mathcal{T} \rightarrow \mathbb{R}$). Although the choice of basis functions is user-defined, the basis functions must be uniformly continuous in order to satisfy Lemma 2.2 of \citet{marcoFunctional}. In this manuscript, we will utilize B-splines for all case studies and simulation studies.

The assumption that $f^{(k)} \in \boldsymbol{\mathcal{S}}$ allows us to turn an infinite-dimensional problem into a finite-dimensional problem, making traditional inference tractable. Although tractable, accounting for covariance and cross-covariance functions separately leads to a model that needs $\mathcal{O}(K^2P^2)$ parameters to represent the overall covariance structure. While the number of clusters, $K$, supported in applications is typically small, the number of basis functions, $P$, 
tends to follow the observational grid resolution, and may therefore be quite large. This is particularly true when one considers higher-dimensional functional data, such as surfaces, which necessitate a substantial number of basis functions, making it computationally expensive to fit models. 
Thus, we will use the multivariate KL decomposition \citep{ramsay2005principal, happ2018multivariate} to reduce the number of parameters needed to estimate the covariance surface to $\mathcal{O}(KPM)$, where $M$ is the number of eigenfunctions used to approximate the covariance structure. A more detailed derivation of a KL decomposition for functions in $\boldsymbol{\mathcal{S}}$ can be found in \citet{marcoFunctional}.

To achieve a concise joint representation of the $K$ latent GPs, we define a multivariate GP, which we denote by $\mathbf{f}(\mathbf{t})$, such that 
\begin{equation}
    \nonumber
    \mathbf{f}(\mathbf{t}) := \left\{f^{(1)}\left(t^{(1)}\right), f^{(2)}\left(t^{(2)}\right), \dots, f^{(K)}\left(t^{(K)}\right)\right\},
\end{equation}
where $\mathbf{t} = \left(t^{(1)}, t^{(2)}, \dots, t^{(K)}\right)$ and $t^{(1)}, t^{(2)}, \dots, t^{(K)} \in \mathcal{T}$. Since $\boldsymbol{\mathcal{S}} \subset L^2$ is a Hilbert space, with the inner product defined as the $L^2$ inner product, we can see that $\mathbf{f} \in \boldsymbol{\mathcal{H}} := \boldsymbol{\mathcal{S}} \times \boldsymbol{\mathcal{S}} \times \dots \times \boldsymbol{\mathcal{S}} := \boldsymbol{\mathcal{S}}^K$, where $\boldsymbol{\mathcal{H}}$ is defined as the direct sum of the Hilbert spaces $\boldsymbol{\mathcal{S}}$, making $\boldsymbol{\mathcal{H}}$ also a Hilbert space. Since $\boldsymbol{\mathcal{H}}$ is a Hilbert space and the covariance operator of $\mathbf{f}$, denoted $\boldsymbol{\mathcal{K}}$, is a positive, bounded, self-adjoint, and compact operator, we know that there exists a complete set of eigenfunctions $\boldsymbol{\Psi}_1, \dots, \boldsymbol{\Psi}_{KP} \in \boldsymbol{\mathcal{H}}$ and the corresponding eigenvalues $\lambda_1 \ge \dots \ge \lambda_{KP} \ge 0$ such that $\boldsymbol{\mathcal{K}}\boldsymbol{\Psi}_p = \lambda_p \boldsymbol{\Psi}_p,$ for $p = 1, \dots, KP$. Using the eigenpairs of the covariance operator, we can rewrite $f^{(k)}$ as
$$ f^{(k)}(t) = \mu^{(k)}(\mathbf{x}, t)  + \sum_{m=1}^{KP} \chi_m \left(\sqrt{\lambda_m}\Psi_m^{(k)}(t)\right),$$
where $\chi_m \sim \mathcal{N}(0,1)$ and $\Psi_m^{(k)}(t)$ is the $k^{th}$ element of $\boldsymbol{\Psi}_m(\mathbf{t})$. Since $\Psi_m^{(k)}(t)\in \boldsymbol{\mathcal{S}}$, we know that there exists $\phi_m \in \mathbb{R}^P$ such that $\sqrt{\lambda_m}\Psi_m^{(k)}(t) = \boldsymbol{\phi}_{km}'\mathbf{B}(t)$, where $\mathbf{B}'(t):= [b_1(t), b_2(t), \dots, b_P(t)]$. Similarly, since $\mu^{(k)}(\mathbf{x}, t) \in \boldsymbol{\mathcal{S}}$, we can introduce a mapping $\mathbf{g}: \mathbb{R}^R \rightarrow \mathbb{R}^P$, such that $\mu^{(k)}(\mathbf{x}, t) = \mathbf{g}_k(\mathbf{x})' \mathbf{B}(t)$. Therefore, we arrive at the general form of our decomposition:
\begin{equation}
    \label{eq: KL_decomp}
    f^{(k)}(t) = \mathbf{g}_k(\mathbf{x})' \mathbf{B}(t)  + \sum_{m=1}^{KP} \chi_m \boldsymbol{\phi}_{km}'\mathbf{B}(t).
\end{equation}
Using this decomposition, our covariance and mean structures can be recovered such that $C^{(k,k')}(s,t) =  \mathbf{B}'(s) \left( \sum_{m=1}^{KP} \boldsymbol{\phi}_{km}\boldsymbol{\phi}'_{k'm}\right)\mathbf{B}(t)$ and $\mu^{(k)}(\mathbf{x}, t) = \mathbf{g}_k(\mathbf{x})' \mathbf{B}(t)$, for $1\le k,k'\le K$ and $s,t \in \mathcal{T}$. To reduce the dimensionality of the problem, we will only use the first $M$ eigenpairs to approximate the $K$ stochastic processes. Although traditional functional principal component analysis (FPCA) will choose the number of eigenfunctions based on the proportion of variance explained, the same strategy cannot be employed in functional mixed membership models because the allocation parameters of the model are typically not known. Therefore, we suggest selecting a large value of $M$, and then proceeding with stable estimation through prior regularization.

\subsection{Model and Prior Specification}
\label{sec: ModelSpecification}
In this section, we fully specify the covariate adjusted functional mixed membership model utilizing a truncated version of the KL decomposition, specified in Equation \ref{eq: KL_decomp}. We start by first specifying how the covariates of interest influence the mean function.  
Under the standard function-on-scalar regression framework \citep{faraway1997regression, brumback1998smoothing}, we have $\mu^{(k)}(\mathbf{x}_i,t) = \beta_{k0} + \sum_{r =1}^R X_{ir}\beta_{kr}(t)$ for $k = 1, \dots, K$. Since we assumed that $\mu^{(k)}(\mathbf{x}_i,t) \in \boldsymbol{\mathcal{S}}$ for $k = 1, \dots, K$, we know that $\beta_{k0}, \dots, \beta_{kR} \in \boldsymbol{\mathcal{S}}$. Therefore, there exist $\boldsymbol{\nu}_k \in \mathbb{R}^P$ and $\boldsymbol{\eta}_k \in \mathbb{R}^{P \times R}$ such that 
\begin{equation}
    \label{eq: CAMean}
    \mu^{(k)}(\mathbf{x}_i,t) = \boldsymbol{\nu}_k'\mathbf{B}(t) + \left(\boldsymbol{\eta}_k\mathbf{x}_i'\right)'\mathbf{B}(t).
\end{equation} 
Under a standardized set of covariates, $\boldsymbol{\nu}_k$ specifies the population-level mean of the $k^{th}$ feature and $\boldsymbol{\eta}_k$ encodes the covariate dependence of the $k^{th}$ feature. Thus, assuming the mean structure specified in Equation \ref{eq: CAMean} and using a truncated version of the KL decomposition specified in Equation \ref{eq: KL_decomp}, we can specify the sampling model of our covariate adjusted functional mixed membership model. 

Let $\{\mathbf{Y}_i(.)\}_{i=1}^N$ be the observed sample paths that we want to model to the $K$ features, $f^{(k)}$, conditionally on the covariates of interest, $\mathbf{x}_i$. Since the observed sample paths are observed at only a finite number of points, we will let $\mathbf{t}_i = [t_{i1}, \dots, t_{in_i}]'$ denote the time points at which the $i^{th}$ function was observed. Without loss of generality, we will define the sampling distribution over the finite-dimensional marginals of $\mathbf{Y}_i(\mathbf{t}_i)$. Using the general form of our proposed model defined in Equation \ref{eq: CAGPMMM}, we have 
\begin{equation}
    \label{eq: likelihood}
    \mathbf{Y}_i(\mathbf{t}_i)\mid \boldsymbol{\Theta}, \mathbf{X} \sim \mathcal{N}\left\{ \sum_{k=1}^K Z_{ik}\left(\mathbf{S}'(\mathbf{t}_i) \left(\boldsymbol{\nu}_k + \boldsymbol{\eta}_k \mathbf{x}_i'\right) + \sum_{m=1}^M\chi_{im}\mathbf{S}'(\mathbf{t}_i) \boldsymbol{\phi}_{km}\right),\; \sigma^2 \mathbf{I}_{n_i}\right\},
\end{equation}
where $\mathbf{S}(\mathbf{t}_i) = [\mathbf{B}(t_1) \cdots \mathbf{B}(t_{n_i})] \in \mathbb{R}^{P \times n_i}$ and $\boldsymbol{\Theta}$ is the collection of the model parameters. As defined in Section \ref{sec: CAFMMM}, $Z_{ik}$ are variables that lie on the unit simplex, such that $Z_{ik} \in (0,1)$ and $\sum_{k=1}^K Z_{ik} = 1$. From this characterization, we can see that each observation is modeled as a convex combination of realizations from the $K$ features with additional Gaussian noise, represented by $\sigma^2$. If we integrate out the $\chi_{im}$ variables, for $i = 1, \dots, N$ and $m = 1, \dots, M$, we arrive at the following likelihood:
\begin{equation}
    \label{eq: likelihoodIntChi}
    \mathbf{Y}_i(\mathbf{t}_i)\mid  \boldsymbol{\Theta}_{-\chi},\mathbf{X} \sim \mathcal{N}\left\{\sum_{k=1}^KZ_{ik}\mathbf{S}'(\mathbf{t}_i) \left(\boldsymbol{\nu}_k + \boldsymbol{\eta}_k \mathbf{x}_i'\right),\; \mathbf{V}(\mathbf{t}_i, \mathbf{z}_i) + \sigma^2\mathbf{I}_{n_i}\right\},
\end{equation}
where $\boldsymbol{\Theta}_{-\chi}$ is the collection of the model parameters excluding the $\chi_{im}$ parameters ($i = 1, \dots, N$ and $m = 1, \dots, M$) and $\mathbf{V}(\mathbf{t}_i, \mathbf{z}_i) =  \sum_{k=1}^K\sum_{k'=1}^K Z_{ik}Z_{ik'}\left\{\mathbf{S}'(\mathbf{t}_i)\sum_{m=1}^{M}\left(\boldsymbol{\phi}_{km}\boldsymbol{\phi}'_{k'm}\right)\mathbf{S}(\mathbf{t}_i)\right\}$. Equation \ref{eq: likelihoodIntChi} illustrates that the proposed covariate adjusted functional mixed membership model can be expressed as an additive model; the mean structure is a convex combination of the feature-specific means, while the covariance can be written as a weighted sum of covariance functions and cross-covariance functions. 

To have an adequately expressive and scalable model, we approximate the covariance surface of the $K$ features using $M$ scaled \textit{pseudo-eigenfunctions}. In this framework, orthogonality will not be imposed on the $\boldsymbol{\phi}_{km}'\mathbf{B}(t)$ parameters, making them pseudo-eigenfunctions instead of the eigenfunctions described in Section \ref{sec: MKL}. From a modeling prospective, this allows us to sample on an unconstrained space, facilitating better Markov chain mixing and easier sampling schemes. Although direct inference on the eigenfunctions is no longer available, a formal analysis can still be conducted by reconstructing the posterior samples of the covariance surface and calculating eigenfunctions from the posterior samples. To avoid overfitting of the covariance functions, we follow \citet{marcoFunctional} by using the multiplicative gamma process shrinkage prior proposed by \citet{bhattacharya2011sparse} to achieve adaptive regularized estimation of the covariance structure. Therefore, letting $\phi_{kpm}$ be the $p^{th}$ element of $\boldsymbol{\phi}_{km}$, we have the following:

$$\phi_{kpm}\mid \gamma_{kpm}, \tilde{\tau}_{mk} \sim \mathcal{N}\left(0, \gamma_{kpm}^{-1}\tilde{\tau}_{mk}^{-1}\right), \;\;\; \gamma_{kpm} \sim \Gamma\left(\nu_\gamma /2 , \nu_\gamma /2\right), \;\;\; \tilde{\tau}_{mk} = \prod_{n=1}^m \delta_{nk},$$
$$ \delta_{1k}\mid a_{1k} \sim \Gamma(a_{1k}, 1), \;\;\; \delta_{jk}\mid a_{2k} \sim \Gamma(a_{2k}, 1), \;\;\; a_{1k} \sim \Gamma(\alpha_1, \beta_1), \;\;\; a_{2k} \sim \Gamma(\alpha_2, \beta_2),$$
where $1 \le k \le K$, $1 \le p \le P$, $1 \le m \le M$, and $2 \le j \le M$. By letting $\alpha_2 > \beta_2$, we can 
show that $\mathbb{E}(\tilde{\tau}_{mk}) > \mathbb{E}(\tilde{\tau}_{m'k})$ for $1 \le m < m' \le M$, leading to the prior on $\phi_{kpm}$ having stochastically decreasing variance as $m$ increases. This will have a regularizing effect on the posterior draws of $\phi_{kpm}$, making $\phi_{kpm}$ more likely to be close to zero as $m$ increases.

In functional data analysis, we often desire smooth mean functions to protect against overfit models. Therefore, we use a first-order random walk penalty proposed by \citet{lang2004bayesian} on the $\boldsymbol{\nu}_k$ and $\boldsymbol{\nu}_k$ parameters to promote adaptive smoothing of the mean function of the features in our model. Therefore, we have that $P(\boldsymbol{\nu}_k\mid \tau_{\boldsymbol{\nu}_k}) \propto \exp\left(-\frac{\tau_{\boldsymbol{\nu}_k}}{2}\sum_{p =1}^{P - 1}\left(\nu_{pk}- {\nu}_{(p+1)k}\right)^2\right),$
 for $k = 1, \dots, K$, where $\tau_{\boldsymbol{\nu}_k} \sim \Gamma(\alpha_{\boldsymbol{\nu}}, \beta_{\boldsymbol{\nu}})$ and $\nu_{pk}$ is the $p^{th}$ element of $\boldsymbol{\nu}_k$. Similarly, we have that 
 $P(\{\eta_{prk}\}_{p=1}^P\mid \tau_{\boldsymbol{\eta}_{rk}}) \propto \exp\left(-\frac{\tau_{\boldsymbol{\eta}_{rk}}}{2}\sum_{p =1}^{P - 1}\left(\eta_{prk}- {\eta}_{(p+1)rk}\right)^2\right),$
 for $k = 1, \dots, K$ and $r = 1, \dots, R$, where $\tau_{\boldsymbol{\eta}_{rk}} \sim \Gamma(\alpha_{\boldsymbol{\eta}}, \beta_{\boldsymbol{\eta}})$ and $\eta_{prk}$ is the $p^{th}$ row and $r^{th}$ column of $\boldsymbol{\eta}_{k}$. Following previous mixed membership models \citep{heller2008statistical, marcoFunctional, marcoMultivariate}, we assume that 
 $\mathbf{z}_i\mid \boldsymbol{\pi}, \alpha_3 \sim_{iid} Dir(\alpha_3\boldsymbol{\pi})$, $\boldsymbol{\pi} \sim Dir(\mathbf{c}_\pi)$, and $\alpha_3 \sim Exp(b)$. Lastly, we assume that $\sigma^2 \sim IG(\alpha_0,  \beta_0)$. The posterior distributions of the parameters in our model, as well as a sampling scheme with tempered transitions, can be found in Section 2 of the Supplementary Materials. A covariate adjusted model where the covariance is also dependent on the covariates of interest can be found in Section 4 of the Supplementary Materials.

\subsection{Model Identifiability}
\label{sec: ModelIdentifiability}

Mixed membership models face multiple identifiability problems due to their increased flexibility over traditional clustering models \citep{chen2022learning, marcoFunctional, marcoMultivariate}. Like traditional clustering models, mixed membership models also face the common \textit{label switching} problem, where an equivalent model can be formulated by permuting the labels or allocation parameters. Although this is one source of non-identifiability, relabelling algorithms can be formulated from the work of \citet{stephens2000dealing}. More complex identifiability problems arise since the allocation parameters are now continuous variables on the unit simplex, rather than binary variables like in clustering. Specifically, an equivalent model can be constructed by rotating and/or changing the volume of the convex polytope constructed by the allocation parameters through linear transformations of the allocation parameters \citep{chen2022learning}. Therefore, geometric assumptions must be made on the allocation parameters to ensure that the mixed membership models are identifiable. 

One way to ensure that we have an identifiable model up to a permutation of the labels is to assume that the \textit{separability condition} holds \citep{pettit1990conditional, donoho2003does, arora2012learning, azar2001spectral, chen2022learning}. The separability condition assumes that at least one observation belongs completely to each of the $K$ features. These observations that belong only to one feature act as ``anchors'', ensuring that the allocation parameters are identifiable up to a permutation of the labels.  Although the separability condition is conceptually simple and relatively easy to implement, it makes strong geometric assumptions on the data-generating process for mixed membership models with three or more features. Weaker geometric assumptions, known as the \textit{sufficiently scattered} condition \citep{chen2022learning}, that ensure an identifiable model in mixed membership models with 3 or more features are discussed in \citet{chen2022learning}. Although the geometric assumptions are relatively weak, implementing these constraints is nontrivial. When extending multivariate mixed membership models to functional data and introducing a covariate-dependent mean structure, ensuring an identifiable mean and covariance structure requires further assumptions. 
 
\begin{lemma}
\label{lemma: identifiable}
Consider a $K$-feature covariate adjusted functional mixed membership model as specified in Equation \ref{eq: likelihoodIntChi}. The parameters $\boldsymbol{\nu}_k$, $\boldsymbol{\eta}_k$, $Z_{ik}$, $\sum_{m=1}^M\left(\boldsymbol{\phi}_{km}\boldsymbol{\phi}'_{k'm}\right)$, and $\sigma^2$ are identifiable up to a permutation of the labels (i.e. label switching), for $k,k' = 1, \dots, K$, $i = 1,\dots, N$, and $m = 1, \dots, M$, given the following assumptions:
\begin{enumerate}
    \item $\mathbf{X}$ is full column rank with $\mathbf{1}$ not in the column space of $\mathbf{X}$.
    \item The separability condition or the sufficiently scattered condition holds on the allocation matrix. Moreover, there exist at least $\frac{K^2 + K}{2}$ observations $(N \ge \frac{K^2 + K}{2})$, with allocation parameters such that the matrix $\mathbf{C} \in \mathbb{R}^{N \times \frac{K^2 + K}{2}}$ is full column rank, where the $i^{th}$ row of $\mathbf{C}$ is specified as $\mathbf{c}_i = [Z_{i1}^2, \dots, Z_{iK}^2, 2Z_{i1}Z_{i2},\dots, 2Z_{i1}Z_{iK}, 2Z_{i2}Z_{i3}, \dots,  2Z_{i(K-1)}Z_{iK}]$.
    \item The sample paths $\mathbf{Y}_i(\mathbf{t}_i)$ are regularly sampled so that $n_i > P$. Moreover, we will assume that the basis functions are B-splines with equidistant knots.
\end{enumerate}
\end{lemma}

Lemma \ref{lemma: identifiable} states a set of sufficient conditions that lead to an identifiable mean and covariance structure up to a permutation of the labels. The proof of Lemma \ref{lemma: identifiable} can be found in Section 1 of the Supporting Materials. 
 The first assumption is similar to the assumption needed in the linear regression setting to ensure identifiability; however, the intercept term is not included in the design matrix, as the $\boldsymbol{\nu}_k$ parameters already account for the intercept. From the second assumption we can see that we assume geometric assumptions on the allocation parameters, provide a lower bound for the number of functional observations, and we assume that the matrix $\mathbf{C}$ is full column rank. The matrix $\mathbf{C}$ is constructed using the allocation parameters, and although the condition on $\mathbf{C}$ may seem restrictive, it often holds in practice. We note that this condition may not hold if there are not enough unique allocation parameter values or if most of your allocation parameters lie on a lower-dimensional subspace of the unit $(K-1)$-simplex. Although the individual $\boldsymbol{\phi}_{km}$ parameters are not identifiable in our model, an eigen analysis can still be performed by constructing posterior draws of the covariance structure and calculating the eigenvalues and eigenfunctions of the posterior draws. Section \ref{sec: simulation_study} provides empirical evidence that the mean and covariance structure converge to the truth as more observations are accrued.

\subsection{Relationship to Function-on-Scalar Regression}
\label{sec: FoS_Regression}
Function-on-scalar regression is a common method in FDA that allows the mean structure of a continuous stochastic process to be dependent on scalar covariates. In function-on-scalar regression, we often assume that the response is a GP, and that the covariates of interest are vector-valued. A comprehensive review of the broader area of functional regression can be found in \citet{ramsay2005functional} and \citet{morris2015functional}. While there have been many advancements and generalizations \citep{krafty2008varying, reiss2010fast, goldsmith2015generalized, kowal2020bayesian} of function-on-scalar regression since the initial papers of \citet{faraway1997regression} and \citet{brumback1998smoothing}, the general form of function-on-scalar regression can be expressed as follows:
\begin{equation}
    \label{eq: FoS_likelihood}
    Y(t) = \mu(t) + \sum_{r=1}^R X_{r} \beta_{r}(t) + \epsilon(t); \;\;\;\; t \in \mathcal{T},
\end{equation}
where $Y(t)$ is the response function evaluated at $t$, $\beta_r(\cdot)$ is the functional coefficient representing the effect that the $r^{th}$ covariate ($X_r$) has on the mean structure, and $\epsilon$ is a mean-zero Gaussian process with covariance function $\mathcal{C}$. The function $\mu:\mathcal{T} \rightarrow \mathbb{R}$ in Equation (\ref{eq: FoS_likelihood}) represents the mean of the GP when all of the covariates, $X_{ir}$ are set to zero. Unlike the traditional setting for multiple linear regression in finite-dimensional vector spaces, function-on-scalar regression requires the estimation of the infinite-dimensional functions $\mu$ and $\beta_1, \dots, \beta_R$ from a finite number of observed sample paths at a finite number of points ($\mathbf{Y}_i(\mathbf{t}_i)$ for $i = 1,\dots, N$ and $\mathbf{t}_i = [t_{i1}, \dots, t_{in_i}]'$).

To make inference tractable, we assume that the data lie in the span of a finite set of basis functions, which will allow us to expand $\mu$ and $\beta_1, \dots, \beta_R$ as a finite sum of the basis functions. The set of basis functions can be specified using data-driven basis functions or by specifying the basis functions \textit{a-priori}. If the basis functions are specified \textit{a-priori}, common choices of basis functions are B-splines and wavelets due to their flexibility, as well as Fourier series for periodic functions. Alternatively, if the use of data-driven basis functions is desired, a common choice is to use the eigenfunctions of the covariance operator as basis functions. In order to estimate the eigenfunctions, functional principal component analysis \citep{shang2014survey} is often performed and the obtained estimates of the eigenfunctions are used. Functional principal component analysis faces a similar problem in that the objective is to estimate eigenfunctions using only a finite number of sample paths observed at a finite number of points. To solve this problem, \citet{rice1991estimating} proposes using a splines basis to estimate smooth eigenfunctions, while \citet{yao2005functional} proposes using local linear smoothers to estimate the smooth eigenfunctions. Therefore, even using data-driven basis functions require smoothing assumptions, suggesting similar results between data-driven basis functions and a reasonable set of \textit{a-priori} specified basis functions paired with a penalty to prevent overfitting.

Specifying the basis functions \textit{a-priori} ($b_1(t),\dots, b_P(t)$), and letting $\mathbf{Y}_i(\mathbf{t}_i)$ be the observed sample paths at points $\mathbf{t}_i = [t_{i1}, \dots, t_{in_i}]'$ ($i =1 ,\dots, N$), we can simplify Equation (\ref{eq: FoS_likelihood}) to get 
\begin{equation}
    \label{eq: FoS_basis}
    \mathbf{Y}_i(\mathbf{t}_i) = \mathbf{S}'(\mathbf{t}_i)\tilde{\boldsymbol{\nu}} + \mathbf{S}'(\mathbf{t}_i)\tilde{\boldsymbol{\eta}} \mathbf{x}_i' + \boldsymbol{\epsilon}_i(\mathbf{t}_i),
\end{equation}
where $\tilde{\boldsymbol{\nu}} \in \mathbb{R}^P$, $\tilde{\boldsymbol{\eta}} \in \mathbb{R}^{P \times R}$, and $\mathbf{S}(\mathbf{t}_i) = [\mathbf{B}(t_1) \cdots \mathbf{B}(t_{n_i})] \in \mathbb{R}^{P \times n_i}$ are the set of basis function evaluated at the time points of interest. As specified in the previous sections, $\mathbf{B}'(t):= [b_1(t), b_2(t), \dots, b_P(t)]$. Equation (\ref{eq: FoS_basis}) shows that the function $\mu(.)$ evaluated at the points $\mathbf{t}_i$ can be represented by $\mathbf{S}'(\mathbf{t}_i)\tilde{\boldsymbol{\nu}}$, and similarly the functional coefficients $\beta_1(\cdot), \dots, \beta_R(\cdot)$ can be represented by $\mathbf{S}'(\mathbf{t}_i)\tilde{\boldsymbol{\eta}}$. Therefore, we are left to estimate $\tilde{\boldsymbol{\nu}}$, $\tilde{\boldsymbol{\eta}}$, and the parameters associated with the covariance function of $\epsilon(\cdot)$, denoted $\mathcal{C}$.

The covariance function $\mathcal{C}$ represents the within-function covariance structure of the data. In the simplest case, we often assume that $\boldsymbol{\epsilon}_i(\mathbf{t}_i) \sim \mathcal{N}(\mathbf{0}, \tilde{\sigma}^2\mathbf{I}_{n_i})$, which means that we only need $\tilde{\sigma}^2$ to specify $\mathcal{C}$. In more complex models \citep{faraway1997regression, krafty2008varying}, one may make less restrictive assumptions and assume that the covariance $\boldsymbol{\epsilon}_i(\mathbf{t}_i) \sim \mathcal{N}(\mathbf{0}_{n_i}, \tilde{V}(\mathbf{t}_i) + \tilde{\sigma}^2\mathbf{I}_{n_i})$, where $\tilde{V}(\cdot)$ is a low-dimensional approximation of a smooth covariance surface using a truncated eigendecomposition. Although functional regression usually assumes that the functions are independent, functional regression models have been proposed to model between-function variation, or cases where observations can be correlated \citep{morris2006wavelet, staicu2010fast}.

Assuming a relatively general covariance structure as in \citet{krafty2008varying}, the function-on-scalar model assumes the following distributional assumptions on our sample paths:
\begin{equation}
    \label{eq: FoS_dist}
    \mathbf{Y}_i(\mathbf{t}_i) \mid \tilde{\boldsymbol{\nu}}, \tilde{\boldsymbol{\eta}}, \tilde{\mathbf{V}}(\mathbf{t}_i),  \tilde{\sigma}^2, \mathbf{X} \sim \mathcal{N}\left\{\mathbf{S}'(\mathbf{t}_i) \left(\tilde{\boldsymbol{\nu}} + \tilde{\boldsymbol{\eta}} \mathbf{x}_i'\right),\; \tilde{\mathbf{V}}(\mathbf{t}_i) + \tilde{\sigma}^2\mathbf{I}_{n_i}\right\}.
\end{equation}
From Equation (\ref{eq: FoS_dist}) it is apparent that the proposed covariate adjusted mixed membership model specified in Equation (\ref{eq: likelihoodIntChi}) is closely related to function-on-scalar regression. The key difference between the two models is that the covariate adjusted mixed membership model does not assume a common mean and covariance structure across all observations conditionally on the covariates of interest. Instead, the covariate adjusted mixed membership model allows each observation to be modeled as a convex combination of $K$ underlying features. Each feature is allowed to have different mean and covariance structures, meaning that the covariates are not assumed to have the same affect on all observations. By allowing this type of heterogeneity in our model, we can conduct a more granular analysis and identify subgroups that interact differently with the covariates of interest. Alternatively, if there are subgroups of the population that interact differently with the covariates of interest, then the results from a function-on-scalar regression model will be confounded, as the effects will likely be averaged out in the analysis. The finite mixture of experts model and the mixture of regressions model offer a reasonably granular level of insight, permitting inference at the sub-population level. However, they do not provide the individual-level inference achievable with the proposed covariate adjusted functional mixed membership model.


\section{Simulation Studies}
\label{sec: simulation_study}
In this section, we will discuss the results of two simulation studies. The first simulation study explores the empirical convergence properties of the mean, covariance, and allocation structures of the proposed covariate adjusted functional mixed membership model. Along with correctly specified models, this simulation study also studies the convergence properties of over-specified and under-specified models. The second simulation study explores the use of information criteria in choosing the number of features in a covariate adjusted functional mixed membership model.

\subsection{Simulation Study 1: Structure Recovery}
\label{sec: sim1}
In the first simulation study, we explore the empirical convergence properties of the proposed covariate adjusted functional mixed membership models. Specifically, we generate data from a covariate adjusted functional mixed membership model and see how well the proposed framework can recover the true mean, covariance, and allocation structures. To evaluate how well we can recover the mean, covariance, and cross-covariance functions, we calculate the relative mean integrated square error (R-MISE), which is defined as $\text{R-MISE} = \frac{\int \{f(t) - \hat{f}(t)\}^2 \text{d}t}{\int f(t)^2 \text{d}t} \times 100\%$ or $\text{R-MISE} = \frac{\int_t\int_x\{f(t,x) - \hat{f}(t,x)\}^2 \text{d}x\text{d}t}{\int_t\int_x f(t,x)^2 \text{d}x\text{d}t} \times 100\%$ in the case of a covariate adjusted model. In this simulation study, $\hat{f}(t)$ will be the posterior median obtained from our posterior samples. To measure how well we recover the allocation structure, $Z_{ik}$, we calculated the root-mean-square error (RMSE). In addition to studying the empirical convergence properties of correctly specified models, we also included a scenario where we fit a covariate adjusted functional mixed membership model, when the generating truth had no covariate dependence. Conversely, we also studied the scenario where we fit a functional mixed membership model with no covariates, when the generating truth was generated from a covariate adjusted functional mixed membership model with one covariate. 

\begin{figure}
    \centering
    \includegraphics[width = .99\textwidth]{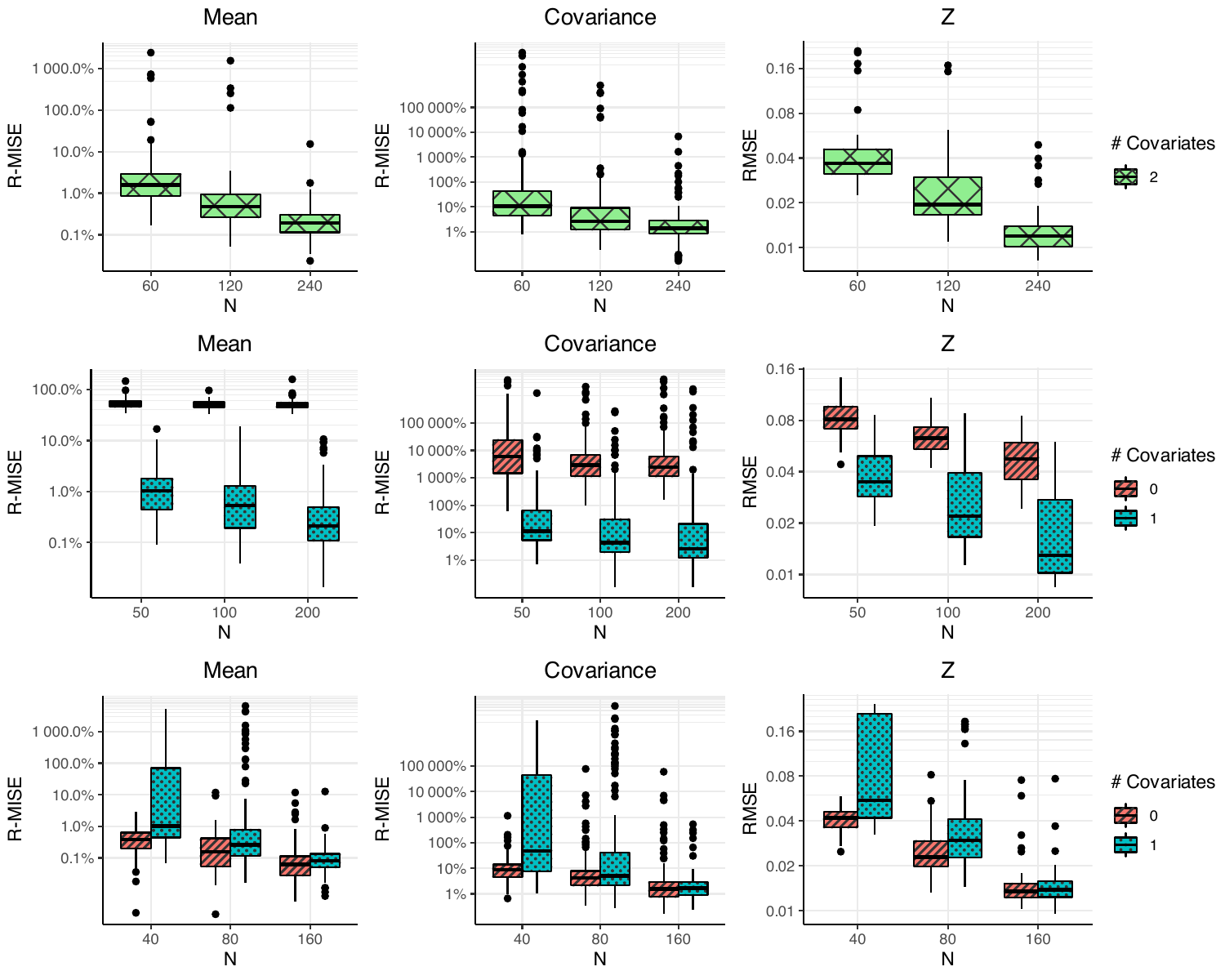}
    \caption{Recovery of the mean, covariance, and allocation structure under the following three data-generating scenarios: (1) data generated from a two-feature covariate adjusted functional mixed membership model with two covariates (Top Row), (2) data generated from a two-feature covariate adjusted functional mixed membership model with one covariate (Middle Row), and (3) data generated from a two-feature functional mixed membership model (Bottom Row).}
    \label{fig: sim1}
\end{figure}

Figure \ref{fig: sim1} contains a visualization of the performance metrics from each of the five scenarios considered in this simulation study. We can see that when the model is specified correctly, it does a good job in recovering the mean structure, even with relatively few observations. On the other hand, a relatively large number of observations are needed to recover the covariance structure when we have a correctly specified model. This simulation study also shows that we pay a penalty in terms of statistical efficiency when we over-specify a model; however, the over-specified model still shows signs of convergence to the true mean, covariance, and allocation structures. Conversely, an under-specified model shows no signs of converging to the true mean and covariance structure as we get more observations. Additional details on how the simulation was conducted, as well as more detailed visualizations of the simulation results, can be found in Section 3.1 of the Supplementary Materials.

\subsection{Simulation Study 2: Information Criteria}
\label{sec: sim2}
When considering the use of a mixed membership model to gain insight into the heterogeneity found in a dataset, a practitioner must specify the number of features to use in the mixed membership model. Correctly specifying the number of features is crucial, as an over-specification will lead to challenges in interpretability of the results, and an under-specification will lead to a model that is not flexible enough to model the heterogeneity in the model. Although information criteria such as BIC and heuristics such as the elbow-method have been shown to be informative in choosing the number of parameters in mixed membership models \citep{marcoFunctional, marcoMultivariate}, it is unclear whether these results extend to covariate adjusted models. In this case, we consider the performance of AIC \citep{akaike1974new}, BIC \citep{schwarz1978estimating}, DIC \citep{spiegelhalter2002bayesian, celeux2006deviance}, and the elbow-method in choosing the number of features in a covariate adjusted functional mixed membership model with one covariate. The information criteria used in this section are specified in Section 3.2 of the Supplementary Materials. As specified, the optimal model should have the smallest AIC, the largest BIC, and the smallest DIC.

\begin{figure}
    \centering
    \includegraphics[width = .99\textwidth]{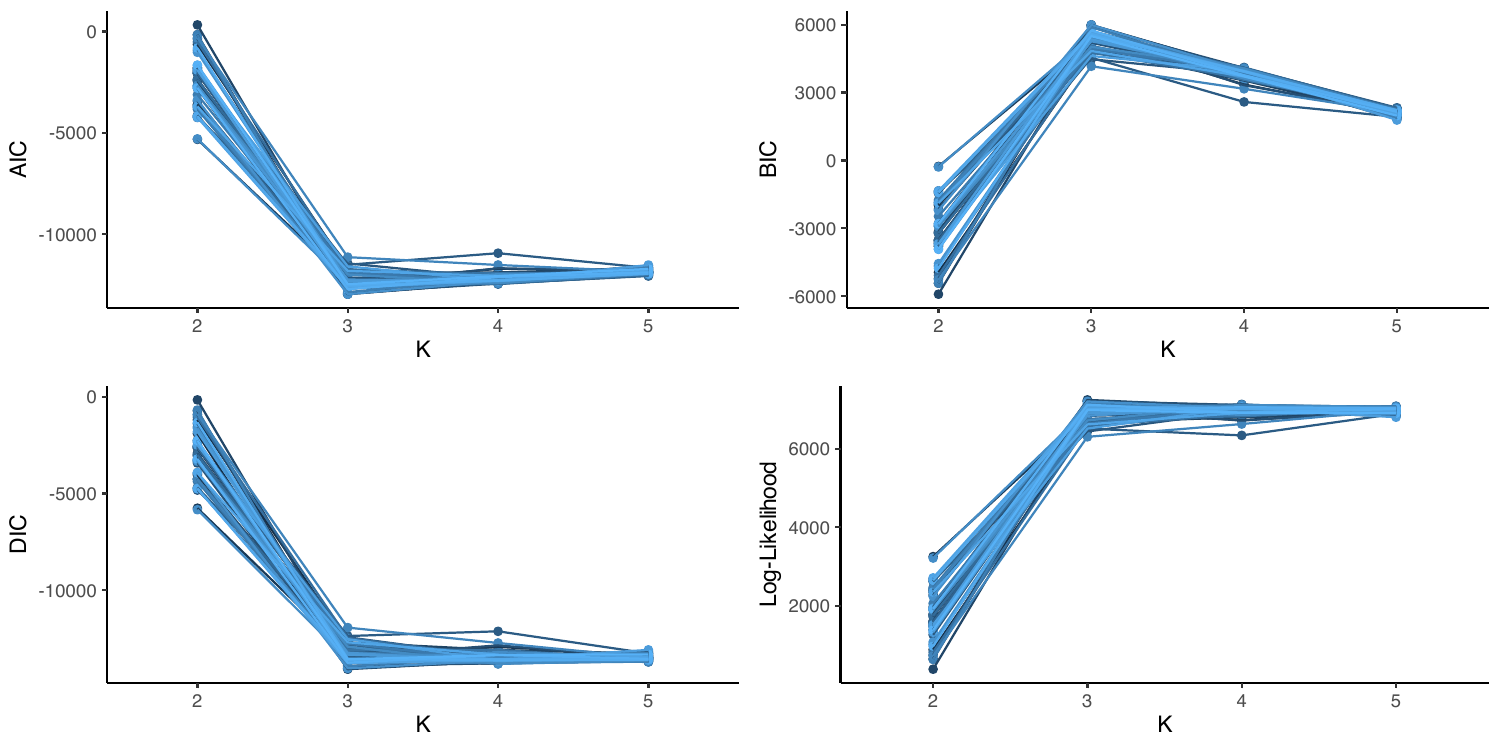}
    \caption{Estimated values of AIC, BIC, DIC, and the log-likelihood for the 50 simulated datasets. The true datasets were generated from a covariate adjusted functional mixed membership with three features.}
    \label{fig: sim2}
\end{figure}

To evaluate the performance of the information criteria and the elbow method, 50 datasets were generated from a covariate adjusted functional mixed membership model with three features and one covariate. Each dataset was analyzed using four covariate adjusted functional mixed membership models; each with a single covariate and a varying number of features ($K = 2, 3, 4, 5$). Figure \ref{fig: sim2} shows the performance of the three information criteria. Similarly to \citet{marcoFunctional} and \citet{marcoMultivariate}, we can see that BIC had the best performance, picking the three-feature model all 50 times. From the simulation study, we found that AIC and DIC tended to choose more complex models, with AIC and DIC choosing the correctly specified model 68\% and 26\% of the time, respectively. Looking at the log-likelihood plot, we can see that there is a distinct elbow at $K =3$, meaning that using heuristic methods such as the elbow method tends to be informative in choosing the number of features in covariate adjusted functional mixed membership models. Thus, we recommend using BIC and the elbow-method in conjunction to choose the number of features, as even though BIC correctly identified the model every time in the simulation, there were times when the BIC between the three-feature model and the four-feature model were similar. Additional details on how this simulation study was conducted can be found in Section 3.2 of the Supplementary Materials.

\section{A Case Study in EEG Imaging for Autism Spectrum Disorder}
\label{sec: ASD}
\begin{figure}
    \centering
    \includegraphics[width = 0.99\textwidth]{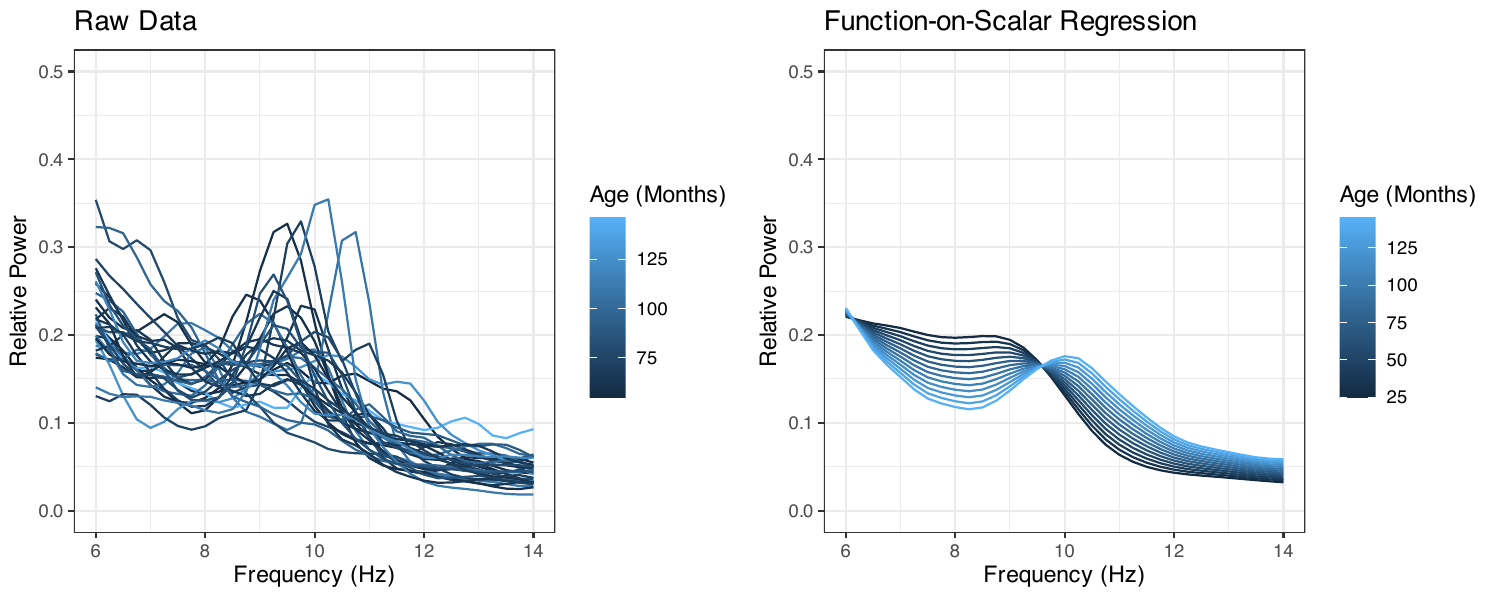}
    \caption{(Left Panel) Alpha frequency patterns averaged over the 25 channels for a sample of EEG recordings from 30 individuals (ASD and TD) with varying ages. (Right Panel) Estimated affects of age on alpha oscillations obtained by fitting a function-on-scalar model.}
    \label{fig: raw_data}
\end{figure}

Autism spectrum disorder (ASD) is a developmental disorder characterized by social communication deficits and restrictive and/or repetitive behaviors \citep{american2013diagnostic}. Although once more narrowly defined, autism is now seen as a spectrum, with some individuals having very mild symptoms, to others that require lifelong support \citep{lord2018autism}. In this case study, we analyze electroencephalogram (EEG) data that were obtained in a resting-state EEG study conducted by \citet{dickinson2018peak}. The study consisted of 58 children who have been diagnosed with ASD between the ages of 2 and 12 years of age, and 39 age-matched typically developing (TD) children, or children who have never been diagnosed with ASD. The children were instructed to view bubbles on a monitor in a dark, sound-attenuated room for 2 minutes, while EEG recordings were taken. The EEG recordings were obtained using a 128-channel HydroCel Geodesic Sensory net, and then interpolated to match the international 10-20 system 25-channel montage. The data were filtered using a band pass of 0.1 to 100 Hz, and then transformed into the frequency domain using a fast Fourier transform. To obtain the relative power, the functions were scaled so that they integrate to 1. Lastly, the relative power was averaged across the 25 channels to obtain a measure of the average relative power. Visualizations of the functional data that characterize the average relative power in the alpha band of frequencies can be seen in Figure \ref{fig: raw_data}. 

In this case study, we will specifically analyze the alpha band of frequencies (neural activity between 6Hz and 14Hz) and compare how alpha oscillations differ between children with ASD and TD children. The alpha band of frequencies has been shown to play a role in neural coordination and communication between distributed brain regions \citep{fries2005mechanism, klimesch2007eeg}. Alpha oscillations are composed of periodic and aperiodic neural activity patterns that coexist in the EEG spectra. Neuroscientists are primarily interested in periodic signals, specifically in the location of a single prominent peak in the spectral density located in the alpha band of frequencies, called the \textit{peak alpha frequency} (PAF). Peak alpha frequency has been shown to be a biomarker of neural development in typically developing children \citep{Rodriguez2017}. Studies have shown that the alpha peak becomes more prominent and shifts to a higher frequency within the first years of life for TD children \citep{Rodriguez2017,scheffler2019covariate}. Compared to TD children, the emergence of a distinct alpha peak and developmental shifts have been shown to be atypical in children with ASD \citep{dickinson2018peak,scheffler2019covariate, marcoFunctional, marcoMultivariate}.

In Section \ref{sec: case_study1}, we analyze how age affects alpha oscillations by fitting a covariate adjusted functional mixed membership model with the log transformation of age as the only covariate. This model provides insight into how alpha frequencies change as children age, regardless of diagnostic group. Additionally, the use of a covariate adjusted functional mixed membership model allows us to directly compare the age-adjusted relative strength of the PAF biomarker between individuals in the study, including between diagnostic groups. This analysis is extended in Section \ref{sec: case_study2} where an analysis on how developmental shifts differ based on diagnostic group. To conduct this analysis, a covariate adjusted functional mixed membership mode was fit with the log transformation of age, diagnostic group, and an interaction between the log transformation of age and diagnostic group as covariates. This analysis provides novel insight into the developmental shifts for children with ASD and TD children. Although these two models may be considered nested models, each model provides unique insight into the data.

\subsection{Modeling Alpha Oscillations Conditionally on Age}
\label{sec: case_study1}

In the first part of this case study, we are primarily interested in modeling the developmental changes in alpha frequencies. As stated in \citet{haegens2014inter}, shifts in PAF can confound the alpha power measures, demonstrating the need for a model that controls for the age of the child. In this subsection, we fit a covariate adjusted functional mixed membership model using the log transformation of age as the covariate. Using conditional predictive ordinates and comparing the pseudomarginal likelihoods, we found that a log transformation of age provided a better fitting model than using a model with age untransformed. Although this model does not use information on the diagnostic group, we are also interested in characterizing the differences in the observed alpha oscillations between individuals across diagnostic groups.  

\begin{figure}
    \centering
    \includegraphics[width = .99\textwidth]{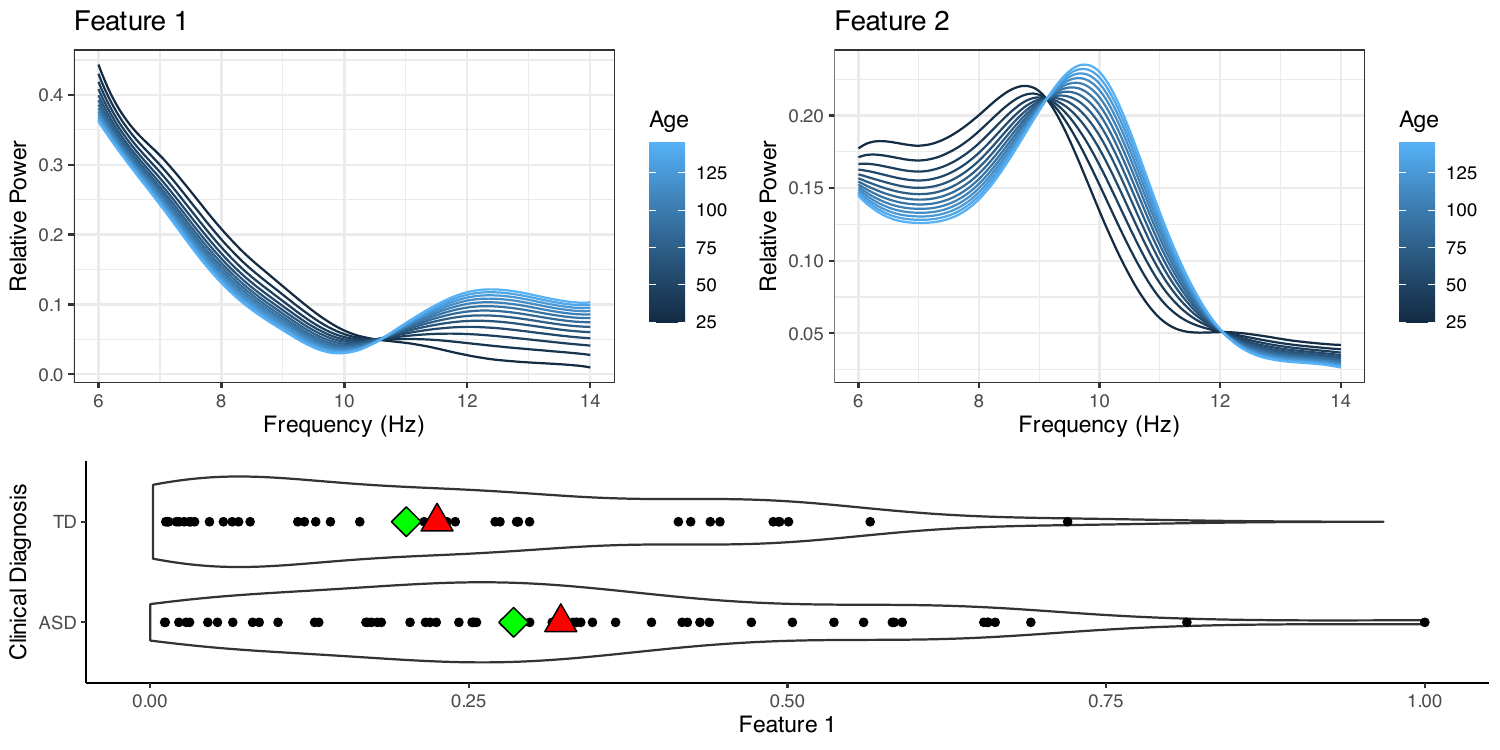}
    \caption{(Top Panels) Estimates of the mean functions of the two functional features conditional on Age. (Bottom Panel) Estimates of the allocation parameters found by fitting a covariate adjusted functional mixed membership model. Diagnostic group level means and medians of the allocation to the first feature is depicted as red triangles and green diamonds, respectively.}
    \label{ASD_means}
\end{figure}

From Figure \ref{ASD_means}, we can see that the first feature consists mainly of aperiodic neural activity patterns, which are commonly referred to as a $1/f$ trend or pink noise. On the other hand, the second feature can be interpreted as a distinct alpha peak, which is considered periodic neural activity. For children that heavily load on the second feature, we can see that as they age, the alpha peak increases in magnitude and the PAF shifts to a higher frequency, which has been observed in many other studies \citep{haegens2014inter, Rodriguez2017, scheffler2019covariate}.  From a clinical perspective, it is also valuable to compare children's alpha power conditional on age, since we know that there are developmental changes in alpha oscillations as children age. From Figure \ref{ASD_means}, we can also see that on average children with ASD have less attenuated alpha peaks compared to their age-adjusted TD counterpart. 

One key feature of this model is that it allows us to directly compare individuals in the TD group and ASD group on the same spectrum. However, to do this, we had to assume that the alpha oscillations of children with ASD and TD children can be represented as a continuous mixture of the same two features shown in Figure \ref{ASD_means}. Just as developmental shifts in PAF can confound the measures of alpha power \citep{haegens2014inter}, the assumption that the alpha oscillations of children with ASD and TD children can be represented by the same two features can also confound the results found in this section if the assumption is shown to be incorrect. In Section \ref{sec: case_study2}, we relax this assumption and allow the functional features to differ according to the diagnostic group.

\subsection{Modeling Alpha Oscillations Conditionally on Age and Diagnostic Group}
\label{sec: case_study2}
Previous studies have shown that the emergence of alpha peaks and the developmental shifts in frequency are atypical in children with ASD \citep{dickinson2018peak, scheffler2019covariate, marcoFunctional, marcoMultivariate}. Therefore, it may not be realistic to assume that the alpha oscillations of children with ASD and TD children can be represented by the same two features. In this section, we will fit a covariate adjusted functional mixed membership model using the log transformation of age, clinical diagnosis, and an interaction between the log transformation of age and clinical diagnosis as covariates of interest. By including the interaction between the log transformation of age and diagnostic group, we allow for differences in the developmental changes of alpha oscillations between diagnostic groups.

\begin{figure}
    \centering
    \includegraphics[width = 0.99\textwidth]{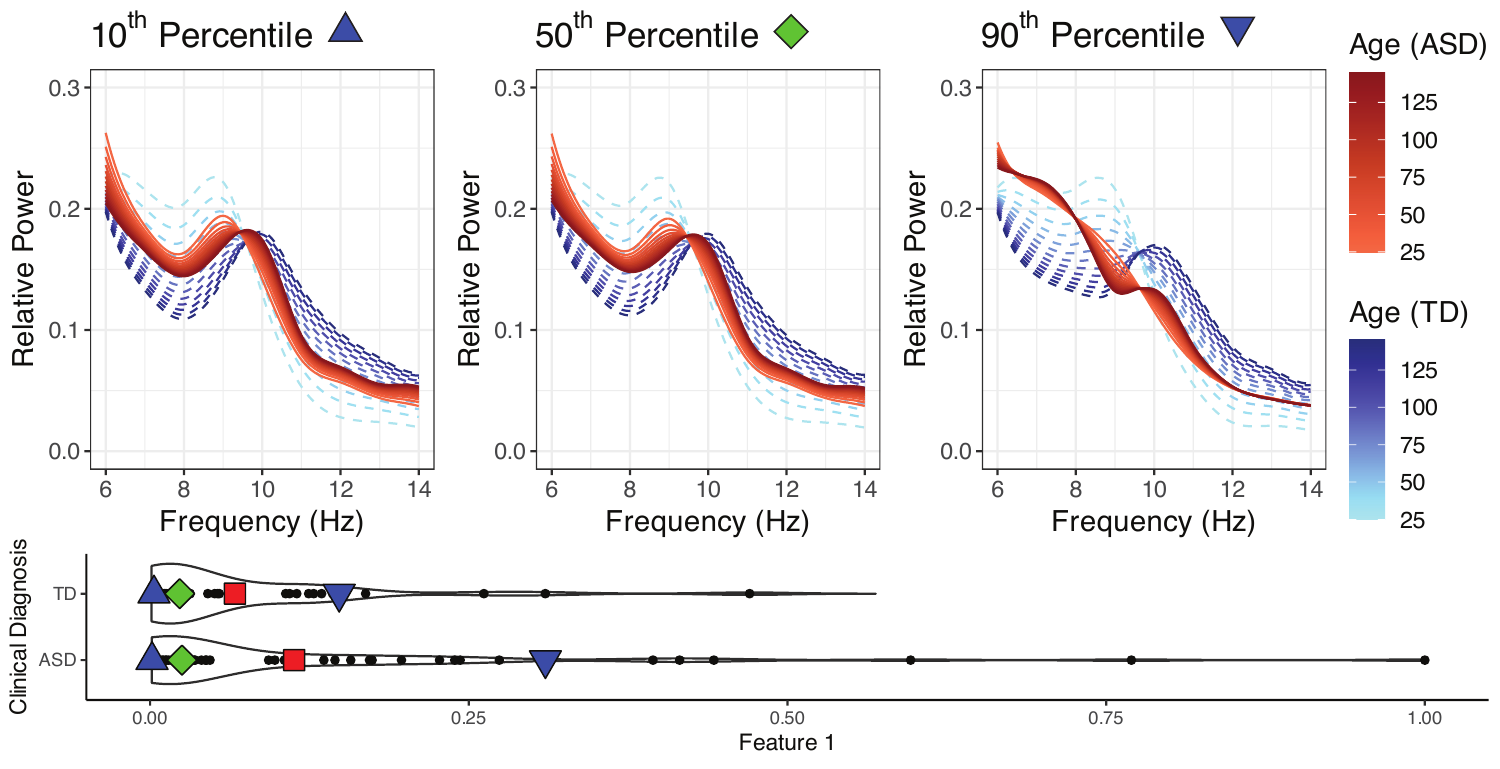}
    \caption{(Top Panels) Estimates of the developmental trajectories conditional on percentiles of the estimated allocation parameters by clinical diagnosis. (Bottom Panel) Estimates of the allocation parameters stratified by clinical diagnosis, with red squares representing the group-specific means, green diamonds representing the group-specific medians, the blue triangles representing the group-specific $10{th}$ percentiles, and the inverted blue triangles representing the group-specific $90^{th}$ percentiles.}
    \label{fig: ASD_TD_percentiles}
\end{figure}

Compared to the model used in Section \ref{sec: case_study1}, this model is more flexible; it allows different features for each diagnostic group and different developmental trajectories for each diagnostic group. Although this model is more expressive, the comparison of individuals across diagnostic groups is no longer straightforward. The model in Section \ref{sec: case_study1} did not use any diagnostic group information, which allowed us to easily compare TD children with ASD children as the distribution of the developmental trajectory depended only on the allocation parameters. On the other hand, comparing individuals across diagnostic groups is not as simple as looking at the inferred allocation parameters, as the underlying features are group-specific. Therefore, two individuals with the same allocation parameters that are in diagnostic groups will likely have completely different developmental trajectories. Moreover, in the model used in Section \ref{sec: case_study1}, we knew that at least one observation belonged to each feature, making the features interpretable as the extreme cases observed. However, for the more flexible model used in this subsection, there will not be at least one observation belonging to each feature for each diagnostic group. We can see that in the TD group, no individual has a loading of more than 50\% on the first feature, meaning that the first TD-specific feature cannot be interpreted as the most extreme case observed in the TD group. Therefore, direct interpretation of the group-specific features is less useful, and instead we will focus on looking at the estimated trajectories of individuals conditional on empirical percentiles of the estimated allocation structure. Visualizations of the mean structure for the two features can be found in the Section 3.3 Supplementary Materials.

From Figure \ref{fig: ASD_TD_percentiles}, we can see the estimated developmental trajectories of TD children and children with ASD, conditional on different empirical percentiles of the estimated allocation structure. Looking at the estimated developmental trajectories of children with a group-specific median allocation structure, we can see that both TD children and children diagnosed with ASD have a discernible alpha peak and that the expected PAF shifts to higher frequencies as the child develops. However, compared to TD children, ASD children have more of a $1/f$ trend, and the shift in PAF is not as large in magnitude. Similar differences between TD children and children diagnosed with ASD can be seen when we look at the $10^{th}$ percentile of observations that belong to the first group-specific feature. However, when looking at the group-specific $90^{th}$ percentile, we can see that children with ASD do not have a discernible alpha peak at 25 months old, and a clear alpha peak never really appears as they develop. Alternatively, TD children in the $90^{th}$ percentile still have a discernible alpha peak that becomes more prominent and shifts to higher frequencies as they develop. However, compared to TD individuals in the $10^{th}$ or $50^{th}$ percentile, there is more aperiodic signal ($1/f$ trend) present in the alpha power of TD children in the $90^{th}$ percentile.

\begin{figure}
    \centering
    \includegraphics[width = 0.99\textwidth]{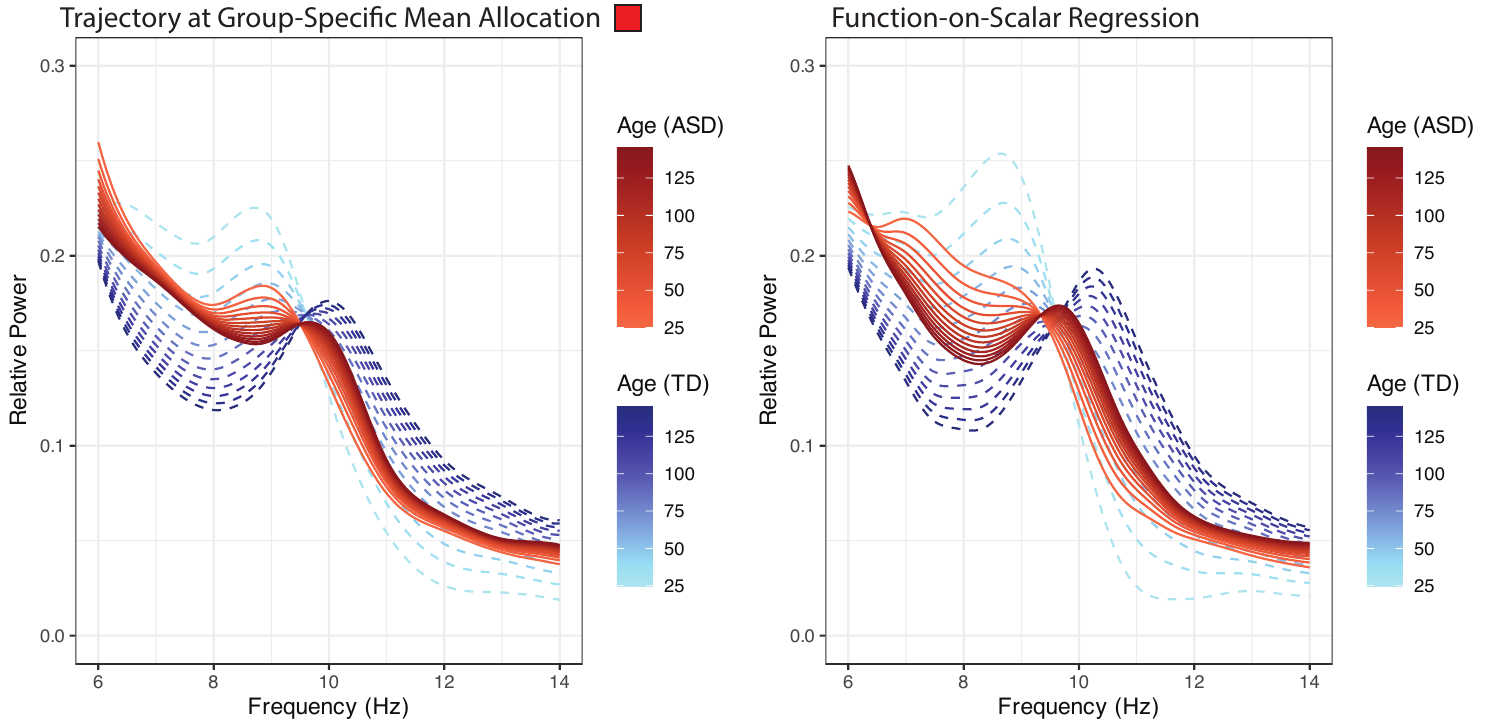}
    \caption{(Left Panel) Estimates of the developmental trajectories stratified by diagnostic group, conditional on the group-specific mean allocation (red squares in Figure \ref{fig: ASD_TD_percentiles}) (Right panel) Estimated population-level development trajectories stratified by diagnostic group, obtained by fitting a function-on-scalar regression model.}
    \label{fig: FoS_TD_vs_ASD}
\end{figure}

As discussed in Section \ref{sec: FoS_Regression}, the proposed covariate adjusted mixed membership model can be thought of as a generalization of function-on-scalar regression. Figure \ref{fig: FoS_TD_vs_ASD} contains the estimated developmental trajectories obtained by fitting a function-on-scalar regression model using the package ``refund'' package in R \citep{goldsmith2016refund}. The results of the function-on-scalar regression coincide with the estimated developmental trajectory of the alpha oscillations obtained from our covariate adjusted mixed membership model, conditional on the group-specific mean allocation (group-specific mean allocations are represented by triangles in the bottom panel of Figure \ref{fig: ASD_TD_percentiles}). As discussed in Section \ref{sec: FoS_Regression}, function-on-scalar regression assumes a common mean for all individuals, controlling for age and diagnostic group. This is a relatively strong assumption in this case, as ASD is relatively broadly defined; with diagnosed individuals having varying degrees of symptoms. Indeed, from the function-on-scalar results, we would only be able to make population-level inference; giving no insight into the heterogeneity of alpha oscillation developmental trajectories for children with ASD (Figure \ref{fig: ASD_TD_percentiles}). Overall, the added flexibility of covariate adjusted mixed membership models allows scientists to have greater insight into the developmental changes of alpha oscillations. Technical details on how the models for Section \ref{sec: case_study1} and Section \ref{sec: case_study2} were fit can be found in Section 3.3 of the Supplementary Materials. Section 3.3 of the Supplementary Materials also discusses model comparison using conditional predictive ordinates \citep{pettit1990conditional, chen2012monte,lewis2014posterior} and pseudomarginal likelihood.

\section{Discussion}
\label{sec: conclusion}

In this manuscript, we extend the framework of the functional mixed membership model proposed by \citet{marcoFunctional} to allow for covariate dependence. This work was primarily motivated by a neurodevelopmental EEG study on alpha oscillations, where alpha oscillations are known to change as children age. Although mixed membership models provided a novel way to quantify the emergence and presence of a developmental biomarker known as an alpha peak \citep{marcoFunctional, marcoMultivariate}, it has been shown that not accounting for developmental shifts in the alpha peak can confound measures of the peak alpha frequency \citep{haegens2014inter}, leading to the need for a covariate adjusted functional mixed membership model. In Section \ref{sec: CAFMMM}, we derive the covariate adjusted functional mixed membership models and compare the framework with common covariate-dependent clustering frameworks, such as the mix of experts model and the mixture of regressions model. After specifying the proposed Bayesian model, we provide a set of sufficient conditions in Lemma \ref{lemma: identifiable} to ensure that the proposed covariate adjusted mixed membership model has identifiable mean, covariance, and allocation structures. Section \ref{sec: simulation_study} contains a simulation study exploring the empirical convergence properties of our model. We conclude by revisiting the neurodevelopmental case study on alpha oscillations in Section \ref{sec: ASD}, where we fit two covariate adjusted functional mixed membership models to gain novel information on the developmental differences of alpha oscillations between children with ASD and TD children. In Section \ref{sec: case_study1}, we fit a covariate adjusted functional mixed membership model adjusting for age, which allowed us to directly compare children with ASD to TD children using the same spectrum while controlling for age. In Section \ref{sec: case_study1}, we fit a covariate adjusted functional mixed membership model adjusting for age and diagnostic group, which no longer allowed us to directly compare children with ASD to TD children; however, we were able to gain novel insight into the developmental trajectory of alpha oscillations for children with ASD.


Although the proposed model was introduced for an arbitrary number of features, and Lemma \ref{lemma: identifiable} holds for an arbitrary number of features, there are computational challenges to working with models with more than three features. As discussed in Section \ref{sec: ModelIdentifiability}, we typically have to assume the separability condition or the sufficiently scattered condition. In a two-feature model, these two assumptions are equivalent and can be enforced by post-processing the Markov chain constructed through MCMC. When considering a model with three or more features, the separability condition becomes a relatively strong geometric assumption on the generating process, making the sufficiently scattered condition more theoretically appealing. For a three-feature model, we can similarly post-process the Markov chain utilizing the work of \citet{parvu2016implementation} to arrive at posterior samples that satisfy the sufficiently scattered assumption. However, ensuring that the sufficiently scattered assumption holds for models with four or more features is still an open problem. Similarly, more work is needed to derive a sampler that obeys the geometric constraints set forth by the sufficiently scattered condition.

As discussed in Section \ref{sec: FoS_Regression}, covariate adjusted functional mixed membership models can be thought of as a more granular version of function-on-scalar regression. Although functional-on-scalar models can typically handle a large number of covariates, with regularized function-on-scalar models \citep{fan2017high, kowal2020bayesian} becoming more popular, the model proposed in this manuscript can handle only a few covariates. Although fitting a covariate adjusted functional mixed membership model with more covariates may be computationally feasible, from the simulation study in Section \ref{sec: sim1}, we can see that we need significantly more observations as we add covariates and lose a lot of statistical efficiency when over-specifying the model. Although one may be tempted to use information criteria to perform variable selection, we urge caution, as the addition of additional covariates can greatly impact the interpretation of the allocation parameters, as seen in Section \ref{sec: case_study2}. Thus, the incorporation of a covariate in the model should be decided based on the modeling goals, prior knowledge of how the covariates affect the observed functions, and computational/statistical efficiency considerations.

As observed in unadjusted mixed membership models \citep{marcoFunctional, marcoMultivariate}, the posterior distribution often has multiple modes, leading to poor posterior exploration using traditional sampling methods. Thus, we use an algorithm similar to Algorithm 1 described in the supplement of \citet{marcoFunctional} to pick a good starting point for our Markov chain. In addition to finding good initial starting points, we also outline a tempered-transition sampling scheme \citep{pritchard2000inference, behrens2012tuning} in Section 2.2 of the Supplementary Materials, which allows us to traverse areas of low posterior probability. An R package for the proposed covariate adjusted functional mixed membership model is available for download at \url{https://github.com/ndmarco/BayesFMMM}.

\section{Funding Acknowledgments}
The authors gratefully acknowledge funding from the NIH/NIMH R01MH122428-01 (DS,DT).

\section{Disclosure}
The authors report that there are no competing interests to declare.

\bibliographystyle{unsrtnat}
\bibliography{CAFMMM}

\pagebreak

\begin{center}
\textbf{\large Supplemental Materials: Covariate Adjusted Functional Mixed Membership Models}
\end{center}
\setcounter{equation}{0}
\setcounter{figure}{0}
\setcounter{table}{0}
\setcounter{page}{1}
\setcounter{section}{0}

\section{Proof of Lemma 2.1}
We will start by defining identifiability and defining some of the notation used in this section. Let $\boldsymbol{\omega} = \{\boldsymbol{\nu}_1, \dots, \boldsymbol{\nu}_K , \boldsymbol{\eta}_1, \dots, \boldsymbol{\eta}_K, \{Z_{j1}, \dots,Z_{jK}\}_{j=1}^N, \{\boldsymbol{\Sigma}_{jk}\}_{1\le j \le k \le K}, \sigma^2\}$, where $\boldsymbol{\Sigma}_{jk} = \sum_{m=1}^M\left(\boldsymbol{\phi}_{jm}\boldsymbol{\phi}'_{km}\right)$. We will say that the parameters $\boldsymbol{\omega}$ are unidentifiable if there exists at least one $\boldsymbol{\omega}^* \ne \boldsymbol{\omega}$  such that $\mathcal{L}(\mathbf{Y}_i(\mathbf{t}_i) \mid \boldsymbol{\omega}, \mathbf{x}_i) = \mathcal{L}(\mathbf{Y}_i(\mathbf{t}_i) \mid \boldsymbol{\omega}^*, \mathbf{x}_i)$ for all sets of observations $\{\mathbf{Y}_i(\mathbf{t}_i)\}_{i=1}^N$ following Assumptions (1)-(3). Otherwise, the parameters $\boldsymbol{\omega}$ are called identifiable. In this case, $\mathcal{L}(\mathbf{Y}_i(\mathbf{t}_i) \mid \boldsymbol{\omega}, \mathbf{x}_i)$ is the likelihood specified in Equation 12 in the main text.

From equation 12 in the main text, we have that
\begin{equation}
\resizebox{.9 \textwidth}{!}{$\mathcal{L}\left(\mathbf{Y}_i(\mathbf{t}_i) \mid \boldsymbol{\omega}, \mathbf{x}_i \right) \propto \text{exp}\left\{-\frac{1}{2} \left(\mathbf{Y}_i(\mathbf{t}_i) - \mu_i(\mathbf{x}_i, \mathbf{t}_i) \right)'\left(\mathbf{V}(\mathbf{t}_i, \mathbf{z}_i) + \sigma^2\mathbf{I}_{n_i} \right)^{-1}\left(\mathbf{Y}_i(\mathbf{t}_i) - \mu_i(\mathbf{x}_i, \mathbf{t}_i) \right) \right\},$}
\end{equation}
where 
$$\mu_i(\mathbf{x}_i, \mathbf{t}_i) = \sum_{k=1}^KZ_{ik}\mathbf{S}'(\mathbf{t}_i) \left(\boldsymbol{\nu}_k + \boldsymbol{\eta}_k \mathbf{x}_i'\right)$$ and $$\mathbf{V}(\mathbf{t}_i, \mathbf{z}_i) =  \sum_{k=1}^K\sum_{k'=1}^K Z_{ik}Z_{ik'}\left\{\mathbf{S}'(\mathbf{t}_i)\sum_{m=1}^{M}\left(\boldsymbol{\phi}_{km}\boldsymbol{\phi}'_{k'm}\right)\mathbf{S}(\mathbf{t}_i)\right\}.$$ Assume that $ \mathcal{L}\left(\mathbf{Y}_i(\mathbf{t}_i) \mid \boldsymbol{\omega}, \mathbf{x}_i \right) = \mathcal{L}\left(\mathbf{Y}_i(\mathbf{t}_i) \mid \boldsymbol{\omega}^*, \mathbf{x}_i \right)$ for all sets of observations $\{\mathbf{Y}_i(\mathbf{t}_i)\}_{i=1}^N$ that follow Assumptions (1)-(3) . Thus we would like to prove that $\boldsymbol{\omega}^* = \boldsymbol{\omega}$ must necessarily be true. Since $\mathcal{L}\left(\mathbf{Y}_i(\mathbf{t}_i) \mid \boldsymbol{\omega}, \mathbf{x}_i \right)$ is written as a quadratic form in $\mathbf{Y}_i(\mathbf{t}_i)$ and $\left(\mathbf{V}(\mathbf{t}_i, \mathbf{z}_i) + \sigma^2\mathbf{I}_{n_i} \right)$ is full rank, we see that the following must necessarily be true:
\begin{enumerate}
    \item \label{equal_mu} $\mu_i^*(\mathbf{x}_i, \mathbf{t}_i) = \mu_i(\mathbf{x}_i, \mathbf{t}_i)$,
    \item \label{equal_var} $\mathbf{V}^*(\mathbf{t}_i, \mathbf{z}_i^*) + (\sigma^2)^*\mathbf{I}_{n_i} = \mathbf{V}(\mathbf{t}_i, \mathbf{z}_i) + \sigma^2\mathbf{I}_{n_i}$,
\end{enumerate}
for $i = 1,\dots, N$. By (\ref{equal_mu}), we have that 
$$\sum_{k=1}^KZ_{ik}\mathbf{S}'(\mathbf{t}_i) \left(\boldsymbol{\nu}_k + \boldsymbol{\eta}_k \mathbf{x}_i'\right) = \sum_{k=1}^KZ_{ik}^*\mathbf{S}'(\mathbf{t}_i) \left(\boldsymbol{\nu}_k^* + \boldsymbol{\eta}_k^* \mathbf{x}_i'\right)\;\;\; (i = 1, \dots, N).$$
Letting $\boldsymbol{\mu}_k = [\boldsymbol{\nu}_k \; \boldsymbol{\eta}_k] \in \mathbb{R}^{P \times (R + 1)}$ and $\mathbf{\tilde{x}}_i = [1 \; \mathbf{x}_i]$ ($\tilde{\mathbf{X}}\in \mathbb{R}^{N \times (R + 1)}$ is the design matrix with the $i^{th}$ row as $\mathbf{\tilde{x}}$), we have
\begin{align}
    \nonumber & \sum_{k=1}^KZ_{ik}\mathbf{S}'(\mathbf{t}_i)\boldsymbol{\mu}_k\mathbf{\tilde{x}}_i'  = \sum_{k=1}^KZ_{ik}^*\mathbf{S}'(\mathbf{t}_i) \boldsymbol{\mu}_k^*\mathbf{\tilde{x}}_i' \;\;\; (i = 1, \dots, N) \\
    \label{eq: equality} \iff & \sum_{k=1}^KZ_{ik}\boldsymbol{\mu}_k\mathbf{\tilde{x}}_i'  = \sum_{k=1}^KZ_{ik}^* \boldsymbol{\mu}_k^*\mathbf{\tilde{x}}_i' \;\;\; (i = 1, \dots, N),
\end{align}
since $n_i \ge P$ by Assumption (3). Since $\tilde{\mathbf{X}}$ is full column rank from Assumption (1), we have that
\begin{equation}
    \label{eq: equality2}
    \sum_{k=1}^K Z_{ik}\boldsymbol{\mu}_k  = \sum_{k=1}^KZ_{ik}^* \boldsymbol{\mu}_k^* \;\;\; (i = 1, \dots, N).
\end{equation}
It is important to note that if Assumption (1) does not hold and $\tilde{\mathbf{X}}$ is not full column rank, we could add any vector in the nullspace of $\tilde{\mathbf{X}}$ to any row of $\boldsymbol{\mu}_k^*$ ($k = 1, \dots, K$) and equation \ref{eq: equality} would still hold. 

We can rewrite Equation \ref{eq: equality2} in matrix form such that
\begin{equation}
    \mathbf{Z} \tilde{\boldsymbol{\mu}} = \mathbf{Z}^* \tilde{\boldsymbol{\mu}}^* 
\end{equation}
where $\tilde{\boldsymbol{\mu}}_r = [\text{vec}(\boldsymbol{\mu}_{1}), \dots, \text{vec}(\boldsymbol{\mu}_{K})]' \in \mathbb{R}^{K \times P(R+1)}$ and $\mathbf{Z} \in \mathbb{R}^{N \times K}$ is the matrix of allocation parameters with $\mathbf{z}_i$ as the $i^{th}$ row of $\mathbf{Z}$. From this we can directly apply the results of \citet{chen2022learning}, to show that $\mathbf{Z} = \mathbf{Z}^*$ and $\tilde{\boldsymbol{\mu}} = \tilde{\boldsymbol{\mu}}^*$ up to a permutation of the labels. Specifically, if the seperability condition holds, then Proposition 1 of \citet{chen2022learning} shows that $\mathbf{Z} = \mathbf{Z}^*$ and $\tilde{\boldsymbol{\mu}} = \tilde{\boldsymbol{\mu}}^*$ up to a permutation of the labels. If the sufficiently scattered condition holds, then Theorem 2 of \citet{chen2022learning} shows that $\mathbf{Z} = \mathbf{Z}^*$ and $\tilde{\boldsymbol{\mu}} = \tilde{\boldsymbol{\mu}}^*$ up to a permutation of the labels. Therefore, assuming that Assumptions (1) - (3) hold, we have $Z_{ik} = Z_{ik}^*$, $\boldsymbol{\nu}_k = \boldsymbol{\nu}_k^*$, and $\boldsymbol{\eta}_k = \boldsymbol{\eta}_k^*$ up to the permutation of the labels, for $k = 1,2$ and $i = 1,\dots, N$.

From (\ref{equal_var}), we have that 
\begin{align}
    \nonumber & \mathbf{V}^*(\mathbf{t}_i, \mathbf{z}_i^*) + (\sigma^2)^*\mathbf{I}_{n_i} = \mathbf{V}(\mathbf{t}_i, \mathbf{z}_i) + \sigma^2\mathbf{I}_{n_i}\\
    \nonumber \iff & \mathbf{V}^*(\mathbf{t}_i, \mathbf{z}_i^*) - \mathbf{V}(\mathbf{t}_i, \mathbf{z}_i) = ((\sigma^2)^* -\sigma^2)\mathbf{I}_{n_i}.
\end{align}
Suppose that $((\sigma^2)^* -\sigma^2) \ne 0$, then we have that $$\text{rank}\left(\mathbf{V}^*(\mathbf{t}_i, \mathbf{z}_i^*) - \mathbf{V}(\mathbf{t}_i, \mathbf{z}_i)\right) = \text{rank}\left(((\sigma^2)^* -\sigma^2)\mathbf{I}_{n_i}\right) > P,$$ by Assumption (3) ($n_i> P$) . Writing $\mathbf{V}(\mathbf{t}_i, \mathbf{z}_i)$ such that 
$$\mathbf{V}(\mathbf{t}_i, \mathbf{z}_i) =  \mathbf{S}'(\mathbf{t}_i)\left\{\sum_{k=1}^K\sum_{k'=1}^K Z_{ik}Z_{ik'}\sum_{m=1}^{M}\left(\boldsymbol{\phi}_{km}\boldsymbol{\phi}'_{k'm}\right)\right\}\mathbf{S}(\mathbf{t}_i),$$
we can see that rank$(\mathbf{V}(\mathbf{t}_i, \mathbf{z}_i)) \le P$, which implies that $\text{rank}\left(\mathbf{V}^*(\mathbf{t}_i, \mathbf{z}_i^*) - \mathbf{V}(\mathbf{t}_i, \mathbf{z}_i)\right) \le P$, leading to a contradiction. Therefore, we have $(\sigma^2)^* = \sigma^2$ and $\mathbf{V}^*(\mathbf{t}_i, \mathbf{z}_i^*)  = \mathbf{V}(\mathbf{t}_i, \mathbf{z}_i)$ up to a permutation of the labels. Assuming no permutation of the labels ($\mathbf{Z} = \mathbf{Z}^*$), we have that
   $$\mathbf{V}(\mathbf{t}_i, \mathbf{z}_i) = \mathbf{V}^*(\mathbf{t}_i, \mathbf{z}_i^*)\iff  \sum_{k=1}^K\sum_{k'=1}^K Z_{ik}Z_{ik'}\boldsymbol{\Sigma}_{kk'} = \sum_{k=1}^K\sum_{k'=1}^K Z_{ik}Z_{ik'}\boldsymbol{\Sigma}_{kk'}^*,$$
since $n_i > P$ for all $i$. Therefore, we have the following system of equations
$$\sum_{k=1}^K Z_{ik}^2 \left(\boldsymbol{\Sigma}_{kk} - \boldsymbol{\Sigma}_{kk}^*\right) + 2\sum_{k=1}^K\sum_{k'>k} Z_{ik}Z_{ik'}\left(\boldsymbol{\Sigma}_{kk'} - \boldsymbol{\Sigma}_{kk'}^*\right) = \mathbf{0},$$
for $i = 1,\dots, N$.
Equality can be proved if we have $\frac{K^2 + K}{2}$ linearly independent equations in our system of equations. Thus, if the coefficient matrix has rank $\frac{K^2 + K}{2}$, then we have $\frac{K^2 + K}{2}$ linearly independent equations in our system of equations. We can see that Assumption (2) gives us that the coefficient matrix, denoted $\mathbf{C}$, has full column rank, meaning that we have $\frac{K^2 + K}{2}$ linearly independent equations in our system of equations. Therefore, we have that $\boldsymbol{\Sigma}_{kk'} = \boldsymbol{\Sigma}_{kk'}^*$ for $1 \le k,k' \le K$, up to a permutation of the labels. Therefore, we have that for $i = 1,\dots, N$ and $k,k' = 1,\dots, K$, the parameters $\boldsymbol{\nu}_k$, $\boldsymbol{\eta}_k$, $Z_{ik}$, $\sum_{m=1}^M\left(\boldsymbol{\phi}_{km}\boldsymbol{\phi}'_{k'm}\right)$, and $\sigma^2$ are identifiable up to a permutation of the labels given Assumptions (1)-(3).

\section{Computation}
\subsection{Posterior Distributions}
\label{sec: posterior_dist}

In this subsection, we will specify the posterior distributions specifically for the functional covariate adjusted mixed membership model proposed in the main manuscript. We will first start with the $\boldsymbol{\phi}_{km}$ parameters, for $j = 1,\dots, K$ and $m = 1, \dots, M$. Let $\mathbf{D}_{\boldsymbol{\phi}_{jm}} = \tilde{\tau}_{\boldsymbol{\phi}_{mj}}^{-1} diag\left(\gamma_{\boldsymbol{\phi}_{j1m}}^{-1}, \dots, \gamma_{\boldsymbol{\phi}_{jPm}}^{-1}\right)$. By letting
$$\begin{aligned}
\mathbf{m}_{\boldsymbol{\phi}_{jm}} = & \frac{1}{\sigma^2} \sum_{i=1}^N \sum_{l = 1}^{n_i}\left(B(t_{il})\chi_{im} \left(y_i(t_{il})Z_{ij} -  Z_{ij}^2 \left(\boldsymbol{\nu}_{j}+ \boldsymbol{\eta}_j\mathbf{x}_i' \right)'B(t_{il}) - Z_{ij}^2\sum_{n \ne m}\chi_{in} \boldsymbol{\phi}_{jn}' B(t_{il})\right. \right.\\
&  \left. \left.  - \sum_{k \ne j} Z_{ij}Z_{ik}\left[\left(\boldsymbol{\nu}_{k}+ \boldsymbol{\eta}_k\mathbf{x}_i' \right)' B(t_{il}) + \sum_{n=1}^M \chi_{in}\boldsymbol{\phi}_{kn}'B(t_{il}) \right] \right) \right),
\end{aligned}$$
and
$$\mathbf{M}_{\boldsymbol{\phi}_{jm}}^{-1} = \frac{1}{\sigma^2}\sum_{i=1}^N \sum_{l = 1 }^{n_i} \left(Z_{ij}^2\chi_{im}^2B(t_{il})B'(t_{il})\right) + \mathbf{D}_{\boldsymbol{\phi}_{jm}}^{-1},$$
 we have that 
$$\boldsymbol{\phi}_{jm} | \boldsymbol{\Theta}_{-\boldsymbol{\phi}_{jm} }, \mathbf{Y}_1, \dots, \mathbf{Y}_N, \mathbf{X} \sim \mathcal{N}\left(\mathbf{M}_{\boldsymbol{\phi}_{jm}}\mathbf{m}_{\boldsymbol{\phi}_{jm}}, \mathbf{M}_{\boldsymbol{\phi}_{jm}}\right).$$

The posterior distribution of $\delta_{{1k}}$, for $k = 1, \dots, K$, is 
$$\begin{aligned}
\delta_{{1k}} | \boldsymbol{\Theta}_{-\delta_{{1k}}}, \mathbf{Y}_1, \dots, \mathbf{Y}_N, \mathbf{X} \sim & \Gamma\left(a_{{1k}} + (PM/2), 1 + \frac{1}{2} \sum_{r=1}^P \gamma_{{k,r,1}}\phi_{k,r,1}^2  \right.  \\
& \left. + \frac{1}{2}\sum_{m=2}^M \sum_{r=1}^P \gamma_{{k,r,m}}\phi_{k,r,m}^2\left( \prod_{j=2}^m \delta_{{jk}} \right)\right).
\end{aligned}.$$
The posterior distribution for $\delta_{{ik}}$, for $i = 2, \dots, M$ and $k = 1, \dots, K$, is 
$$\begin{aligned}
\delta_{{ik}} | \boldsymbol{\Theta}_{-\delta_{{ik}}}, \mathbf{Y}_1, \dots, \mathbf{Y}_N, \mathbf{X}  \sim & \Gamma\Bigg(a_{{2k}} + (P(M - i + 1)/2), 1   \\
& \left. +\frac{1}{2}\sum_{m = i}^M \sum_{r=1}^P \gamma_{\boldsymbol{\xi}_{k,r,m}}\phi_{k,r,m}^2\left( \prod_{j=1; j \ne i}^m \delta_{{jk}} \right)\right).
\end{aligned}$$

The posterior distribution for $a_{{1k}}$ ($k = 1, \dots, K$) is not a commonly known distribution, however we have that
$$P(a_{{1k}}|\boldsymbol{\Theta}_{-a_{{1k}}}, \mathbf{Y}_1, \dots, \mathbf{Y}_N, \mathbf{X}) \propto \frac{1}{\Gamma(a_{{1k}})}\delta_{{1k}}^{a_{{1k}} -1} a_{{1k}}^{\alpha_{1} -1} exp \left\{-a_{{1k}}\beta_{1} \right\}.$$
Since this is not a known kernel of a distribution, we will have to use Metropolis-Hastings algorithm. Consider the proposal distribution $Q(a_{{1k}}'| a_{{1k}}) = \mathcal{N}\left(a_{{1k}}, \epsilon_1\beta_{1}^{-1}, 0, + \infty\right)$ (Truncated Normal) for some small $\epsilon_1 > 0$. Thus the probability of accepting any step is
$$A(a_{{1k}}',a_{{1k}}) = \min \left\{1, \frac{P\left(a_{{1k}}'| \boldsymbol{\Theta}_{-a_{{1k}}'}, \mathbf{Y}_1, \dots, \mathbf{Y}_N, \mathbf{X}\right)}{P\left(a_{{1k}}| \boldsymbol{\Theta}_{-a_{{1k}}}, \mathbf{Y}_1, \dots, \mathbf{Y}_N, \mathbf{X}\right)} \frac{Q\left(a_{{1k}}|a_{{1k}}'\right)}{Q\left(a_{{1k}}'|a_{{1k}}\right)}\right\}.$$

Similarly for $a_{{2k}}$ ($k = 1, \dots, K$), we have
$$P(a_{{2k}} | \boldsymbol{\Theta}_{-a_{{2k}}}, \mathbf{Y}_1, \dots, \mathbf{Y}_N, \mathbf{X}) \propto \frac{1}{\Gamma(a_{{2k}})^{M-1}}\left(\prod_{i=2}^M\delta_{{ik}}^{a_{{2k}} -1}\right) a_{{2k}}^{\alpha_{{2k}} -1} exp \left\{-a_{{2k}}\beta_{2} \right\}.$$
We will use a similar proposal distribution, such that $Q(a_{{2k}}'| a_{{2k}}) = \mathcal{N}\left(a_{{2k}}, \epsilon_2\beta_{2}^{-1}, 0, + \infty\right)$ for some small $\epsilon_2 > 0$. Thus the probability of accepting any step is
$$A(a_{{2k}}',a_{{2k}}) = \min \left\{1, \frac{P\left(a_{{2k}}'| \boldsymbol{\Theta}_{-a_{{2k}}'}, \mathbf{Y}_1, \dots, \mathbf{Y}_N, \mathbf{X}\right)}{P\left(a_{{2k}}| \boldsymbol{\Theta}_{-a_{{2k}}}, \mathbf{Y}_1, \dots, \mathbf{Y}_N, \mathbf{X}\right)} \frac{Q\left(a_{{2k}}|a_{{2k}}'\right)}{Q\left(a_{{2k}}'|a_{{2k}}\right)}\right\}.$$

The posterior distribution for the $\mathbf{z}_i$ parameters are not a commonly known distribution, so we will use the Metropolis-Hastings algorithm. We know that
$$\begin{aligned}
p(\mathbf{z}_i| \boldsymbol{\Theta}_{-\mathbf{z}_i}, \mathbf{Y}_1, \dots, \mathbf{Y}_N, \mathbf{X}) & \propto \prod_{k=1}^K Z_{ik}^{\alpha_3\pi_k - 1}\\
& \times \prod_{l=1}^{n_i} exp\left\{-\frac{1}{2\sigma^2}\left(y_i(t_{il}) -  \sum_{k=1}^K Z_{ik}\left(\left(\boldsymbol{\nu}_k + \boldsymbol{\eta}_k \mathbf{x}_i'\right)'B(t_{il}) \right. \right.\right.\\
& \left. \left. \left.+ \sum_{m=1}^M\chi_{im}\boldsymbol{\phi}_{km}' B(t_{il})\right)\right)^2\right\}.
\end{aligned}$$
We will use $Q(\mathbf{z}_i'| \mathbf{z}_i) = Dir(a_{\mathbf{z}} \mathbf{z}_i)$ for some large $a_{\mathbf{z}} \in \mathbb{R}^+$ as the proposal distribution. Thus the probability of accepting a proposed step is 
$$A(\mathbf{z}_i', \mathbf{z}_i) = \min \left\{1, \frac{P\left(\mathbf{z}_i'| \boldsymbol{\Theta}_{-\mathbf{z}_i}, \mathbf{Y}_1, \dots, \mathbf{Y}_N, \mathbf{X} \right)}{P\left(\mathbf{z}_i| \boldsymbol{\Theta}_{-\mathbf{z}_i}, \mathbf{Y}_1, \dots, \mathbf{Y}_N, \mathbf{X}\right)} \frac{Q\left(\mathbf{z}_i|\mathbf{z}_i'\right)}{Q\left(\mathbf{z}_i'|\mathbf{z}_i\right)}\right\}.$$

Similarly, a Gibbs update is not available for an update of the $\boldsymbol{\pi}$ parameters. We have that 
$$\begin{aligned}
p(\boldsymbol{\pi}|\boldsymbol{\Theta}_{-\boldsymbol{\pi}}, \mathbf{Y}_1,\dots, \mathbf{Y}_N, \mathbf{X}) & \propto \prod_{k=1}^K \pi_k^{c_k - 1} \\
& \times \prod_{i=1}^N\frac{1}{B(\alpha_3\boldsymbol{\pi})}\prod_{k=1}^K Z_{ik}^{\alpha_3\pi_k - 1}.
\end{aligned}$$
Letting out proposal distribution be such that $Q(\boldsymbol{\pi}'| \boldsymbol{\pi}) = Dir(a_{\boldsymbol{\pi}} \boldsymbol{\pi})$, for some large $a_{\boldsymbol{\pi}} \in \mathbb{R}^+$, we have that our probability of accepting any proposal is
$$A(\boldsymbol{\pi}', \boldsymbol{\pi}) = \min \left\{1, \frac{P\left(\boldsymbol{\pi}'| \boldsymbol{\Theta}_{-\boldsymbol{\pi}'}, \mathbf{Y}_1, \dots, \mathbf{Y}_N, \mathbf{X}\right)}{P\left(\boldsymbol{\pi}| \boldsymbol{\Theta}_{-\boldsymbol{\pi}}, \mathbf{Y}_1, \dots, \mathbf{Y}_N, \mathbf{X}\right)} \frac{Q\left(\boldsymbol{\pi}|\boldsymbol{\pi}'\right)}{Q\left(\boldsymbol{\pi}'|\boldsymbol{\pi}\right)}\right\}.$$
The posterior distribution of $\alpha_3$ is also not a commonly known distribution, so we will use the Metropolis-Hastings algorithm to sample from the posterior distribution. We have that 
$$\begin{aligned}
p(\alpha_3|\boldsymbol{\Theta}_{-\alpha_3}, \mathbf{Y}_1, \dots, \mathbf{Y}_N, \mathbf{X}) & \propto e^{-b\alpha_3} \\
& \times \prod_{i=1}^N\frac{1}{B(\alpha_3\boldsymbol{\pi})}\prod_{k=1}^K Z_{ik}^{\alpha_3\pi_k - 1}.
\end{aligned}$$
Using a proposal distribution such that $Q(\alpha_3'|\alpha_3) = \mathcal{N}(\alpha_3, \sigma^2_{\alpha_3}, 0, +\infty)$ (Truncated Normal), we are left with the probability of accepting a proposed state as
$$A(\alpha_3',\alpha_3) = \min \left\{1, \frac{P\left(\alpha_3'| \boldsymbol{\Theta}_{-\alpha_3'}, \mathbf{Y}_1, \dots, \mathbf{Y}_N, \mathbf{X}\right)}{P\left(\alpha_3| \boldsymbol{\Theta}_{-\alpha_3}, \mathbf{Y}_1, \dots, \mathbf{Y}_N, \mathbf{X}\right)} \frac{Q\left(\alpha_3|\alpha_3'\right)}{Q\left(\alpha_3'|\alpha_3\right)}\right\}.$$

Let $\mathbf{P}$ be the following tridiagonal matrix:
$$\mathbf{P}= \begin{bmatrix}
1 & -1 & 0 &  & \\
-1 & 2 & -1 &  &  \\
 & \ddots & \ddots & \ddots&  \\
 &  & -1 & 2 & -1  \\
 &  & 0 & -1 & 1 \\
\end{bmatrix}.$$
Thus, letting
$$\mathbf{B}_{\boldsymbol{\nu}_j} = \left( \tau_{\boldsymbol{\nu}_j}\mathbf{P} + \frac{1}{\sigma^2} \sum_{i =1}^N \sum_{l=1}^{n_i}Z_{ij}^2B(t_{il})B'(t_{il})  \right)^{-1}$$
and
$$\begin{aligned} \mathbf{b}_{\boldsymbol{\nu}_j} & = \frac{1}{\sigma^2}\sum_{i=1}^N\sum_{l=1}^{n_i}Z_{ij}B(t_{il})\left[y_i(t_{il}) - \left(\sum_{k\ne j}Z_{ik}\boldsymbol{\nu}'_{k}B(t_{il})\right) \right.  \\ 
& - \left.\left(\sum_{k=1}^K Z_{ik} \left[\mathbf{x}_i \boldsymbol{\eta}_k' B(t_{il}) + \sum_{m=1}^M\chi_{im} \boldsymbol{\phi}_{kn}'B(t_{il}) \right]\right)\right],
\end{aligned}$$
we have that 
$$\boldsymbol{\nu}_j| \boldsymbol{\Theta}_{-\boldsymbol{\nu}_j}, \mathbf{Y}_1, \dots, \mathbf{Y}_N, \mathbf{X} \sim \mathcal{N}\left(\mathbf{B}_{\boldsymbol{\nu}_j}\mathbf{b}_{\boldsymbol{\nu}_j}, \mathbf{B}_{\boldsymbol{\nu}_j}\right).$$

Let $\boldsymbol{\eta}_{jd}$ denote the $d^{th}$ column of the matrix $\boldsymbol{\eta}_j$. Thus, letting
$$\mathbf{B}_{\boldsymbol{\eta}_{jd}} = \left( \tau_{\boldsymbol{\eta}_{jd}}\mathbf{P} + \frac{1}{\sigma^2} \sum_{i =1}^N \sum_{l=1}^{n_i}Z_{ij}^2 x_{id}^2 B(t_{il})B'(t_{il})  \right)^{-1}$$
and 
$$\begin{aligned} \mathbf{b}_{\boldsymbol{\eta}_{jd}} = & \frac{1}{\sigma^2}\sum_{i=1}^N\sum_{l=1}^{n_i}Z_{ij}x_{id}B(t_{il})\left[y_i(t_{il}) - \left(\sum_{r\ne d}Z_{ij}x_{ir} \boldsymbol{\eta}_{jr}'B(t_{il})\right) - \left(\sum_{k \ne j} Z_{ik} \mathbf{x}_i\boldsymbol{\eta}_k' B(t_{il}) \right) \right.  \\ 
& - \left.\left(\sum_{k=1}^K Z_{ik}\left[\boldsymbol{\nu}_k' B(t_{il}) + \sum_{m=1}^M\chi_{im}\boldsymbol{\phi}_{kn}'B(t_{il}) \right]\right)\right],
\end{aligned}$$
we have that 
$$\boldsymbol{\eta}_{jd}| \boldsymbol{\Theta}_{-\boldsymbol{\eta}_{jd}}, \mathbf{Y}_1, \dots, \mathbf{Y}_N, \mathbf{X} \sim \mathcal{N}\left(\mathbf{B}_{\boldsymbol{\eta}_{jd}}\mathbf{b}_{\boldsymbol{\eta}_{jd}}, \mathbf{B}_{\boldsymbol{\eta}_{jd}}\right).$$

Thus we can see that we can draw samples from the posterior of the parameters controlling the mean structure using a Gibbs sampler. Similarly, we can use a Gibbs sampler to draw samples from the posterior distribution of $\tau_{\boldsymbol{\eta}_{jd}}$ and $\tau_{\boldsymbol{\nu}_j}$. We have that the posterior distributions are
$$\tau_{\boldsymbol{\nu}_j}| \boldsymbol{\Theta}_{-\tau_{\boldsymbol{\nu}_j}}, \mathbf{Y}_1, \dots, \mathbf{Y}_N, \mathbf{X} \sim \Gamma\left(\alpha_{\boldsymbol{\nu}} + P/2, \beta_{\boldsymbol{\nu}} + \frac{1}{2}\boldsymbol{\nu}'_j\mathbf{P}\boldsymbol{\nu}_j\right)$$
and
$$\tau_{\boldsymbol{\eta}_{jd}}| \boldsymbol{\Theta}_{-\tau_{\boldsymbol{\eta}_{jd}}}, \mathbf{Y}_1, \dots, \mathbf{Y}_N, \mathbf{X} \sim \Gamma\left(\alpha_{\boldsymbol{\eta}} + P/2, \beta_{\boldsymbol{\eta}} + \frac{1}{2}\boldsymbol{\eta}'_{jd}\mathbf{P}\boldsymbol{\eta}_{jd}\right),$$
for $ j = 1, \dots, K$ and $d = 1, \dots, R$. The parameter $\sigma^2$ can be updated by using a Gibbs update. If we let 
$$\beta_{\sigma} =\frac{1}{2}\sum_{i=1}^N\sum_{l=1}^{n_i}\left(y_i(t_{il}) -  \sum_{k=1}^K Z_{ik}\left(\left(\boldsymbol{\nu}_k + \boldsymbol{\eta}_k \mathbf{x}_i'\right)'B(t_{il}) + \sum_{n=1}^M\chi_{in}\boldsymbol{\phi}_{kn}'B(t_{il})\right)\right)^2,$$
then we have
$$\sigma^2| \boldsymbol{\Theta}_{-\sigma^2}, \mathbf{Y}_1, \dots, \mathbf{Y}_N, \mathbf{X}  \sim  IG\left(\alpha_0 + \frac{\sum_{i=1}^N n_i}{2} , \beta_0 +\beta_{\sigma}\right).$$
Lastly, we can update the $\chi_{im}$ parameters, for $i = 1, \dots, N$ and $m = 1, \dots, M$, using a Gibbs update. If we let 
$$\begin{aligned}
\mathbf{w}_{im} = & \frac{1}{\sigma^2}\left[\sum_{l=1}^{n_i} \left(\sum_{k = 1}^K Z_{ik}\boldsymbol{\phi}_{km}'B(t_{il})\right)\right. \\
& \left.\left(y_i(t_{il}) - \sum_{k = 1}^K Z_{ik}\left(\left(\boldsymbol{\nu}_k + \boldsymbol{\eta}_k \mathbf{x}_i'\right)'B(t_{il})  + \sum_{n\ne m}\chi_{in}\boldsymbol{\phi}_{kn}'B(t_{il})\right)\right)\right]
\end{aligned}$$

and 
$$\mathbf{W}_{im}^{-1} = 1 + \frac{1}{\sigma^2} \sum_{l=1}^{n_i}\left(\sum_{k = 1}^K Z_{ik}\boldsymbol{\phi}_{km}'B(t_{il})\right)^2,$$
then we have that 
$$\chi_{im}| \boldsymbol{\zeta}_{-\chi_{im}}, \mathbf{Y}_1, \dots, \mathbf{Y}_N, \mathbf{X} \sim \mathcal{N}(\mathbf{W}_{im}\mathbf{w}_{im}, \mathbf{W}_{im}).$$

\subsection{Tempered Transitions}
One of the main computational problems we face in these flexible, unsupervised models is a multi-modal posterior distribution. In order to help the Markov chain move across modes, or traverse areas of low posterior probability, we can utilize tempered transitions.

In this paper, we will be following the works of \citet{behrens2012tuning} and \citet{pritchard2000inference} and only temper the likelihood. The target distribution that we want to temper is usually assumed to be written as
$$p(x) \propto \pi(x)exp\left(-\beta_h h(x)\right),$$
where $\beta_h$ controls how much the distribution is tempered ($1 = \beta_0 < \dots < \beta_h < \dots < \beta_{N_t}$). In this setting, we will assume that the hyperparameters $N_t$ and $\beta_{N_t}$ are user specified, and will depend on the complexity of the model. For more complex or larger models, we will need to set $N_t$ relatively high. In this implementation, we assume the $\beta_h$ parameters to follow a geometric scheme, but in more complex models, $\beta_{N_t}$ may need to be relatively small.

We can rewrite our likelihood for the functional covariate adjusted model to fit the above form:
$$\begin{aligned}
p_h(y_i(t)|\boldsymbol{\Theta}, \mathbf{X}) \propto & exp\left\{- \beta_h\left(\frac{1}{2}log(\sigma^2) + \frac{1}{2\sigma^2} \left(y_i(t) -  \sum_{k=1}^K Z_{ik}\Bigg(\left(\boldsymbol{\nu}_k + \boldsymbol{\eta}_k \mathbf{x}_i'\right)'B(t)\right.\right.\right. \\ 
& + \left. \left.\left. \sum_{n=1}^M\chi_{in}\boldsymbol{\phi}_{k'n}'B(t)\Bigg)\right)^2\right)\right\} \\
 = & \left(\sigma^2\right)^{-\beta_h / 2}exp\left\{-\frac{\beta_h}{2\sigma^2}\left(y_i(t) -  \sum_{k=1}^K Z_{ik}\Bigg(\left(\boldsymbol{\nu}_k + \boldsymbol{\eta}_k \mathbf{x}_i'\right)'B(t) \right. \right. \\ 
 & +  \left. \left. \sum_{n=1}^M\chi_{in}\boldsymbol{\phi}_{k'n}'B(t)\Bigg)\right)^2\right\}.
\end{aligned}$$

Let $\boldsymbol{\Theta}_h$ be the set of parameters generated from the model using the tempered likelihood associated with $\beta_h$. The tempered transition algorithm can be summarized by the following steps:
\begin{enumerate}
    \item Start with initial state $\boldsymbol{\Theta}_0$.
    \item Transition from $\boldsymbol{\Theta}_0$ to $\boldsymbol{\Theta}_1$ using the tempered likelihood associated with $\beta_1$.
    \item Continue in this manner until we transition from $\boldsymbol{\Theta}_{N_t - 1}$ to $\boldsymbol{\Theta}_{N_t}$ using the tempered likelihood associated with $\beta_{N_t}$.
    \item Transition from $\boldsymbol{\Theta}_{N_t}$ to $\boldsymbol{\Theta}_{N_t +1}$ using the tempered likelihood associated with $\beta_{N_t}$.
    \item Continue in this manner until we transition from $\boldsymbol{\Theta}_{2N_t -1}$ to $\boldsymbol{\Theta}_{2N_t}$ using $\beta_1$.
    \item Accept transition from $\boldsymbol{\Theta}_0$ to $\boldsymbol{\Theta}_{2N_t}$ with probability 
    $$\min \left\{1, \prod_{h=0}^{N_t - 1} \frac{\prod_{i=1}^N\prod_{l=1}^{n_i} p_{h+1}(y_i(t_{il})|\boldsymbol{\Theta}_h, \mathbf{X}_i)}{\prod_{i=1}^N\prod_{l=1}^{n_i} p_{h}(y_i(t_{il})|\boldsymbol{\Theta}_h, \mathbf{X}_i)} \prod_{h=N_t + 1}^{2N_t} \frac{\prod_{i=1}^N\prod_{l=1}^{n_i} p_{h}(y_i(t_{il})|\boldsymbol{\Theta}_h, \mathbf{X}_i)}{\prod_{i=1}^N\prod_{l=1}^{n_i} p_{h+1}(y_i(t_{il})|\boldsymbol{\Theta}_h,\mathbf{X}_i)}\right\}$$
    in the functional case, or 
    $$\min \left\{1, \prod_{h=0}^{N_t - 1} \frac{\prod_{i=1}^N\prod_{l=1}^{n_i} p_{h+1}(\mathbf{y}_i|\boldsymbol{\Theta}_h, \mathbf{X}_i)}{\prod_{i=1}^N\prod_{l=1}^{n_i} p_{h}(\mathbf{y}_i)|\boldsymbol{\Theta}_h, \mathbf{X}_i)} \prod_{h=N_t + 1}^{2N_t} \frac{\prod_{i=1}^N\prod_{l=1}^{n_i} p_{h}(\mathbf{y}_i|\boldsymbol{\Theta}_h,\mathbf{X}_i)}{\prod_{i=1}^N\prod_{l=1}^{n_i} p_{h+1}(\mathbf{y}_i|\boldsymbol{\Theta}_h, \mathbf{X}_i)}\right\}$$
    in the multivariate case.
\end{enumerate}
Since we only temper the likelihood, many of the posterior distributions derived in section \ref{sec: posterior_dist} can be utilized. Thus the following posteriors are the only ones that change due to the tempering of the likelihood.
Starting with the $\boldsymbol{\Phi}$ parameters, we have 
$$\begin{aligned}
\left(\mathbf{m}_{\boldsymbol{\phi}_{jm}}\right)_h = & \frac{\beta_h}{(\sigma^2)_h} \sum_{i=1}^N \sum_{l = 1}^{n_i}\left(B(t_{il})(\chi_{im})_h \left(y_i(t_{il})(Z_{ij})_h -  (Z_{ij})_h^2 \left((\boldsymbol{\nu}_{j})_h+ (\boldsymbol{\eta}_j)_h\mathbf{x}_i' \right)'B(t_{il}) \right. \right.\\
&  \left. \left. - (Z_{ij})_h^2\sum_{n \ne m}(\chi_{in})_h (\boldsymbol{\phi}_{jn})_h' B(t_{il}) \right. \right.\\
& \left. \left. - \sum_{k \ne j} Z_{ij}Z_{ik}\left[\left((\boldsymbol{\nu}_{k})_h+ (\boldsymbol{\eta}_k)_h\mathbf{x}_i' \right)' B(t_{il}) + \sum_{n=1}^M \chi_{in} (\boldsymbol{\phi}_{kn})_h'B(t_{il}) \right] \right) \right),
\end{aligned}$$
and
$$\left(\mathbf{M}_{\boldsymbol{\phi}_{jm}}\right)_h^{-1} = \frac{\beta_h}{(\sigma^2)_h}\sum_{i=1}^N \sum_{l = 1 }^{n_i} \left((Z_{ij})_h^2(\chi_{im})_h^2B(t_{il})B'(t_{il})\right) + \left(\mathbf{D}_{\boldsymbol{\phi}_{jm}}\right)_h^{-1},$$
 we have that 
$$\left(\boldsymbol{\phi}_{jm}\right)_h | \boldsymbol{\Theta}_{-\left(\boldsymbol{\phi}_{jm}\right)_h}, \mathbf{Y}_1, \dots, \mathbf{Y}_N, \mathbf{X} \sim \mathcal{N}\left(\left(\mathbf{M}_{\boldsymbol{\phi}_{jm}}\right)_h \left(\mathbf{m}_{\boldsymbol{\phi}_{jm}}\right)_h, \left(\mathbf{M}_{\boldsymbol{\phi}_{jm}}\right)_h\right).$$

As in the untempered case, we have that the posterior distribution $\mathbf{Z}$ parameters under the tempered likelihood is not a commonly known distribution. Therefore, we will use the Metropolis-Hastings algorithm. We have that
$$\begin{aligned}
p((\mathbf{z}_i)_h| \boldsymbol{\Theta}_{-(\mathbf{z}_i)_h}, \mathbf{Y}_1, \dots, \mathbf{Y}_N, \mathbf{X}) & \propto \prod_{k=1}^K (Z_{ik})_h^{(\alpha_3)_h(\pi_k)_h - 1}\\
& \times \prod_{l=1}^{n_i} exp\left\{-\frac{\beta_h}{2(\sigma^2)_h}\left(y_i(t_{il}) -  \sum_{k=1}^K (Z_{ik})_h\left(\left((\boldsymbol{\nu}_k)_h + (\boldsymbol{\eta}_k)_h \mathbf{x}_i'\right)'B(t_{il}) \right. \right.\right.\\
& \left. \left. \left.+ \sum_{m=1}^M(\chi_{im})_h(\boldsymbol{\phi}_{km})_h'B(t_{il})\right)\right)^2\right\}.
\end{aligned}$$
We will use $Q((\mathbf{z}_i)_h'| (\mathbf{z}_i)_h) = Dir(a_{\mathbf{z}} (\mathbf{z}_i)_h)$ for some large $a_{\mathbf{z}} \in \mathbb{R}^+$ as the proposal distribution. Thus the probability of accepting a proposed step is 
$$A((\mathbf{z}_i)_h', (\mathbf{z}_i)_h) = \min \left\{1, \frac{P\left((\mathbf{z}_i)_h'| \boldsymbol{\Theta}_{-(\mathbf{z}_i)_h'}, \mathbf{Y}_1, \dots, \mathbf{Y}_N, \mathbf{X} \right)}{P\left((\mathbf{z}_i)_h| \boldsymbol{\Theta}_{-(\mathbf{z}_i)_h}, \mathbf{Y}_1, \dots, \mathbf{Y}_N, \mathbf{X}\right)} \frac{Q\left((\mathbf{z}_i)_h|(\mathbf{z}_i)_h'\right)}{Q\left((\mathbf{z}_i)_h'|(\mathbf{z}_i)_h\right)}\right\}.$$

Letting
$$\left(\mathbf{B}_{\boldsymbol{\nu}_j}\right)_h = \left( \left(\tau_{\boldsymbol{\nu}_j}\right)_h\mathbf{P} + \frac{\beta_h}{(\sigma^2)_h} \sum_{i =1}^N \sum_{l=1}^{n_i}(Z_{ij})_h^2B(t_{il})B'(t_{il}) \right)^{-1}$$
and
$$\begin{aligned} \left(\mathbf{b}_{\boldsymbol{\nu}_j}\right)_h & = \frac{\beta_h}{(\sigma^2)_h}\sum_{i=1}^N\sum_{l=1}^{n_i}(Z_{ij})_h B(t_{il})\left[y_i(t_{il}) - \left(\sum_{k\ne j}(Z_{ik})_h(\boldsymbol{\nu}'_{k})_hB(t_{il})\right) \right.  \\ 
& - \left.\left(\sum_{k=1}^K (Z_{ik}) \left[\mathbf{x}_i (\boldsymbol{\eta}_k)_h' B(t_{il}) + \sum_{m=1}^M(\chi_{im})_h (\boldsymbol{\phi}_{kn})_h'B(t_{il}) \right]\right)\right],
\end{aligned}$$
we have that 
$$\left(\boldsymbol{\nu}_j\right)_h| \boldsymbol{\Theta}_{-\left(\boldsymbol{\nu}_j\right)_h}, \mathbf{Y}_1, \dots, \mathbf{Y}_N, \mathbf{X} \sim \mathcal{N}\left(\left(\mathbf{B}_{\boldsymbol{\nu}_j}\right)_h\left(\mathbf{b}_{\boldsymbol{\nu}_j}\right)_h, \left(\mathbf{B}_{\boldsymbol{\nu}_j}\right)_h\right).$$

Let $\left(\boldsymbol{\eta}_{jd}\right)_h$ denote the $d^{th}$ column of the matrix $(\boldsymbol{\eta}_j)_h$. Thus, letting
$$\left(\mathbf{B}_{\boldsymbol{\eta}_{jd}}\right)_h = \left( \left(\tau_{\boldsymbol{\eta}_{jd}}\right)_h\mathbf{P} + \frac{\beta_h}{(\sigma^2)_h} \sum_{i =1}^N \sum_{l=1}^{n_i}(Z_{ij})_h^2 x_{id}^2 B(t_{il})B'(t_{il}) \right)^{-1}$$
and 
$$\begin{aligned} \left(\mathbf{b}_{\boldsymbol{\eta}_{jd}}\right)_h = & \frac{\beta_h}{(\sigma^2)_h}\sum_{i=1}^N\sum_{l=1}^{n_i}(Z_{ij})_hx_{id}B(t_{il})\left[y_i(t_{il}) - \left(\sum_{r\ne d}(Z_{ij})_hx_{ir} (\boldsymbol{\eta}_{jr})_h'B(t_{il})\right)  \right.  \\ 
& - \left(\sum_{k \ne j} (Z_{ik})_h \mathbf{x}_i(\boldsymbol{\eta}_k)_h' B(t_{il}) \right) \\
& - \left.\left(\sum_{k=1}^K (Z_{ik})_h\left[(\boldsymbol{\nu}_k)_h' B(t_{il}) + \sum_{m=1}^M(\chi_{im})_h (\boldsymbol{\phi}_{kn})_h'B(t_{il})\right] \right)\right],
\end{aligned}$$
we have that 
$$\left(\boldsymbol{\eta}_{jd}\right)_h| \boldsymbol{\Theta}_{-\left(\boldsymbol{\eta}_{jd}\right)_h}, \mathbf{Y}_1, \dots, \mathbf{Y}_N, \mathbf{X} \sim \mathcal{N}\left(\left(\mathbf{B}_{\boldsymbol{\eta}_{jd}}\right)_h\left(\mathbf{b}_{\boldsymbol{\eta}_{jd}}\right)_h, \left(\mathbf{B}_{\boldsymbol{\eta}_{jd}}\right)_h\right).$$

If we let 
$$\begin{aligned}
\left(\beta_{\sigma}\right)_h =\frac{\beta_h}{2}\sum_{i=1}^N\sum_{l=1}^{n_i}\left(y_i(t_{il}) -  \sum_{k=1}^K (Z_{ik})_h\Bigg(\left((\boldsymbol{\nu}_k)_h + (\boldsymbol{\eta}_k)_h \mathbf{x}_i'\right)'B(t_{il}) \right. \\+ \left. \left. \sum_{n=1}^M(\chi_{in})_h(\boldsymbol{\phi}_{kn})_h'B(t_{il})\right)\right)^2,
\end{aligned}$$
then we have
$$(\sigma^2)_h| \boldsymbol{\Theta}_{-(\sigma^2)_h}, \mathbf{Y}_1, \dots, \mathbf{Y}_N, \mathbf{X}  \sim  IG\left(\alpha_0 + \frac{\beta_h\sum_{i=1}^N n_i}{2} , \beta_0 +\left(\beta_{\sigma}\right)_h\right).$$
Lastly, we can update the $\chi_{im}$ parameters, for $i = 1, \dots, N$ and $m = 1, \dots, M$, using a Gibbs update. If we let 
$$\begin{aligned}
\left(\mathbf{w}_{im}\right)_h = & \frac{\beta_h}{(\sigma^2)_h}\left[\sum_{l=1}^{n_i} \left(\sum_{k = 1}^K (Z_{ik})_h(\boldsymbol{\phi}_{km})_h'B(t_{il})\right)\right. \Bigg(y_i(t_{il})\\
& \left. - \sum_{k = 1}^K (Z_{ik})_h\left(\left((\boldsymbol{\nu}_k)_h + (\boldsymbol{\eta}_k)_h \mathbf{x}_i'\right)'B(t_{il})  + \sum_{n\ne m}(\chi_{in})_h(\boldsymbol{\phi}_{kn})_h'B(t_{il})\right)\Bigg)\right]
\end{aligned}$$

and 
$$\left(\mathbf{W}_{im}\right)_h^{-1} = 1 + \frac{\beta_h}{\sigma^2} \sum_{l=1}^{n_i}\left(\sum_{k = 1}^K (Z_{ik})_h(\boldsymbol{\phi}_{km})_h'B(t_{il})\right)^2,$$
then we have that 
$$(\chi_{im})_h| \boldsymbol{\zeta}_{-(\chi_{im})_h}, \mathbf{Y}_1, \dots, \mathbf{Y}_N, \mathbf{X} \sim \mathcal{N}\left(\left(\mathbf{W}_{im}\right)_h \left(\mathbf{w}_{im}\right)_h, \left(\mathbf{W}_{im}\right)_h\right).$$

\section{Simulation Study and Case Studies}
\label{sim_study and Case Study}
\subsection{Simulation Study 1}
This subsection contains detailed information on how the simulation study in Section 3 of the main text was conducted. This simulation study primarily looked at how well we could recover the true mean structure, the covariance structure, and the allocation structure. In this simulation study, we simulated datasets from 3 scenarios at 3 different sample sizes for each scenario. Once the datasets were generated, we fit a variety of covariate adjusted functional mixed membership models, as well as unadjusted functional mixed membership models, on the datasets to see how well we could recover the mean, covariance, and allocation structures. 

The first scenario we considered was a covariate adjusted functional mixed membership model with 2 true covariates. To generate all of the datasets, we assumed that the observations were in the span of B-spline basis with 8 basis functions. For this scenario, we generated 3 datasets with sample sizes of 60, 120, and 240 functional observations, all observed on a grid of 25 time points. The data was generated by first generating the model parameters (as discussed below) and then generating data from the likelihood specified in Equation 11 of the main text. The model parameters for this dataset were generated as follows:
$$\boldsymbol{\nu}_1 \sim \mathcal{N}\left((6, 4, \dots, -6, -8)', 4\mathbf{P} \right),$$
$$\boldsymbol{\nu}_2 \sim \mathcal{N}\left((-8, -6, \dots, 4, 6)', 4\mathbf{P} \right),$$
$$\boldsymbol{\eta}_{k1} \sim \mathcal{N}\left(\mathbf{1}, \mathbf{P}\right) \;\;\; k = 1,2,$$
$$\boldsymbol{\eta}_{k2} \sim \mathcal{N}\left((3, 2, \dots, -4)', \mathbf{P}\right) \;\;\; k = 1,2.$$
The $\boldsymbol{\Phi}$ parameters were drawn according to the following distributions:
$$\boldsymbol{\phi}_{km}  = \mathbf{q}_{km}\;\;\; k = 1, 2 \;\;\; m = 1,2,3,$$
where $\mathbf{q}_{k1} \sim \mathcal{N}(\mathbf{0}_{8}, 2.25\mathbf{I}_{8})$, $\mathbf{q}_{k2} \sim \mathcal{N}(\mathbf{0}_{8}, \mathbf{I}_{8})$, $\mathbf{q}_{k3} \sim \mathcal{N}(\mathbf{0}_{8}, 0.49\mathbf{I}_{8})$. The $\chi_{im}$ parameters were drawn from a standard normal distribution. The $\mathbf{z}_i$ parameters were drawn from a mixture of Dirichlet distributions. Roughly 30\% of the $\mathbf{z}_i$ parameters were drawn from a Dirichlet distribution with $\alpha_1 = 10$ and $\alpha_2 = 1$. Another roughly 30\% of the $\mathbf{z}_i$ parameters were drawn from a Dirichlet distribution where $\alpha_1 = 1$ and $\alpha_2 = 10$. The rest of the $\mathbf{z}_i$ parameters were drawn from a Dirichlet distribution with $\alpha_1 = \alpha_2 = 1$. The covariates, $\mathbf{X}$, were drawn from a standard normal distribution. Models in this scenario were run for 500,000 MCMC iterations.

For the second scenario, we considered data drawn from a covariate adjusted functional mixed membership model with one covariate. We considered three sample sizes of 50, 100, and 200 functional samples observed on a grid of 25 time points. The model parameters for this dataset were generated as follows:
$$\boldsymbol{\nu}_1 \sim \mathcal{N}\left((6, 4, \dots, -6, -8)', 4\mathbf{P} \right),$$
$$\boldsymbol{\nu}_2 \sim \mathcal{N}\left((-8, -6, \dots, 4, 6)', 4\mathbf{P} \right),$$
$$\boldsymbol{\eta}_{11} \sim \mathcal{N}\left(\mathbf{2}, \mathbf{P}\right),$$
$$\boldsymbol{\eta}_{21} \sim \mathcal{N}\left(-\mathbf{2}, \mathbf{P}\right).$$
The $\boldsymbol{\Phi}$ parameters were drawn according to the following distributions:
$$\boldsymbol{\phi}_{km}  = \mathbf{q}_{km}\;\;\; k = 1, 2 \;\;\; m = 1,2,3,$$
where $\mathbf{q}_{k1} \sim \mathcal{N}(\mathbf{0}_{8}, 2.25\mathbf{I}_{8})$, $\mathbf{q}_{k2} \sim \mathcal{N}(\mathbf{0}_{8}, \mathbf{I}_{8})$, $\mathbf{q}_{k3} \sim \mathcal{N}(\mathbf{0}_{8}, 0.49\mathbf{I}_{8})$. The $\chi_{im}$ parameters were drawn from a standard normal distribution. The $\mathbf{z}_i$ parameters were drawn from a mixture of Dirichlet distributions. Roughly 30\% of the $\mathbf{z}_i$ parameters were drawn from a Dirichlet distribution with $\alpha_1 = 10$ and $\alpha_2 = 1$. Another roughly 30\% of the $\mathbf{z}_i$ parameters were drawn from a Dirichlet distribution where $\alpha_1 = 1$ and $\alpha_2 = 10$. The rest of the $\mathbf{z}_i$ parameters were drawn from a Dirichlet distribution with $\alpha_1 = \alpha_2 = 1$. The covariates, $\mathbf{X}$, were drawn from a normal distribution with variance of nine and mean of zero. Models in this scenario were run for 500,000 MCMC iterations.

For the third scenario, we generated data from an unadjusted functional mixed membership model. We considered three sample sizes of 40, 80, and 160 functional samples observed on a grid of 25 time points. The model parameters for this dataset were generated as follows:
$$\boldsymbol{\nu}_1 \sim \mathcal{N}\left((6, 4, \dots, -6, -8)', 4\mathbf{P} \right),$$
$$\boldsymbol{\nu}_2 \sim \mathcal{N}\left((-8, -6, \dots, 4, 6)', 4\mathbf{P} \right),$$

The $\boldsymbol{\Phi}$ parameters were drawn according to the following distributions:
$$\boldsymbol{\phi}_{km}  = \mathbf{q}_{km} \;\;\; k = 1, 2 \;\;\; m = 1,2,$$
where $\mathbf{q}_{k1} \sim \mathcal{N}(\mathbf{0}_{8}, 2.25\mathbf{I}_{8})$, $\mathbf{q}_{k2} \sim \mathcal{N}(\mathbf{0}_{8}, \mathbf{I}_{8})$, $\mathbf{q}_{k3} \sim \mathcal{N}(\mathbf{0}_{8}, 0.49\mathbf{I}_{8})$. The $\chi_{im}$ parameters were drawn from a standard normal distribution. The $\mathbf{z}_i$ parameters were drawn from a mixture of Dirichlet distributions. Approximately 30\% of the $\mathbf{z}_i$ parameters were drawn from a Dirichlet distribution with $\alpha_1 = 10$ and $\alpha_2 = 1$. Another roughly 30\% of the $\mathbf{z}_i$ parameters were drawn from a Dirichlet distribution where $\alpha_1 = 1$ and $\alpha_2 = 10$. The rest of the $\mathbf{z}_i$ parameters were drawn from a Dirichlet distribution with $\alpha_1 = \alpha_2 = 1$. The models in this scenario were run for 500,000 MCMC iterations. The code for running this simulation study can be found on Github. 

The following plots are more detailed visualizations of the results obtained in the first simulation study.

\begin{figure}
    \centering
    \includegraphics[width = 0.99\textwidth]{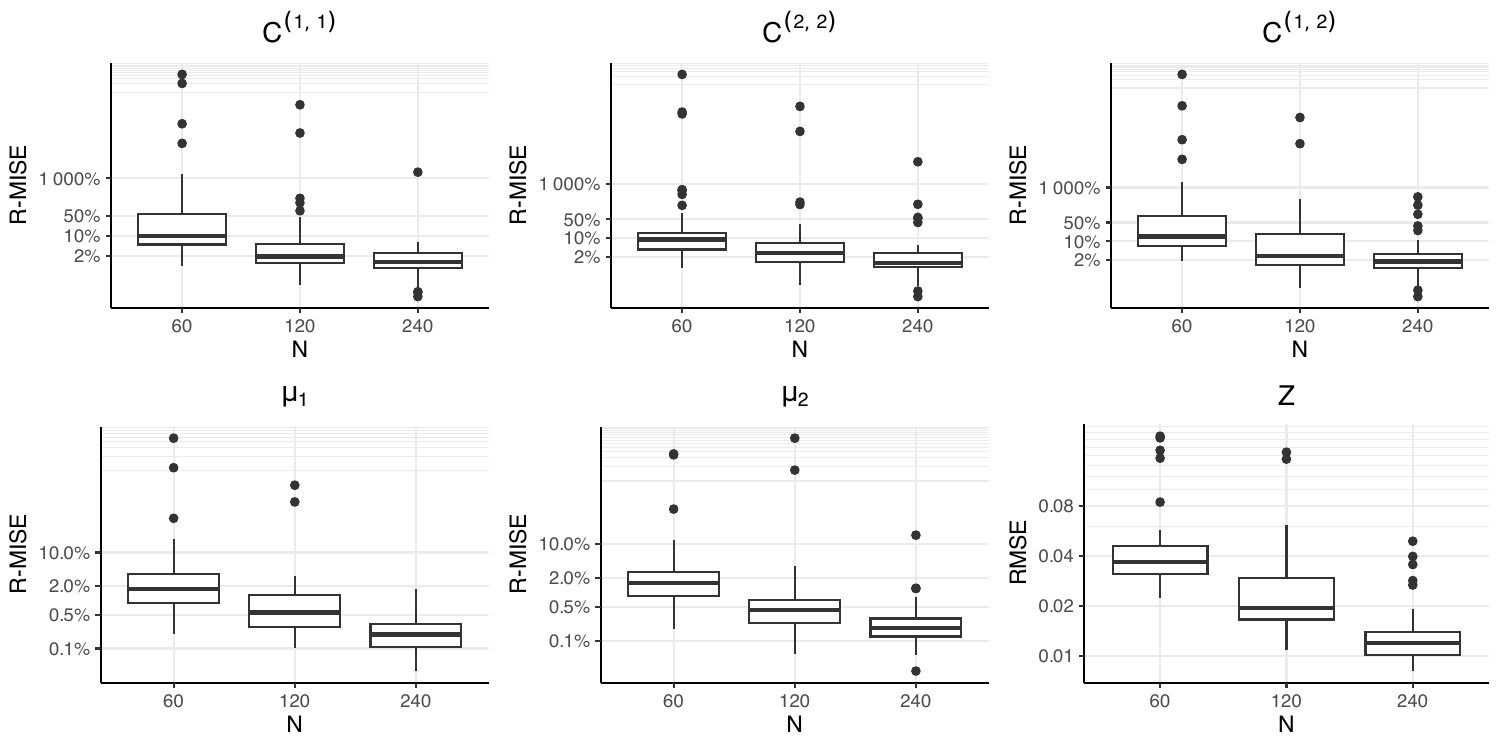}
    \caption{RMISE and RMSE results from the simulations models fit with two covariates, where the true data was generated with two covariates.}
    \label{fig:two_param}
\end{figure}
\begin{figure}
    \centering
    \includegraphics[width = 0.99\textwidth]{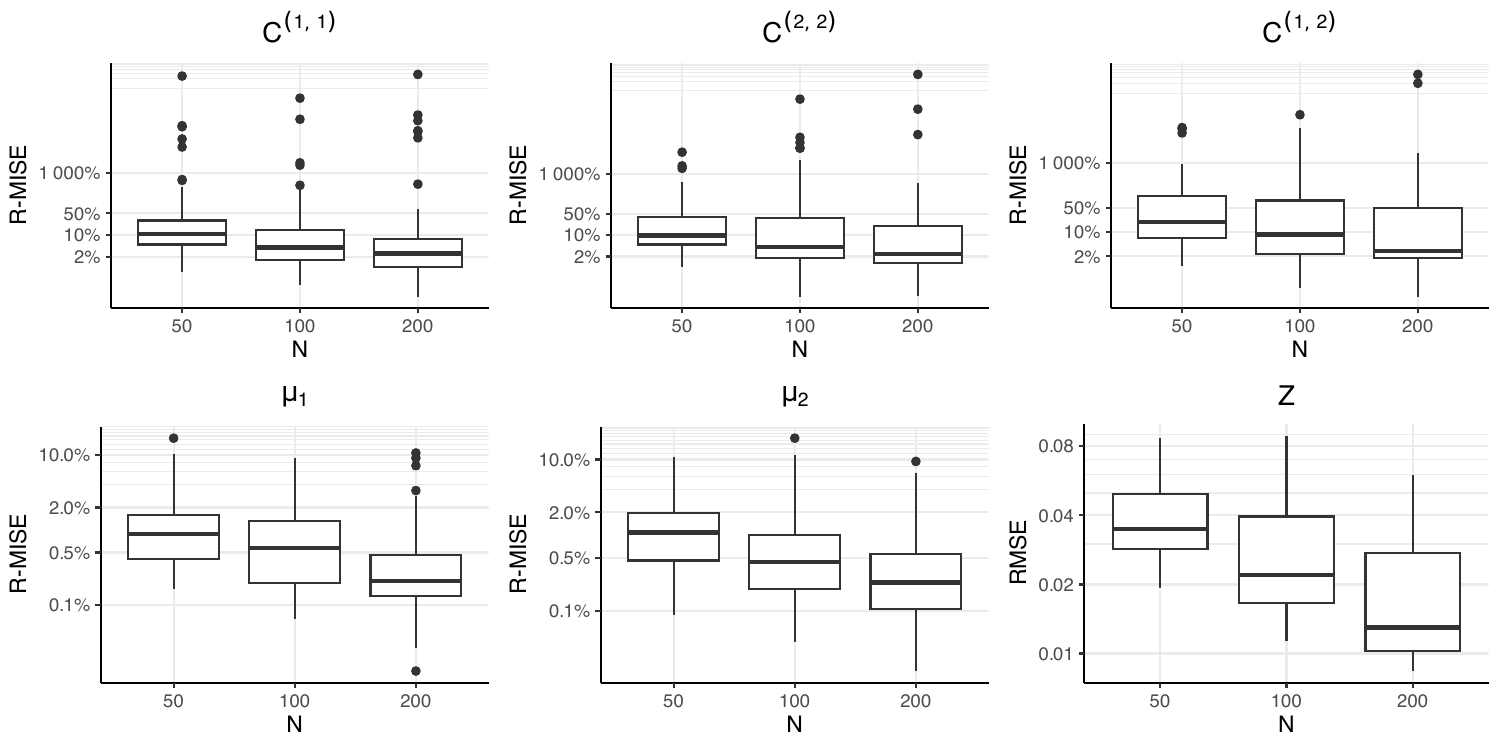}
    \caption{RMISE and RMSE results from the simulations models fit with one covariate, where the true data was generated with one covariate.}
    \label{fig:one_param_truth_one}
\end{figure}
\begin{figure}
    \centering
    \includegraphics[width = 0.99\textwidth]{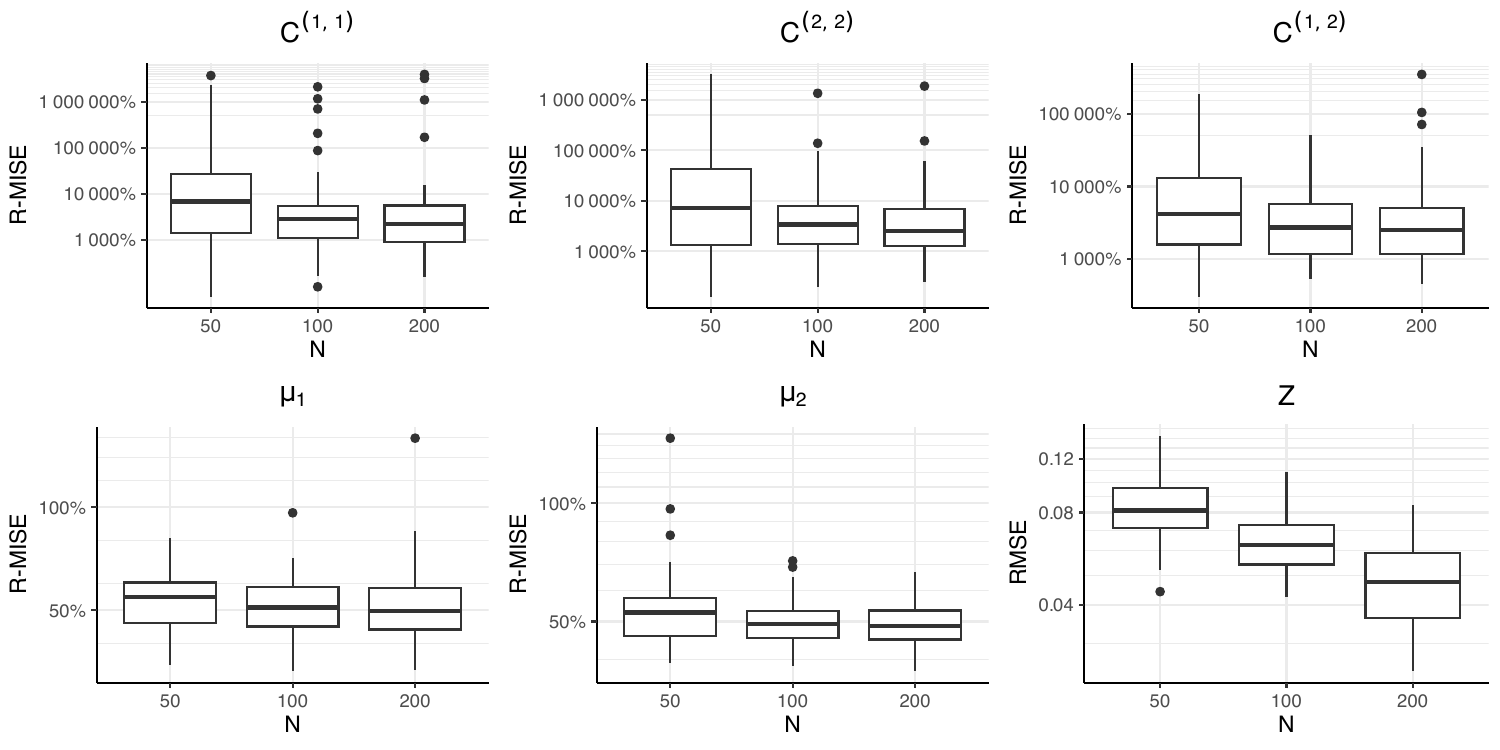}
    \caption{RMISE and RMSE results from the simulations models fit with no covariates, where the true data was generated with one covariate.}
    \label{fig:zero_param_truth_one}
\end{figure}
\begin{figure}
    \centering
    \includegraphics[width = 0.99\textwidth]{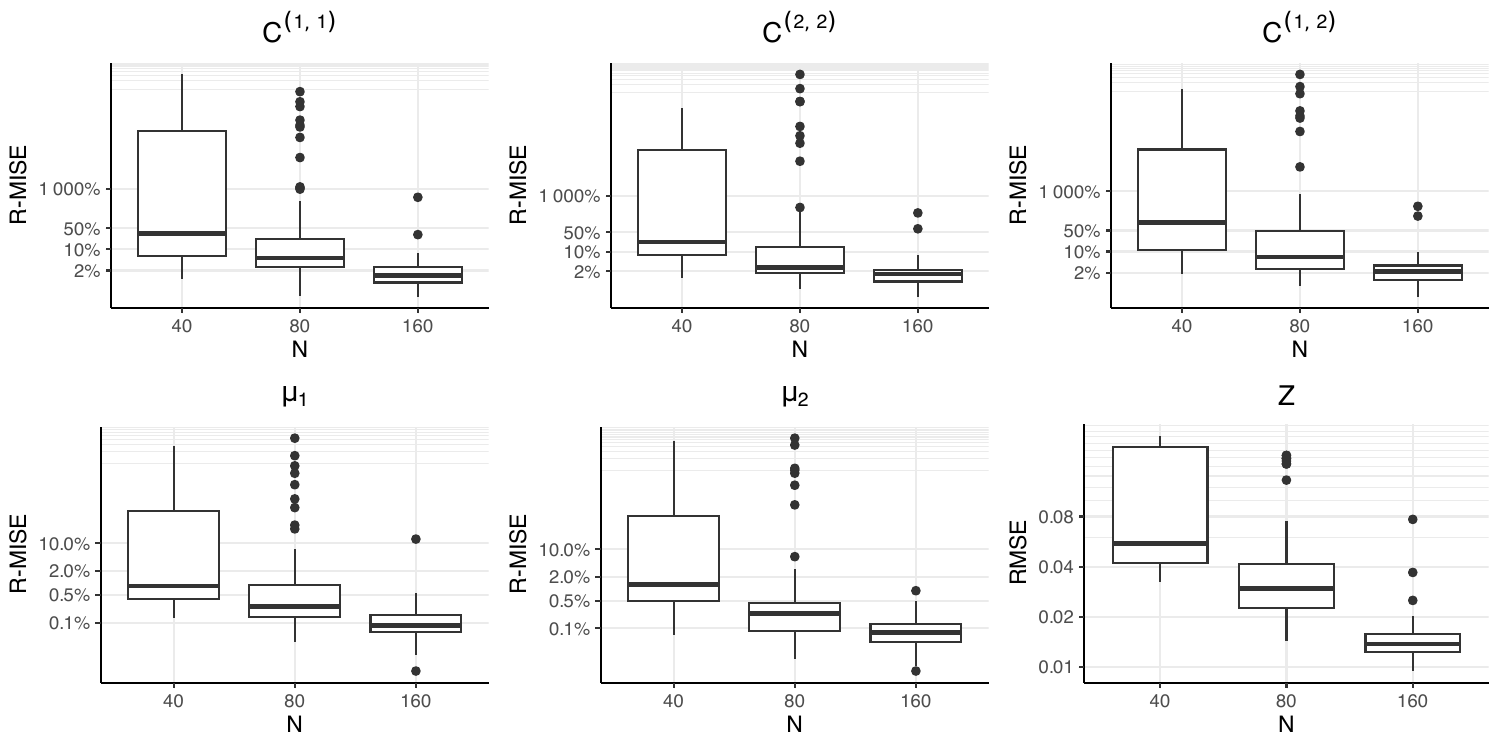}
    \caption{RMISE and RMSE results from the simulations models fit with one covariate, where the true data was generated with no covariates.}
    \label{fig:one_param_truth_zero}
\end{figure}
\begin{figure}
    \centering
    \includegraphics[width = 0.99\textwidth]{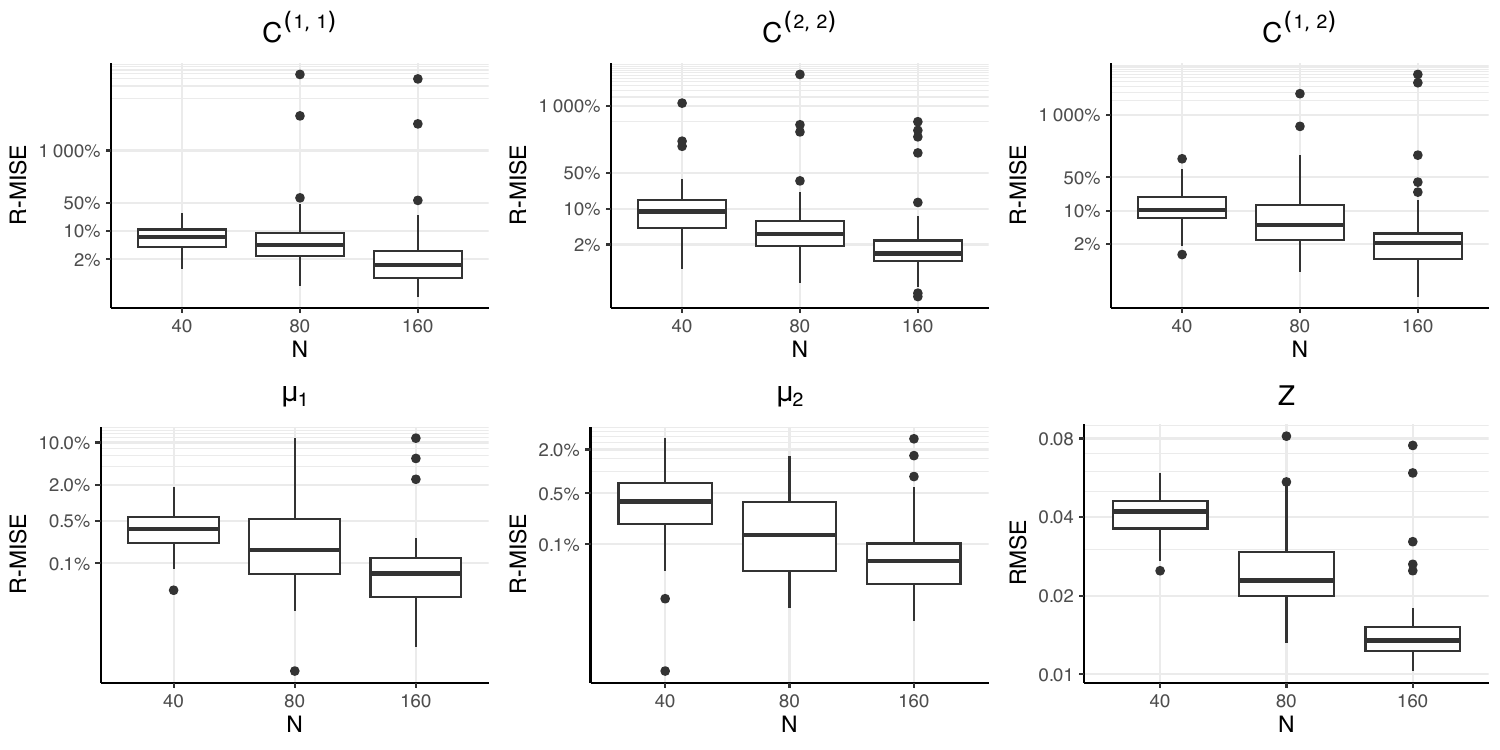}
    \caption{RMISE and RMSE results from the simulations models fit with no covariates, where the true data was generated with no covariates.}
    \label{fig:zero_param_truth_zero}
\end{figure}
\subsection{Simulation Study 2}

In this simulation study, we evaluated the performance of AIC, BIC, DIC, and the elbow method in choosing the number of features in a covariate adjusted mixed membership model. In this simulation study, we considered a covariate adjusted functional mixed membership model with only one continuous covariate. We generated 50 different data sets with 150 functional observations, observed along a uniform grid of 25 time points. The data was generated from a covariate adjusted mixed membership model with model parameters generated as follows:

$$\boldsymbol{\nu}_1 \sim \mathcal{N}\left((6, 4, \dots, -6, -8)', 4\mathbf{P} \right),$$
$$\boldsymbol{\nu}_2 \sim \mathcal{N}\left((-8, -6, \dots, 4, 6)', 4\mathbf{P} \right),$$
$$\boldsymbol{\eta}_{11} \sim \mathcal{N}\left(\mathbf{2}, \mathbf{P}\right),$$
$$\boldsymbol{\eta}_{21} \sim \mathcal{N}\left(-\mathbf{2}, \mathbf{P}\right),$$
$$\boldsymbol{\eta}_{21} \sim \mathcal{N}\left(\mathbf{1}, \mathbf{P}\right).$$

he $\boldsymbol{\Phi}$ parameters were drawn according to the following distributions:
$$\boldsymbol{\phi}_{km}  = \mathbf{q}_{km} \;\;\; k = 1, 2 \;\;\; m = 1,2,$$
where $\mathbf{q}_{k1} \sim \mathcal{N}(\mathbf{0}_{8}, 2.25\mathbf{I}_{8})$, $\mathbf{q}_{k2} \sim \mathcal{N}(\mathbf{0}_{8}, \mathbf{I}_{8})$, $\mathbf{q}_{k3} \sim \mathcal{N}(\mathbf{0}_{8}, 0.49\mathbf{I}_{8})$. The $\chi_{im}$ parameters were drawn from a standard normal distribution. The $\mathbf{z}_i$ parameters were drawn from a mixture of Dirichlet distributions. Approximately 20\% of the $\mathbf{z}_i$ parameters were drawn from a Dirichlet distribution with $\alpha_1 = 30$, $\alpha_2 = 1$, and $\alpha_3 = 1$. Another roughly 20\% of the $\mathbf{z}_i$ parameters were drawn from a Dirichlet distribution where $\alpha_1 = 1$, $\alpha_2 = 30$, and $\alpha_3 = 1$. Another roughly 20\% of the $\mathbf{z}_i$ parameters were drawn from a Dirichlet distribution where $\alpha_1 = 1$, $\alpha_2 = 1$, and $\alpha_3 = 30$. The rest of the $\mathbf{z}_i$ parameters were drawn from a Dirichlet distribution with $\alpha_1 = \alpha_2 = \alpha_3 = 1$. Four models were then fit for each dataset, with $K = 2, 3, 4, 5$. The models in this scenario were run for 200,000 MCMC iterations. 

The Bayesian Information Criterion (BIC), proposed by \citet{schwarz1978estimating}, is defined as:
$$\text{BIC} = 2\text{log}P\left(\mathbf{Y}|\hat{\boldsymbol{\Theta}}, \mathbf{X}\right) - d\text{log}(n)$$
where $d$ is the number of parameters and $\hat{\boldsymbol{\Theta}}$ are the maximum likelihood estimators (MLE) of our parameters. In the case of our proposed model, we have that
\begin{equation}
\text{BIC} = 2\text{log}P\left(\mathbf{Y}|\hat{\boldsymbol{\nu}}_{1:K}, \hat{\boldsymbol{\eta}}_{1:K}, \hat{\boldsymbol{\Phi}}_{1:KM}, \hat{\sigma}^2, \hat{\mathbf{Z}}, \hat{\boldsymbol{\chi}}, \mathbf{X}\right) - d\text{log}(\tilde{N})
    \label{BIC}
\end{equation}
where $\tilde{N} = \sum_{i}n_i$ (where $n_i$ is the number of observed time points observed for the $i^{th}$ function), and $d = (N + P)K + 2MKP + 4K + (N + K)M + 2 + PRK + KR$.

Similarly, the AIC, proposed by \citet{akaike1974new}, can be written as
\begin{equation}
    \text{AIC} = -2\text{log}P\left(\mathbf{Y}|\hat{\boldsymbol{\nu}}_{1:K}, \hat{\boldsymbol{\eta}}_{1:K}, \hat{\boldsymbol{\Phi}}_{1:KM}, \hat{\sigma}^2, \hat{\mathbf{Z}}, \hat{\boldsymbol{\chi}}, \mathbf{X}\right) + 2d.
    \label{AIC}
\end{equation}

Following the work of \citet{roeder1997practical}, we will use the posterior mean instead of the MLE for our estimates of BIC and AIC.

The modified DIC, proposed by \citet{celeux2006deviance}, is advantageous to the original DIC proposed by \citet{spiegelhalter2002bayesian} when we have a posterior distribution with multiple modes and when identifiability may be a problem. The modified DIC (referred to as $\text{DIC}_3$ in \citet{celeux2006deviance}) is specified as
\begin{equation}
    \text{DIC} = -4 \mathbb{E}_{\boldsymbol{\Theta}}[\text{log} f(\mathbf{Y}|\boldsymbol{\Theta}), \mathbf{X} |\mathbf{Y}] + 2 log \hat{f}(\mathbf{Y})
    \label{DIC}
\end{equation}
where $\hat{f}(y_{ij}) = \frac{1}{N_{MC}}\sum_{l=1}^{N_{MC}}P\left(y_{ij}|\boldsymbol{\nu}^{(l)}_{1:K}, \boldsymbol{\eta}^{(l)}_{1:K}, \boldsymbol{\Phi}^{(l)}_{1:KM}, \left(\sigma^2\right)^{(l)}, \mathbf{Z}^{(l)}, \mathbf{x}_i\right)$, $\hat{f}(\mathbf{Y}) = \prod_{i=1}^{N}\prod_{j=1}^{n_i}\hat{f}(y_{ij})$, and $N_{MC}$ is the number of MCMC samples used for estimating $\hat{f}(y_{ij})$. We can approximate $\mathbb{E}_{\boldsymbol{\Theta}}[\text{log} f(\mathbf{Y}|\boldsymbol{\Theta})|\mathbf{Y}]$ by using the MCMC samples, such that
$$\mathbb{E}_{\boldsymbol{\Theta}}[\text{log} f(\mathbf{Y}|\boldsymbol{\Theta}, \mathbf{X})|\mathbf{Y}] \approx \frac{1}{N_{MC}} \sum_{l=1}^{N_{MC}}\sum_{i=1}^{N}\sum_{j=1}^{n_i}\text{log}\left[P\left(y_{ij}|\boldsymbol{\nu}^{(l)}_{1:K}, \boldsymbol{\eta}^{(l)}_{1:K}, \boldsymbol{\Phi}^{(l)}_{1:KM}, \left(\sigma^2\right)^{(l)}, \mathbf{Z}^{(l)}, \mathbf{x}_i\right)\right].$$

\subsection{Case Study}
As discussed in the main manuscript, we fit two covariate adjusted functional mixed membership models using resting-state EEG data from \citet{dickinson2018peak}. The first model used only the log transformation of age as the covariate, and the second model used the log transformation of age, diagnostic group (ASD vs TD), and an interaction between the log transformation of age and diagnostic group as the covariates.

\begin{figure}
    \centering
    \includegraphics[width = 0.99\textwidth]{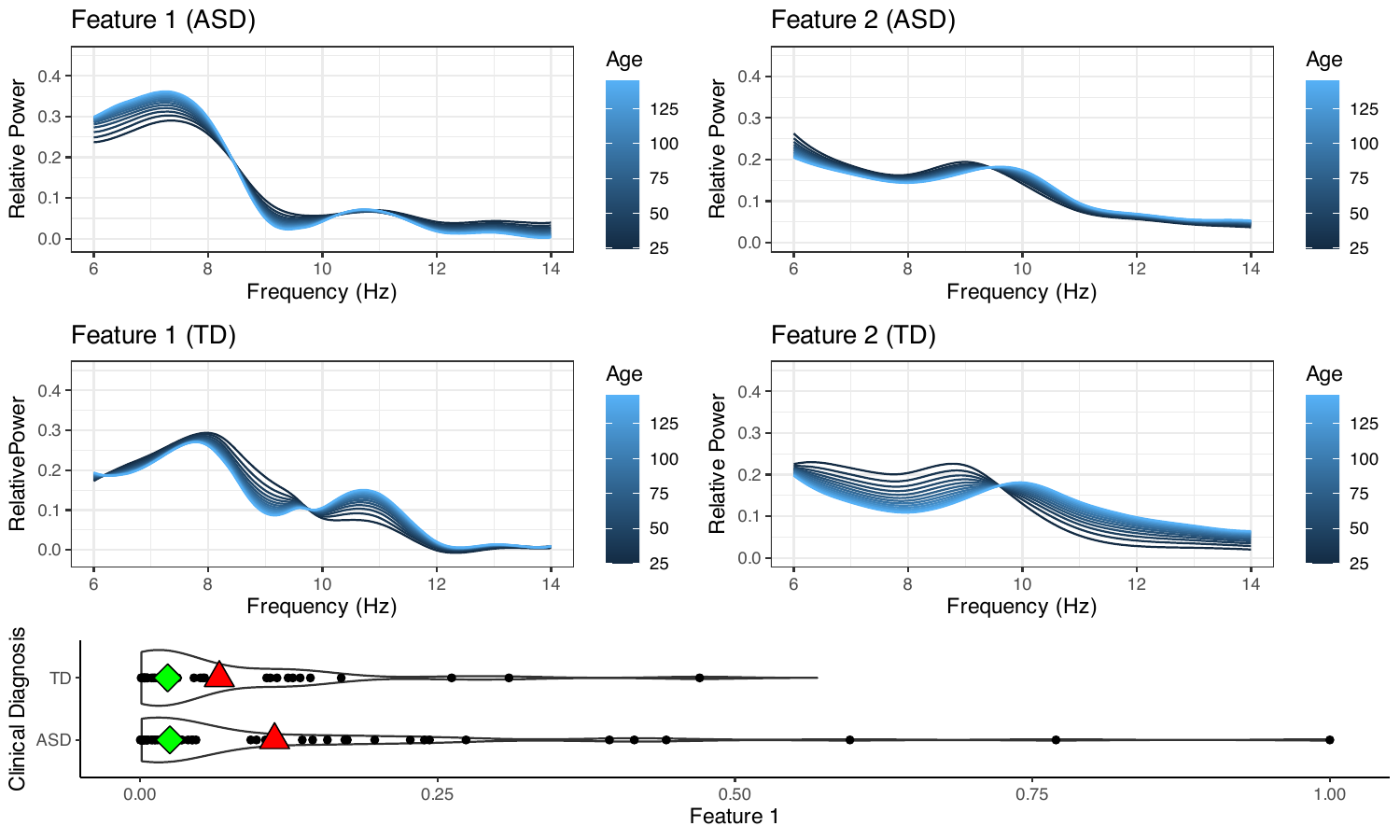}
    \caption{(Top 4 Panels)Estimated mean structures for ASD and TD individuals, for each of the two features used in the covariate adjusted functional mixed membership model fit in Section 4.2 in the main manuscript. (Bottom Panel) Estimates of the allocation parameters stratified by clinical Diagnosis, where the red triangles depict the group level means and the green diamonds depict the group level medians.}
    \label{fig:Feature_means}
\end{figure}

The group-specific mean structures for each of the two features can be seen in Figure \ref{fig:Feature_means}. As stated in the manuscript, these feature means cannot be interpreted as the expected trajectories of the most extreme observations and are thus harder to directly interpret. 

\subsection{Comparison Between Mixed Membership Models}
In the main manuscript, we extended the analysis on alpha oscillations conducted by \citet{marcoFunctional} to allow for a covariate-dependent mixed membership model. While previous studies \citep{haegens2014inter, Rodriguez2017, scheffler2019covariate} have shown that alpha oscillations are dependent on age, little is known about how alpha oscillations differ between children with ASD and TD children conditional on age. Although it is attractive to add more covariates to our model, the small sample sizes often found in neurodevelopmental studies limit our ability to fit models with a large amount of covariates. Thus, to avoid having overfit models, we can perform cross-validated methods such as conditional predictive ordinates (CPO) \citep{pettit1990conditional, chen2012monte,lewis2014posterior}. CPO for our model can be defined as $P\left(\mathbf{Y}_i(\mathbf{t}_i)\mid \{\mathbf{Y}_j(\mathbf{t}_j)\}_{j\ne i} \right)$. Unlike traditional cross-validation methods, CPO requires no extra sampling to be conducted. Following \citet{chen2012monte} and \citet{lewis2014posterior}, an estimate of CPO for our model can be obtained using the following MCMC approximation:
\begin{equation}
    \hat{CPO}_i = \left(\frac{1}{N_{MC}}\sum_{r=1}^{N_{MC}} \frac{1}{P\left(\mathbf{Y}_i(\mathbf{t}_i)\mid \hat{\boldsymbol{\Theta}}_{-\chi}^r,\mathbf{x}_i\right)} \right)^{-1},
\end{equation}
where $\hat{\boldsymbol{\Theta}}_{-\chi}^r$ are the samples from the $r^{th}$ MCMC iteration, $N_{MC}$ are the number of MCMC iterations (not including burn-in), and $P\left(\mathbf{Y}_i(\mathbf{t}_i)\mid \hat{\boldsymbol{\Theta}}_{-\chi}^r,\mathbf{x}_i\right)$ is specified in Equation 12 in the main manuscript. While CPO is a measure of how well the model fits each individual observation, the pseudomarginal likelihood (PML), defined as $\hat{PML} = \prod_{i=1}^N\hat{CPO}_i$, is an overall measure of how well the model fits the entire dataset. Using CPO and PML, we will compare the two covariate adjusted functional mixed membership models fit in this section.

\begin{figure}
    \centering
    \includegraphics[width = 0.95\textwidth]{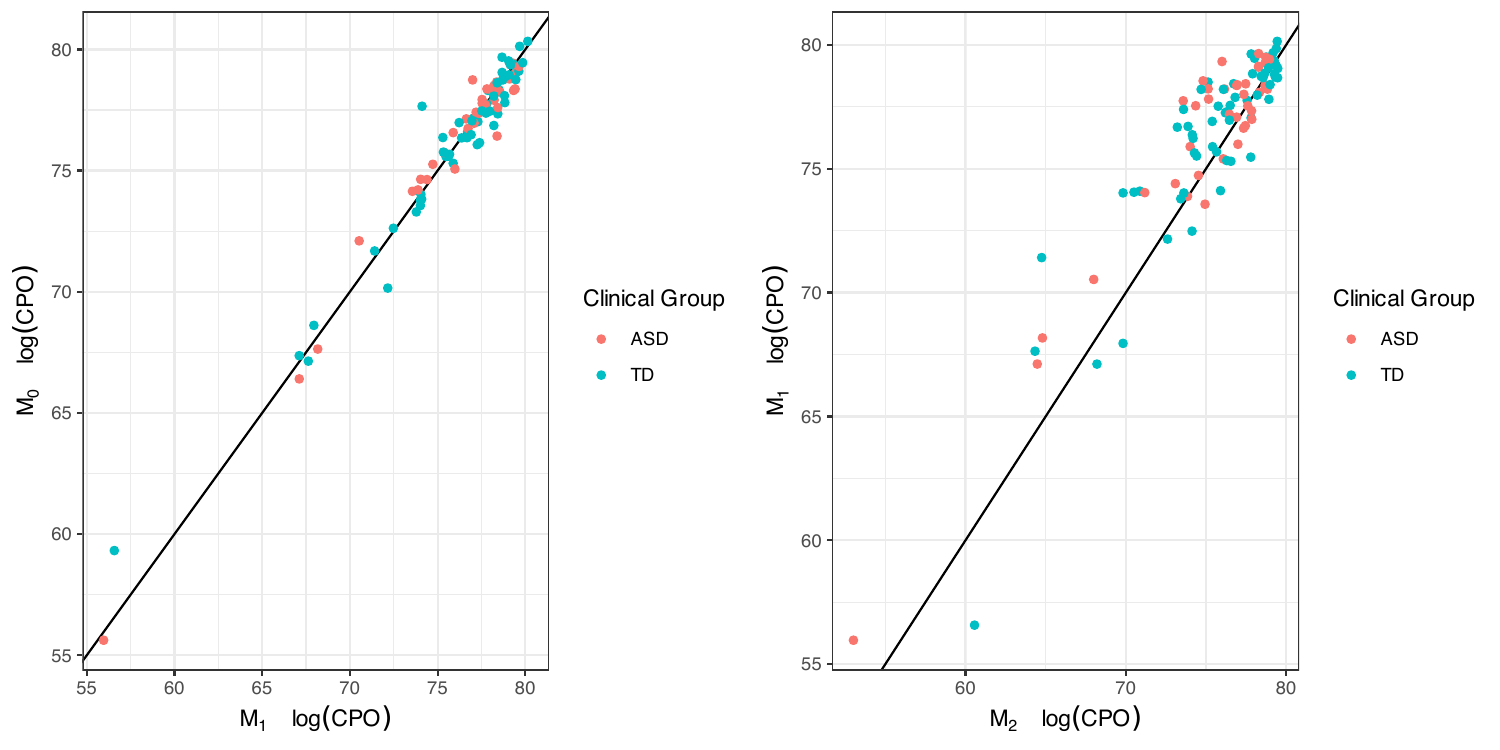}
    \caption{CPO comparisons between different models on the log scale. $M_0$ denotes the covariate adjusted model with age as the covariate, $M_1$ denotes the covariate adjusted functional mixed membership model using the log transform of age as the covariate, and $M_2$ denotes the covariate adjusted functional mixed membership model using the log transform of age, diagnostic group, and the interaction between the log transform of age and diagnostic group as the covariates.}
    \label{fig: CPO}
\end{figure}

In this section, we will let $M_0$ denote the covariate adjusted model with age as the covariate, $M_1$ denote the covariate adjusted functional mixed membership model using the log transform of age as the covariate, and $M_2$ denote the covariate adjusted functional mixed membership model using the log transform of age, diagnostic group, and the interaction between the log transform of age and diagnostic group as the covariates. Figure \ref{fig: CPO} contains the CPO values from all 3 of the models considered. Although the fit is similar between $M_0$ and $M_1$, both the pseudomarginal likelihood and the likelihood were higher in the log transformed age model ($M_1:$ $\log{(\hat{\text{PML}})}=  7391.9$, $\log L = 7833.2$; $M_0:$ $\log{(\hat{\text{PML}})}=  7390.6$, $\log L = 7826.3$). Thus, the analysis in the main manuscript was performed using the log of age as the covariate. From Figure \ref{fig: CPO}, we can see that $M_1$ tends to fit the data slightly better than $M_2$ ($M_1:$ $\log{(\hat{\text{PML}})}=  7391.9$, $M_2:$ $\log{(\hat{\text{PML}})}= 7303.9$). Although the fit may be slightly worse for the covariate adjusted model with age and diagnostic group as covariates, this model gives us useful information about how the two features differ between children with ASD and TD children. 

\section{Mean and Covariance Covariate-dependent Mixed Membership Model}
\subsection{Model Specification}
In this section, we completely specify a mixed membership model where the mean and covariance structures depend on the covariates of interest. As in the main text of this manuscript, we will let $\{\mathbf{Y}_i(.)\}_{i=1}^N$ be the observed sample paths and $\mathbf{t}_i = [t_{i1}, \dots, t_{in_i}]'$ denote the time points at which the $i^{th}$ function was observed. We will also let $\mathbf{X} \in \mathbb{R}^{N \times R}$ denote the design matrix and $\mathbf{x}_i = [X_{i1} \dots X_{iR}]$ denote the $i^{th}$ row of the design matrix (or the covariates associated with the $i^{th}$ observation). By introducing covariate-dependent pseudo-eigenfunctions, we arrive at the likelihood of our mixed membership model where the mean and covariance structures are dependent on the covariates of interest:
\begin{equation}
    \resizebox{.9 \textwidth}{!}{$\mathbf{Y}_i(\mathbf{t}_i)\mid\boldsymbol{\Theta}, \mathbf{X} \sim  \mathcal{N}\left\{ \sum_{k=1}^K Z_{ik}\left(\mathbf{S}'(\mathbf{t}_i) \left(\boldsymbol{\nu}_k + \boldsymbol{\eta}_k \mathbf{x}_i'\right) + \sum_{m=1}^{M}\chi_{im}\mathbf{S}'({\bf t}_i) \left(\boldsymbol{\phi}_{km} + \boldsymbol{\xi}_{km}\mathbf{x}_i'\right)\right),\; \sigma^2 \mathbf{I}_{n_i}\right\}.$}
    \label{cov_adj_full_func}
\end{equation}
From equation \ref{cov_adj_full_func}, we can see that $\boldsymbol{\xi}_{km} \in \mathbb{R}^{P \times R}$, directly controls the effect that the covariates have on the pseudo-eigenfunctions for $k = 1,\dots, K$ and $m = 1, \dots, M$. By integrating out the $\chi_{im}$ parameters ($i = 1,\dots, N$ and $m = 1,\dots, M$), we get a model of the following form:
\begin{equation}
    \mathbf{Y}_i(\mathbf{t}_i)\mid\boldsymbol{\Theta}_{-\chi}, \mathbf{X} \sim \mathcal{N}\left\{ \sum_{k=1}^K Z_{ik}\mathbf{S}'(\mathbf{t}_i) \left(\boldsymbol{\nu}_k + \boldsymbol{\eta}_k \mathbf{x}_i'\right),\; \mathbf{V}(\mathbf{t}_i, \mathbf{z}_i) + \sigma^2\mathbf{I}_{n_i} \right\},
    \label{int_likelihood_func}
\end{equation}
where $\boldsymbol{\Theta}_{-\chi}$ is the collection of our model parameters excluding the $\chi_{im}$ variables, and the error-free mixed membership covariance is  
\begin{equation}
     \mathbf{V}(\mathbf{t}_i, \mathbf{z}_i) =  \sum_{k=1}^K\sum_{k'=1}^K Z_{ik}Z_{ik'}\left\{\mathbf{S}'(\mathbf{t}_i)\sum_{m=1}^{M}\left[\left(\boldsymbol{\phi}_{km} + \boldsymbol{\xi}_{km}\mathbf{x}_i' \right)\left(\boldsymbol{\phi}_{k'm} + \boldsymbol{\xi}_{k'm}\mathbf{x}_i'\right)'\right]\mathbf{S}(\mathbf{t}_i)\right\}.
    \label{fmean_model_Cov}
\end{equation}
As with the pseudo-eigenfunctions in the unadjusted model, we will utilize the multiplicative gamma process prior as our prior on the $\boldsymbol{\xi}_{km}$ variables. Letting $\xi_{(krm)_p}$ denote the element in the $p^{th}$ row and $r^{th}$ column of $\boldsymbol{\xi}_{km}$. Thus we have:

$$\xi_{(krm)_p}\mid\gamma_{\boldsymbol{\xi}_{krmp}}, \tilde{\tau}_{\boldsymbol{\xi}_{mkr}} \sim \mathcal{N}\left(0, \gamma_{\boldsymbol{\xi}_{krmp}}^{-1}\tilde{\tau}_{\boldsymbol{\xi}_{mkr}}^{-1}\right), \;\;\; \gamma_{\xi_{krmp}} \sim \Gamma\left(\nu_\gamma /2 , \nu_\gamma /2\right), \;\;\; \tilde{\tau}_{\boldsymbol{\xi}_{mkr}} = \prod_{n=1}^m \delta_{\boldsymbol{\xi}_{nkr}},$$
$$ \delta_{\boldsymbol{\xi}_{1kr}}\mid a_{\boldsymbol{\xi}_{1kr}} \sim \Gamma(a_{\boldsymbol{\xi}_{1kr}}, 1), \;\;\; \delta_{\boldsymbol{\xi}_{jkr}}\mid a_{\boldsymbol{\xi}_{2kr}} \sim \Gamma(a_{\boldsymbol{\xi}_{2kr}}, 1), \;\;\; a_{\boldsymbol{\xi}_{1kr}} \sim \Gamma(\alpha_1, \beta_1), \;\;\; a_{\boldsymbol{\xi}_{2kr}} \sim \Gamma(\alpha_2, \beta_2),$$
for $k = 1, \dots, K$, $r = 1, \dots, R$, $m = 1,\dots, M$, and $p = 1, \dots P$. The rest of the parameters in the model have the same prior distributions as the model with the covariate-dependence on the mean structure only in the main text. Specifically, we have
$$\phi_{kpm}|\gamma_{kpm}, \tilde{\tau}_{mk} \sim \mathcal{N}\left(0, \gamma_{kpm}^{-1}\tilde{\tau}_{mk}^{-1}\right), \;\;\; \gamma_{kpm} \sim \Gamma\left(\nu_\gamma /2 , \nu_\gamma /2\right), \;\;\; \tilde{\tau}_{mk} = \prod_{n=1}^m \delta_{nk},$$
$$ \delta_{1k}|a_{1k} \sim \Gamma(a_{1k}, 1), \;\;\; \delta_{jk}|a_{2k} \sim \Gamma(a_{2k}, 1), \;\;\; a_{1k} \sim \Gamma(\alpha_1, \beta_1), \;\;\; a_{2k} \sim \Gamma(\alpha_2, \beta_2),$$
for $k = 1, \dots, K$, $m = 1,\dots, M$, and $p = 1, \dots P$. Similarly, we have
$$P(\boldsymbol{\nu}_k|\tau_{\boldsymbol{\nu}_k}) \propto exp\left(-\frac{\tau_{\boldsymbol{\nu}_k}}{2}\sum_{p =1}^{P - 1}\left(\nu_{pk}'- {\nu}_{(p+1)k}\right)^2\right),$$
 for $k = 1, \dots, K$, where $\tau_{\boldsymbol{\nu}_k} \sim \Gamma(\alpha_{\boldsymbol{\nu}}, \beta_{\boldsymbol{\nu}})$ and $\nu_{pk}$ is the $p^{th}$ element of $\boldsymbol{\nu}_k$. Likewise, we have that 
 $$P(\{\eta_{prk}\}_{p=1}^P|\tau_{\boldsymbol{\eta}_{rk}}) \propto exp\left(-\frac{\tau_{\boldsymbol{\eta}_{rk}}}{2}\sum_{p =1}^{P - 1}\left(\eta_{prk}'- {\eta}_{(p+1)rk}\right)^2\right),$$
 for $k = 1, \dots, K$ and $r = 1, \dots, R$, where $\tau_{\boldsymbol{\eta}_{rk}} \sim \Gamma(\alpha_{\boldsymbol{\eta}}, \beta_{\boldsymbol{\eta}})$ and $\eta_{prk}$ is the $p^{th}$ row and $r^{th}$ column of $\boldsymbol{\eta}_{k}$. Lastly, we assume that $\mathbf{z}_i\mid \boldsymbol{\pi}, \alpha_3 \sim_{iid} Dir(\alpha_3\boldsymbol{\pi})$, $\boldsymbol{\pi} \sim Dir(\mathbf{c}_\pi)$, $\alpha_3 \sim Exp(b)$, and $\sigma^2 \sim IG(\alpha_0,  \beta_0)$.

\subsection{Posterior Distributions}
\label{posterior_xi}
In this subsection, we will specify the posterior distributions specifically for the functional covariate adjusted mixed membership model where the covariance is covariate-dependent. We will first start with the $\boldsymbol{\phi}_{km}$ parameters, for $j = 1,\dots, K$ and $m = 1, \dots, M$. Let $\mathbf{D}_{\boldsymbol{\phi}_{jm}} = \tilde{\tau}_{\boldsymbol{\phi}_{mj}}^{-1} diag\left(\gamma_{\boldsymbol{\phi}_{j1m}}^{-1}, \dots, \gamma_{\boldsymbol{\phi}_{jPm}}^{-1}\right)$. By letting
$$\begin{aligned}
\mathbf{m}_{\boldsymbol{\phi}_{jm}} = & \frac{1}{\sigma^2} \sum_{i=1}^N \sum_{l = 1}^{n_i}\left(B(t_{il})\chi_{im} \left(y_i(t_{il})Z_{ij} -  Z_{ij}^2 \left(\boldsymbol{\nu}_{j}+ \boldsymbol{\eta}_j\mathbf{x}_i' \right)'B(t_{il}) - Z_{ij}^2\sum_{n \ne m}\chi_{in} \boldsymbol{\phi}_{jn}' B(t_{il})\right. \right.\\
&  \left. \left. - Z_{ij}^2\sum_{n=1}^M \chi_{in}\mathbf{x}_i\boldsymbol{\xi}_{jn}'B(t_{il})  - \sum_{k \ne j} Z_{ij}Z_{ik}\left[\left(\boldsymbol{\nu}_{k}+ \boldsymbol{\eta}_k\mathbf{x}_i' \right)' B(t_{il}) + \sum_{n=1}^M \chi_{in} \left(\boldsymbol{\phi}_{kn} + \boldsymbol{\xi}_{kn}\mathbf{x}_i'\right)'B(t_{il}) \right] \right) \right),
\end{aligned}$$
and
$$\mathbf{M}_{\boldsymbol{\phi}_{jm}}^{-1} = \frac{1}{\sigma^2}\sum_{i=1}^N \sum_{l = 1 }^{n_i} \left(Z_{ij}^2\chi_{im}^2B(t_{il})B'(t_{il})\right) + \mathbf{D}_{\boldsymbol{\phi}_{jm}}^{-1},$$
 we have that 
$$\boldsymbol{\phi}_{jm} | \boldsymbol{\Theta}_{-\boldsymbol{\phi}_{jm} }, \mathbf{Y}_1, \dots, \mathbf{Y}_N, \mathbf{X} \sim \mathcal{N}\left(\mathbf{M}_{\boldsymbol{\phi}_{jm}}\mathbf{m}_{\boldsymbol{\phi}_{jm}}, \mathbf{M}_{\boldsymbol{\phi}_{jm}}\right).$$

Let $\boldsymbol{\xi}_{krm}$ be the $r^{th}$ column of the matrix $\boldsymbol{\xi}_{km}$. We will let $\mathbf{D}_{\boldsymbol{\xi}_{krm}} = \tilde{\tau}_{\boldsymbol{\xi}_{mjr}}^{-1} diag\left(\gamma_{\boldsymbol{\xi}_{jrm1}}^{-1}, \dots, \gamma_{\boldsymbol{\phi}_{\boldsymbol{\xi}_{jrmP}}}^{-1}\right)$. We will also let $x_{ir}$ denote the $r^{th}$ element of $\mathbf{x}_{i}$. Thus, letting 
$$\begin{aligned}
\mathbf{m}_{\boldsymbol{\xi}_{kdm}} = & \frac{1}{\sigma^2} \sum_{i=1}^N \sum_{l = 1}^{n_i}\left(B(t_{il})\chi_{im}x_{id}Z_{ik}\left(y_i(t_{il}) -  \sum_{j=1}^K Z_{ij}\left[\left(\boldsymbol{\nu}_{j}+ \boldsymbol{\eta}_j\mathbf{x}_i' \right)' B(t_{il}) + \sum_{n=1}^M \chi_{in} \boldsymbol{\phi}_{jn}'B(t_{il}) \right]\right. \right.\\
& \left. \left. - \sum_{(j,n,r) \ne (k,m,d)} Z_{ij} \chi_{in}x_{ir} \boldsymbol{\xi}_{krn}'B(t_{il}) \right) \right)
\end{aligned}$$

$$\mathbf{M}_{\boldsymbol{\xi}_{kdm}}^{-1} = \frac{1}{\sigma^2}\sum_{i=1}^N \sum_{l = 1 }^{n_i} \left(Z_{ik}^2\chi_{im}^2 x_{id}^2B(t_{il})B'(t_{il})\right) + \mathbf{D}_{\boldsymbol{\xi}_{kdm}}^{-1},$$
 we have that 
$$\boldsymbol{\xi}_{kdm} | \boldsymbol{\Theta}_{-\boldsymbol{\xi}_{kdm}}, \mathbf{Y}_1, \dots, \mathbf{Y}_N, \mathbf{X} \sim \mathcal{N}\left(\mathbf{M}_{\boldsymbol{\xi}_{kdm}}\mathbf{m}_{\boldsymbol{\xi}_{kdm}}, \mathbf{M}_{\boldsymbol{\xi}_{kdm}}\right).$$

The posterior distribution of $\delta_{\boldsymbol{\phi}_{1k}}$, for $k = 1, \dots, K$, is 
$$\begin{aligned}
\delta_{\boldsymbol{\phi}_{1k}} | \boldsymbol{\Theta}_{-\delta_{\boldsymbol{\phi}_{1k}}}, \mathbf{Y}_1, \dots, \mathbf{Y}_N, \mathbf{X} \sim & \Gamma\left(a_{\boldsymbol{\phi}_{1k}} + (PM/2), 1 + \frac{1}{2} \sum_{r=1}^P \gamma_{\boldsymbol{\phi}_{k,r,1}}\phi_{k,r,1}^2  \right.  \\
& \left. + \frac{1}{2}\sum_{m=2}^M \sum_{r=1}^P \gamma_{\boldsymbol{\phi}_{k,r,m}}\phi_{k,r,m}^2\left( \prod_{j=2}^m \delta_{\boldsymbol{\phi}_{jk}} \right)\right).
\end{aligned}.$$
The posterior distribution for $\delta_{\boldsymbol{\phi}_{ik}}$, for $i = 2, \dots, M$ and $k = 1, \dots, K$, is 
$$\begin{aligned}
\delta_{\boldsymbol{\phi}_{ik}} | \boldsymbol{\Theta}_{-\delta_{\boldsymbol{\phi}_{ik}}}, \mathbf{Y}_1, \dots, \mathbf{Y}_N, \mathbf{X}  \sim & \Gamma\Bigg(a_{\boldsymbol{\phi}_{2k}} + (P(M - i + 1)/2), 1   \\
& \left. +\frac{1}{2}\sum_{m = i}^M \sum_{r=1}^P \gamma_{\boldsymbol{\xi}_{k,r,m}}\phi_{k,r,m}^2\left( \prod_{j=1; j \ne i}^m \delta_{\boldsymbol{\phi}_{jk}} \right)\right).
\end{aligned}$$

The posterior distribution of $\delta_{\boldsymbol{\xi}_{1kd}}$, for $k = 1, \dots, K$ and $d = 1, \dots, R$, is 
$$\begin{aligned}
\delta_{\boldsymbol{\xi}_{1kd}} | \boldsymbol{\Theta}_{-\delta_{\boldsymbol{\xi}_{1kd}}}, \mathbf{Y}_1, \dots, \mathbf{Y}_N, \mathbf{X} \sim & \Gamma\left(a_{\boldsymbol{\xi}_{1kd}} + (PM/2), 1 + \frac{1}{2} \sum_{r=1}^P \gamma_{\boldsymbol{\xi}_{kdr1}}\xi_{kdr1}^2  \right.  \\
& \left. + \frac{1}{2}\sum_{m=2}^M \sum_{r=1}^P \gamma_{\boldsymbol{\xi}_{kdrm}}\xi_{kdrm}^2\left( \prod_{j=2}^m \delta_{\boldsymbol{\xi}_{jkd}} \right)\right).
\end{aligned}.$$
The posterior distribution for $\delta_{\boldsymbol{\xi}_{ikd}}$, for $i = 2, \dots, M$, $k = 1, \dots, K$, and $d = 1, \dots, D$ is 
$$\begin{aligned}
\delta_{\boldsymbol{\xi}_{ikd}} | \boldsymbol{\Theta}_{-\delta_{\boldsymbol{\xi}_{ikd}}}, \mathbf{Y}_1, \dots, \mathbf{Y}_N, \mathbf{X}  \sim & \Gamma\Bigg(a_{\boldsymbol{\xi}_{2kd}} + (P(M - i + 1)/2), 1   \\
& \left. +\frac{1}{2}\sum_{m = i}^M \sum_{r=1}^P \gamma_{\boldsymbol{\xi}_{kdrm}}\phi_{kdrm}^2\left( \prod_{j=1; j \ne i}^m \delta_{\boldsymbol{\xi}_{jkd}} \right)\right).
\end{aligned}$$

The posterior distribution for $a_{\boldsymbol{\phi}_{1k}}$ ($k = 1, \dots, K$) is not a commonly known distribution, however we have that
$$P(a_{\boldsymbol{\phi}_{1k}}|\boldsymbol{\Theta}_{-a_{\boldsymbol{\phi}_{1k}}}, \mathbf{Y}_1, \dots, \mathbf{Y}_N, \mathbf{X}) \propto \frac{1}{\Gamma(a_{\boldsymbol{\phi}_{1k}})}\delta_{\boldsymbol{\phi}_{1k}}^{a_{\boldsymbol{\phi}_{1k}} -1} a_{\boldsymbol{\phi}_{1k}}^{\alpha_{1} -1} exp \left\{-a_{\boldsymbol{\phi}_{1k}}\beta_{1} \right\}.$$
Since this is not a known kernel of a distribution, we will have to use Metropolis-Hastings algorithm. Consider the proposal distribution $Q(a_{\boldsymbol{\phi}_{1k}}'| a_{\boldsymbol{\phi}_{1k}}) = \mathcal{N}\left(a_{\boldsymbol{\phi}_{1k}}, \epsilon_1\beta_{1}^{-1}, 0, + \infty\right)$ (Truncated Normal) for some small $\epsilon_1 > 0$. Thus the probability of accepting any step is
$$A(a_{\boldsymbol{\phi}_{1k}}',a_{\boldsymbol{\phi}_{1k}}) = \min \left\{1, \frac{P\left(a_{\boldsymbol{\phi}_{1k}}'| \boldsymbol{\Theta}_{-a_{\boldsymbol{\phi}_{1k}}'}, \mathbf{Y}_1, \dots, \mathbf{Y}_N, \mathbf{X}\right)}{P\left(a_{\boldsymbol{\phi}_{1k}}| \boldsymbol{\Theta}_{-a_{\boldsymbol{\phi}_{1k}}}, \mathbf{Y}_1, \dots, \mathbf{Y}_N, \mathbf{X}\right)} \frac{Q\left(a_{\boldsymbol{\phi}_{1k}}|a_{\boldsymbol{\phi}_{1k}}'\right)}{Q\left(a_{\boldsymbol{\phi}_{1k}}'|a_{\boldsymbol{\phi}_{1k}}\right)}\right\}.$$

Similarly for $a_{\boldsymbol{\phi}_{2k}}$ ($k = 1, \dots, K$), we have
$$P(a_{\boldsymbol{\phi}_{2k}} | \boldsymbol{\Theta}_{-a_{\boldsymbol{\phi}_{2k}}}, \mathbf{Y}_1, \dots, \mathbf{Y}_N, \mathbf{X}) \propto \frac{1}{\Gamma(a_{\boldsymbol{\phi}_{2k}})^{M-1}}\left(\prod_{i=2}^M\delta_{\boldsymbol{\phi}_{ik}}^{a_{\boldsymbol{\phi}_{2k}} -1}\right) a_{\boldsymbol{\phi}_{2k}}^{\alpha_{\boldsymbol{\phi}_{2k}} -1} exp \left\{-a_{\boldsymbol{\phi}_{2k}}\beta_{2} \right\}.$$
We will use a similar proposal distribution, such that $Q(a_{\boldsymbol{\phi}_{2k}}'| a_{\boldsymbol{\phi}_{2k}}) = \mathcal{N}\left(a_{\boldsymbol{\phi}_{2k}}, \epsilon_2\beta_{2}^{-1}, 0, + \infty\right)$ for some small $\epsilon_2 > 0$. Thus the probability of accepting any step is
$$A(a_{\boldsymbol{\phi}_{2k}}',a_{\boldsymbol{\phi}_{2k}}) = \min \left\{1, \frac{P\left(a_{\boldsymbol{\phi}_{2k}}'| \boldsymbol{\Theta}_{-a_{\boldsymbol{\phi}_{2k}}'}, \mathbf{Y}_1, \dots, \mathbf{Y}_N, \mathbf{X}\right)}{P\left(a_{\boldsymbol{\phi}_{2k}}| \boldsymbol{\Theta}_{-a_{\boldsymbol{\phi}_{2k}}}, \mathbf{Y}_1, \dots, \mathbf{Y}_N, \mathbf{X}\right)} \frac{Q\left(a_{\boldsymbol{\phi}_{2k}}|a_{\boldsymbol{\phi}_{2k}}'\right)}{Q\left(a_{\boldsymbol{\phi}_{2k}}'|a_{\boldsymbol{\phi}_{2k}}\right)}\right\}.$$

Similarly, the posterior distribution for $a_{\boldsymbol{\xi}_{1kd}}$ ($k = 1, \dots, K$ and $d = 1, \dots, R$) is not a commonly known distribution, however we have that
$$P(a_{\boldsymbol{\xi}_{1kd}}|\boldsymbol{\Theta}_{-a_{\boldsymbol{\xi}_{1kd}}}, \mathbf{Y}_1, \dots, \mathbf{Y}_N, \mathbf{X}) \propto \frac{1}{\Gamma(a_{\boldsymbol{\xi}_{1kd}})}\delta_{\boldsymbol{\xi}_{1kd}}^{a_{\boldsymbol{\xi}_{1kd}} -1} a_{\boldsymbol{\xi}_{1kd}}^{\alpha_{1} -1} exp \left\{-a_{\boldsymbol{\xi}_{1kd}}\beta_{1} \right\}.$$

We will use a similar proposal distribution, such that $Q(a_{\boldsymbol{\xi}_{1kd}}'| a_{\boldsymbol{\xi}_{1kd}}) = \mathcal{N}\left(a_{\boldsymbol{\xi}_{1kd}}, \epsilon_1\beta_{1}^{-1}, 0, + \infty\right)$ for some small $\epsilon_1 > 0$. Thus the probability of accepting any step is
$$A(a_{\boldsymbol{\xi}_{1kd}}',a_{\boldsymbol{\xi}_{1kd}}) = \min \left\{1, \frac{P\left(a_{\boldsymbol{\xi}_{1kd}}'| \boldsymbol{\Theta}_{-a_{\boldsymbol{\xi}_{1kd}}'}, \mathbf{Y}_1, \dots, \mathbf{Y}_N, \mathbf{X}\right)}{P\left(a_{\boldsymbol{\xi}_{1kd}}| \boldsymbol{\Theta}_{-a_{\boldsymbol{\phi}_{1k}}}, \mathbf{Y}_1, \dots, \mathbf{Y}_N, \mathbf{X}\right)} \frac{Q\left(a_{\boldsymbol{\xi}_{1kd}}|a_{\boldsymbol{\xi}_{1kd}}'\right)}{Q\left(a_{\boldsymbol{\xi}_{1kd}}'|a_{\boldsymbol{\xi}_{1kd}}\right)}\right\}.$$

Similarly for $a_{\boldsymbol{\xi}_{2kd}}$ ($k = 1, \dots, K$ and $d = 1, \dots, R$), we have
$$P(a_{\boldsymbol{\xi}_{2kd}} | \boldsymbol{\Theta}_{-a_{\boldsymbol{\xi}_{2kd}}}, \mathbf{Y}_1, \dots, \mathbf{Y}_N, \mathbf{X}) \propto \frac{1}{\Gamma(a_{\boldsymbol{\xi}_{2kd}})^{M-1}}\left(\prod_{i=2}^M\delta_{\boldsymbol{\xi}_{ikd}}^{a_{\boldsymbol{\xi}_{2kd}} -1}\right) a_{\boldsymbol{\xi}_{2kd}}^{\alpha_{\boldsymbol{\xi}_{2kd}} -1} exp \left\{-a_{\boldsymbol{\xi}_{2kd}}\beta_{2} \right\}.$$
We will use a similar proposal distribution, such that $Q(a_{\boldsymbol{\xi}_{2kd}}'| a_{\boldsymbol{\xi}_{2kd}}) = \mathcal{N}\left(a_{\boldsymbol{\xi}_{2kd}}, \epsilon_2\beta_{2}^{-1}, 0, + \infty\right)$ for some small $\epsilon_2 > 0$. Thus the probability of accepting any step is
$$A(a_{\boldsymbol{\xi}_{2kd}}',a_{\boldsymbol{\xi}_{2kd}}) = \min \left\{1, \frac{P\left(a_{\boldsymbol{\xi}_{2kd}}'| \boldsymbol{\Theta}_{-a_{\boldsymbol{\xi}_{2kd}}'}, \mathbf{Y}_1, \dots, \mathbf{Y}_N, \mathbf{X}\right)}{P\left(a_{\boldsymbol{\xi}_{2kd}}| \boldsymbol{\Theta}_{-a_{\boldsymbol{\xi}_{2kd}}}, \mathbf{Y}_1, \dots, \mathbf{Y}_N, \mathbf{X}\right)} \frac{Q\left(a_{\boldsymbol{\xi}_{2kd}}|a_{\boldsymbol{\xi}_{2kd}}'\right)}{Q\left(a_{\boldsymbol{\xi}_{2kd}}'|a_{\boldsymbol{\xi}_{2kd}}\right)}\right\}.$$
For the $\gamma_{\phi_{jrm}}$ parameters, for $j = 1, \dots K$, $p = 1, \dots, P$, and $m = 1, \dots, M$, we have 
$$\gamma_{\phi_{jpm}}| \boldsymbol{\Theta}_{-\gamma_{\phi_{jpm}}}, \mathbf{Y}_1, \dots, \mathbf{Y}_N, \mathbf{X} \sim \Gamma\left(\frac{\nu_\gamma + 1}{2},\frac{\phi_{jpm}^2\tilde{\tau}_{\boldsymbol{\phi}_{mj}} + \nu_\gamma}{2} \right).$$
Similarly, for the $\gamma_{\boldsymbol{\xi}_{jdpm}}$ parameters, we have
$$\gamma_{\xi_{jrpm}}| \boldsymbol{\Theta}_{-\gamma_{\xi_{jrpm}}}, \mathbf{Y}_1, \dots, \mathbf{Y}_N, \mathbf{X} \sim \Gamma\left(\frac{\nu_\gamma + 1}{2},\frac{\xi_{jrpm}^2\tilde{\tau}_{\boldsymbol{\xi}_{mjr}} + \nu_\gamma}{2} \right),$$
for $j = 1,\dots,K$, $r = 1,\dots, R$, $p = 1, \dots, P$, and $m = 1, \dots, M$. 
The posterior distribution for the $\mathbf{z}_i$ parameters are not a commonly known distribution, so we will use the Metropolis-Hastings algorithm. We know that
$$\begin{aligned}
p(\mathbf{z}_i| \boldsymbol{\Theta}_{-\mathbf{z}_i}, \mathbf{Y}_1, \dots, \mathbf{Y}_N, \mathbf{X}) & \propto \prod_{k=1}^K Z_{ik}^{\alpha_3\pi_k - 1}\\
& \times \prod_{l=1}^{n_i} exp\left\{-\frac{1}{2\sigma^2}\left(y_i(t_{il}) -  \sum_{k=1}^K Z_{ik}\left(\left(\boldsymbol{\nu}_k + \boldsymbol{\eta}_k \mathbf{x}_i'\right)'B(t_{il}) \right. \right.\right.\\
& \left. \left. \left.+ \sum_{m=1}^M\chi_{im}\left(\boldsymbol{\phi}_{km} + \boldsymbol{\xi}_{km}\mathbf{x}_i'\right)'B(t_{il})\right)\right)^2\right\}.
\end{aligned}$$
We will use $Q(\mathbf{z}_i'| \mathbf{z}_i) = Dir(a_{\mathbf{z}} \mathbf{z}_i)$ for some large $a_{\mathbf{z}} \in \mathbb{R}^+$ as the proposal distribution. Thus the probability of accepting a proposed step is 
$$A(\mathbf{z}_i', \mathbf{z}_i) = \min \left\{1, \frac{P\left(\mathbf{z}_i'| \boldsymbol{\Theta}_{-\mathbf{z}_i}, \mathbf{Y}_1, \dots, \mathbf{Y}_N, \mathbf{X} \right)}{P\left(\mathbf{z}_i| \boldsymbol{\Theta}_{-\mathbf{z}_i}, \mathbf{Y}_1, \dots, \mathbf{Y}_N, \mathbf{X}\right)} \frac{Q\left(\mathbf{z}_i|\mathbf{z}_i'\right)}{Q\left(\mathbf{z}_i'|\mathbf{z}_i\right)}\right\}.$$

Similarly, a Gibbs update is not available for an update of the $\boldsymbol{\pi}$ parameters. We have that 
$$\begin{aligned}
p(\boldsymbol{\pi}|\boldsymbol{\Theta}_{-\boldsymbol{\pi}}, \mathbf{Y}_1,\dots, \mathbf{Y}_N, \mathbf{X}) & \propto \prod_{k=1}^K \pi_k^{c_k - 1} \\
& \times \prod_{i=1}^N\frac{1}{B(\alpha_3\boldsymbol{\pi})}\prod_{k=1}^K Z_{ik}^{\alpha_3\pi_k - 1}.
\end{aligned}$$
Letting out proposal distribution be such that $Q(\boldsymbol{\pi}'| \boldsymbol{\pi}) = Dir(a_{\boldsymbol{\pi}} \boldsymbol{\pi})$, for some large $a_{\boldsymbol{\pi}} \in \mathbb{R}^+$, we have that our probability of accepting any proposal is
$$A(\boldsymbol{\pi}', \boldsymbol{\pi}) = \min \left\{1, \frac{P\left(\boldsymbol{\pi}'| \boldsymbol{\Theta}_{-\boldsymbol{\pi}'}, \mathbf{Y}_1, \dots, \mathbf{Y}_N, \mathbf{X}\right)}{P\left(\boldsymbol{\pi}| \boldsymbol{\Theta}_{-\boldsymbol{\pi}}, \mathbf{Y}_1, \dots, \mathbf{Y}_N, \mathbf{X}\right)} \frac{Q\left(\boldsymbol{\pi}|\boldsymbol{\pi}'\right)}{Q\left(\boldsymbol{\pi}'|\boldsymbol{\pi}\right)}\right\}.$$
The posterior distribution of $\alpha_3$ is also not a commonly known distribution, so we will use the Metropolis-Hastings algorithm to sample from the posterior distribution. We have that 
$$\begin{aligned}
p(\alpha_3|\boldsymbol{\Theta}_{-\alpha_3}, \mathbf{Y}_1, \dots, \mathbf{Y}_N, \mathbf{X}) & \propto e^{-b\alpha_3} \\
& \times \prod_{i=1}^N\frac{1}{B(\alpha_3\boldsymbol{\pi})}\prod_{k=1}^K Z_{ik}^{\alpha_3\pi_k - 1}.
\end{aligned}$$
Using a proposal distribution such that $Q(\alpha_3'|\alpha_3) = \mathcal{N}(\alpha_3, \sigma^2_{\alpha_3}, 0, +\infty)$ (Truncated Normal), we are left with the probability of accepting a proposed state as
$$A(\alpha_3',\alpha_3) = \min \left\{1, \frac{P\left(\alpha_3'| \boldsymbol{\Theta}_{-\alpha_3'}, \mathbf{Y}_1, \dots, \mathbf{Y}_N, \mathbf{X}\right)}{P\left(\alpha_3| \boldsymbol{\Theta}_{-\alpha_3}, \mathbf{Y}_1, \dots, \mathbf{Y}_N, \mathbf{X}\right)} \frac{Q\left(\alpha_3|\alpha_3'\right)}{Q\left(\alpha_3'|\alpha_3\right)}\right\}.$$

Let $\mathbf{P}$ be the following tridiagonal matrix:
$$\mathbf{P}= \begin{bmatrix}
1 & -1 & 0 &  & \\
-1 & 2 & -1 &  &  \\
 & \ddots & \ddots & \ddots&  \\
 &  & -1 & 2 & -1  \\
 &  & 0 & -1 & 1 \\
\end{bmatrix}.$$
Thus, letting
$$\mathbf{B}_{\boldsymbol{\nu}_j} = \left( \tau_{\boldsymbol{\nu}_j}\mathbf{P} + \frac{1}{\sigma^2} \sum_{i =1}^N \sum_{l=1}^{n_i}Z_{ij}^2B(t_{il})B'(t_{il})  \right)^{-1}$$
and
$$\begin{aligned} \mathbf{b}_{\boldsymbol{\nu}_j} & = \frac{1}{\sigma^2}\sum_{i=1}^N\sum_{l=1}^{n_i}Z_{ij}B(t_{il})\left[y_i(t_{il}) - \left(\sum_{k\ne j}Z_{ik}\boldsymbol{\nu}'_{k}B(t_{il})\right) \right.  \\ 
& - \left.\left(\sum_{k=1}^K Z_{ik} \left[\mathbf{x}_i \boldsymbol{\eta}_k' B(t_{il}) + \sum_{m=1}^M\chi_{im} \left(\boldsymbol{\phi}_{kn} + \boldsymbol{\xi}_{kn}\mathbf{x}_i'\right)'B(t_{il}) \right]\right)\right],
\end{aligned}$$
we have that 
$$\boldsymbol{\nu}_j| \boldsymbol{\Theta}_{-\boldsymbol{\nu}_j}, \mathbf{Y}_1, \dots, \mathbf{Y}_N, \mathbf{X} \sim \mathcal{N}\left(\mathbf{B}_{\boldsymbol{\nu}_j}\mathbf{b}_{\boldsymbol{\nu}_j}, \mathbf{B}_{\boldsymbol{\nu}_j}\right).$$

Let $\boldsymbol{\eta}_{jd}$ denote the $d^{th}$ column of the matrix $\boldsymbol{\eta}_j$. Thus, letting
$$\mathbf{B}_{\boldsymbol{\eta}_{jd}} = \left( \tau_{\boldsymbol{\eta}_{jd}}\mathbf{P} + \frac{1}{\sigma^2} \sum_{i =1}^N \sum_{l=1}^{n_i}Z_{ij}^2 x_{id}^2 B(t_{il})B'(t_{il})  \right)^{-1}$$
and 
$$\begin{aligned} \mathbf{b}_{\boldsymbol{\eta}_{jd}} = & \frac{1}{\sigma^2}\sum_{i=1}^N\sum_{l=1}^{n_i}Z_{ij}x_{id}B(t_{il})\left[y_i(t_{il}) - \left(\sum_{r\ne d}Z_{ij}x_{ir} \boldsymbol{\eta}_{jr}'B(t_{il})\right) - \left(\sum_{k \ne j} Z_{ik} \mathbf{x}_i\boldsymbol{\eta}_k' B(t_{il}) \right) \right.  \\ 
& - \left.\left(\sum_{k=1}^K Z_{ik}\left[\boldsymbol{\nu}_k' B(t_{il}) + \sum_{m=1}^M\chi_{im} \left(\boldsymbol{\phi}_{kn} + \boldsymbol{\xi}_{kn}\mathbf{x}_i'\right)'B(t_{il}) \right]\right)\right],
\end{aligned}$$
we have that 
$$\boldsymbol{\eta}_{jd}| \boldsymbol{\Theta}_{-\boldsymbol{\eta}_{jd}}, \mathbf{Y}_1, \dots, \mathbf{Y}_N, \mathbf{X} \sim \mathcal{N}\left(\mathbf{B}_{\boldsymbol{\eta}_{jd}}\mathbf{b}_{\boldsymbol{\eta}_{jd}}, \mathbf{B}_{\boldsymbol{\eta}_{jd}}\right).$$

Thus we can see that we can draw samples from the posterior of the parameters controlling the mean structure using a Gibbs sampler. Similarly, we can use a Gibbs sampler to draw samples from the posterior distribution of $\tau_{\boldsymbol{\eta}_{jd}}$ and $\tau_{\boldsymbol{\nu}_j}$. We have that the posterior distributions are
$$\tau_{\boldsymbol{\nu}_j}| \boldsymbol{\Theta}_{-\tau_{\boldsymbol{\nu}_j}}, \mathbf{Y}_1, \dots, \mathbf{Y}_N, \mathbf{X} \sim \Gamma\left(\alpha_{\boldsymbol{\nu}} + P/2, \beta_{\boldsymbol{\nu}} + \frac{1}{2}\boldsymbol{\nu}'_j\mathbf{P}\boldsymbol{\nu}_j\right)$$
and
$$\tau_{\boldsymbol{\eta}_{jd}}| \boldsymbol{\Theta}_{-\tau_{\boldsymbol{\eta}_{jd}}}, \mathbf{Y}_1, \dots, \mathbf{Y}_N, \mathbf{X} \sim \Gamma\left(\alpha_{\boldsymbol{\eta}} + P/2, \beta_{\boldsymbol{\eta}} + \frac{1}{2}\boldsymbol{\eta}'_{jd}\mathbf{P}\boldsymbol{\eta}_{jd}\right),$$
for $ j = 1, \dots, K$ and $d = 1, \dots, R$. The parameter $\sigma^2$ can be updated by using a Gibbs update. If we let 
$$\beta_{\sigma} =\frac{1}{2}\sum_{i=1}^N\sum_{l=1}^{n_i}\left(y_i(t_{il}) -  \sum_{k=1}^K Z_{ik}\left(\left(\boldsymbol{\nu}_k + \boldsymbol{\eta}_k \mathbf{x}_i'\right)'B(t_{il}) + \sum_{n=1}^M\chi_{in}\left(\boldsymbol{\phi}_{kn} + \boldsymbol{\xi}_{kn}\mathbf{x}_i'\right)'B(t_{il})\right)\right)^2,$$
then we have
$$\sigma^2| \boldsymbol{\Theta}_{-\sigma^2}, \mathbf{Y}_1, \dots, \mathbf{Y}_N, \mathbf{X}  \sim  IG\left(\alpha_0 + \frac{\sum_{i=1}^N n_i}{2} , \beta_0 +\beta_{\sigma}\right).$$
Lastly, we can update the $\chi_{im}$ parameters, for $i = 1, \dots, N$ and $m = 1, \dots, M$, using a Gibbs update. If we let 
$$\begin{aligned}
\mathbf{w}_{im} = & \frac{1}{\sigma^2}\left[\sum_{l=1}^{n_i} \left(\sum_{k = 1}^K Z_{ik}\left(\boldsymbol{\phi}_{km} + \boldsymbol{\xi}_{km}\mathbf{x}_i'\right)'B(t_{il})\right)\right. \\
& \left.\left(y_i(t_{il}) - \sum_{k = 1}^K Z_{ik}\left(\left(\boldsymbol{\nu}_k + \boldsymbol{\eta}_k \mathbf{x}_i'\right)'B(t_{il})  + \sum_{n\ne m}\chi_{in}\left(\boldsymbol{\phi}_{kn} + \boldsymbol{\xi}_{kn}\mathbf{x}_i'\right)'B(t_{il})\right)\right)\right]
\end{aligned}$$

and 
$$\mathbf{W}_{im}^{-1} = 1 + \frac{1}{\sigma^2} \sum_{l=1}^{n_i}\left(\sum_{k = 1}^K Z_{ik}\left(\boldsymbol{\phi}_{km} + \boldsymbol{\xi}_{km}\mathbf{x}_i'\right)'B(t_{il})\right)^2,$$
then we have that 
$$\chi_{im}| \boldsymbol{\zeta}_{-\chi_{im}}, \mathbf{Y}_1, \dots, \mathbf{Y}_N, \mathbf{X} \sim \mathcal{N}(\mathbf{W}_{im}\mathbf{w}_{im}, \mathbf{W}_{im}).$$

\subsection{Tempered Transitions}
Since we only temper the likelihood, many of the posterior distributions derived in Section \ref{posterior_xi} can be utilized.
Starting with the $\boldsymbol{\Phi}$ parameters, we have 
$$\begin{aligned}
\left(\mathbf{m}_{\boldsymbol{\phi}_{jm}}\right)_h = & \frac{\beta_h}{(\sigma^2)_h} \sum_{i=1}^N \sum_{l = 1}^{n_i}\left(B(t_{il})(\chi_{im})_h \left(y_i(t_{il})(Z_{ij})_h -  (Z_{ij})_h^2 \left((\boldsymbol{\nu}_{j})_h+ (\boldsymbol{\eta}_j)_h\mathbf{x}_i' \right)'B(t_{il}) \right. \right.\\
&  \left. \left. - (Z_{ij})_h^2\sum_{n \ne m}(\chi_{in})_h (\boldsymbol{\phi}_{jn})_h' B(t_{il})- (Z_{ij})_h^2\sum_{n=1}^M (\chi_{in})\mathbf{x}_i(\boldsymbol{\xi}_{jn})_h'B(t_{il})  \right. \right.\\
& \left. \left. - \sum_{k \ne j} Z_{ij}Z_{ik}\left[\left((\boldsymbol{\nu}_{k})_h+ (\boldsymbol{\eta}_k)_h\mathbf{x}_i' \right)' B(t_{il}) + \sum_{n=1}^M \chi_{in} \left((\boldsymbol{\phi}_{kn})_h + (\boldsymbol{\xi}_{kn})_h\mathbf{x}_i'\right)'B(t_{il}) \right] \right) \right),
\end{aligned}$$
and
$$\left(\mathbf{M}_{\boldsymbol{\phi}_{jm}}\right)_h^{-1} = \frac{\beta_h}{(\sigma^2)_h}\sum_{i=1}^N \sum_{l = 1 }^{n_i} \left((Z_{ij})_h^2(\chi_{im})_h^2B(t_{il})B'(t_{il})\right) + \left(\mathbf{D}_{\boldsymbol{\phi}_{jm}}\right)_h^{-1},$$
 we have that 
$$\left(\boldsymbol{\phi}_{jm}\right)_h | \boldsymbol{\Theta}_{-\left(\boldsymbol{\phi}_{jm}\right)_h}, \mathbf{Y}_1, \dots, \mathbf{Y}_N, \mathbf{X} \sim \mathcal{N}\left(\left(\mathbf{M}_{\boldsymbol{\phi}_{jm}}\right)_h \left(\mathbf{m}_{\boldsymbol{\phi}_{jm}}\right)_h, \left(\mathbf{M}_{\boldsymbol{\phi}_{jm}}\right)_h\right).$$
Letting 
$$\begin{aligned}
\left(\mathbf{m}_{\boldsymbol{\xi}_{kdm}}\right)_h = & \frac{\beta_h}{(\sigma^2)_h} \sum_{i=1}^N \sum_{l = 1}^{n_i}\Bigg(B(t_{il})(\chi_{im})_h x_{id}(Z_{ik})_h\Bigg(y_i(t_{il}) -\\
& \sum_{j=1}^K (Z_{ij})_h\left[\left((\boldsymbol{\nu}_{j})_h + (\boldsymbol{\eta}_j)_h\mathbf{x}_i' \right)' B(t_{il}) + \sum_{n=1}^M (\chi_{in})_h (\boldsymbol{\phi}_{jn})_h'B(t_{il}) \right]\\
& \left. \left. - \sum_{(j,n,r) \ne (k,m,d)} (Z_{ij})_h (\chi_{in})_h x_{ir} (\boldsymbol{\xi}_{krn})_h'B(t_{il}) \right) \right)
\end{aligned}$$

$$\left(\mathbf{M}_{\boldsymbol{\xi}_{kdm}}\right)_h^{-1} = \frac{\beta_h}{(\sigma^2)_h}\sum_{i=1}^N \sum_{l = 1 }^{n_i} \left((Z_{ik})_h^2(\chi_{im})_h^2 x_{id}^2B(t_{il})B'(t_{il})\right) + \left(\mathbf{D}_{\boldsymbol{\xi}_{kdm}}\right)_h^{-1},$$
 we have that 
$$\left(\boldsymbol{\xi}_{kdm}\right)_h | \boldsymbol{\Theta}_{-\left(\boldsymbol{\xi}_{kdm}\right)_h}, \mathbf{Y}_1, \dots, \mathbf{Y}_N, \mathbf{X} \sim \mathcal{N}\left(\left(\mathbf{M}_{\boldsymbol{\xi}_{kdm}}\right)_h \left(\mathbf{m}_{\boldsymbol{\xi}_{kdm}}\right)_h, \left(\mathbf{M}_{\boldsymbol{\xi}_{kdm}}\right)_h\right).$$

As in the untempered case, we have that the posterior distribution $\mathbf{Z}$ parameters under the tempered likelihood is not a commonly known distribution. Therefore, we will use the Metropolis-Hastings algorithm. We have that
$$\begin{aligned}
p((\mathbf{z}_i)_h| \boldsymbol{\Theta}_{-(\mathbf{z}_i)_h}, \mathbf{Y}_1, \dots, \mathbf{Y}_N, \mathbf{X}) & \propto \prod_{k=1}^K (Z_{ik})_h^{(\alpha_3)_h(\pi_k)_h - 1}\\
& \times \prod_{l=1}^{n_i} exp\left\{-\frac{\beta_h}{2(\sigma^2)_h}\left(y_i(t_{il}) -  \sum_{k=1}^K (Z_{ik})_h\left(\left((\boldsymbol{\nu}_k)_h + (\boldsymbol{\eta}_k)_h \mathbf{x}_i'\right)'B(t_{il}) \right. \right.\right.\\
& \left. \left. \left.+ \sum_{m=1}^M(\chi_{im})_h\left((\boldsymbol{\phi}_{km})_h + (\boldsymbol{\xi}_{km})_h\mathbf{x}_i'\right)'B(t_{il})\right)\right)^2\right\}.
\end{aligned}$$
We will use $Q((\mathbf{z}_i)_h'| (\mathbf{z}_i)_h) = Dir(a_{\mathbf{z}} (\mathbf{z}_i)_h)$ for some large $a_{\mathbf{z}} \in \mathbb{R}^+$ as the proposal distribution. Thus the probability of accepting a proposed step is 
$$A((\mathbf{z}_i)_h', (\mathbf{z}_i)_h) = \min \left\{1, \frac{P\left((\mathbf{z}_i)_h'| \boldsymbol{\Theta}_{-(\mathbf{z}_i)_h'}, \mathbf{Y}_1, \dots, \mathbf{Y}_N, \mathbf{X} \right)}{P\left((\mathbf{z}_i)_h| \boldsymbol{\Theta}_{-(\mathbf{z}_i)_h}, \mathbf{Y}_1, \dots, \mathbf{Y}_N, \mathbf{X}\right)} \frac{Q\left((\mathbf{z}_i)_h|(\mathbf{z}_i)_h'\right)}{Q\left((\mathbf{z}_i)_h'|(\mathbf{z}_i)_h\right)}\right\}.$$

Letting
$$\left(\mathbf{B}_{\boldsymbol{\nu}_j}\right)_h = \left( \left(\tau_{\boldsymbol{\nu}_j}\right)_h\mathbf{P} + \frac{\beta_h}{(\sigma^2)_h} \sum_{i =1}^N \sum_{l=1}^{n_i}(Z_{ij})_h^2B(t_{il})B'(t_{il}) \right)^{-1}$$
and
$$\begin{aligned} \left(\mathbf{b}_{\boldsymbol{\nu}_j}\right)_h & = \frac{\beta_h}{(\sigma^2)_h}\sum_{i=1}^N\sum_{l=1}^{n_i}(Z_{ij})_h B(t_{il})\left[y_i(t_{il}) - \left(\sum_{k\ne j}(Z_{ik})_h(\boldsymbol{\nu}'_{k})_hB(t_{il})\right) \right.  \\ 
& - \left.\left(\sum_{k=1}^K (Z_{ik}) \left[\mathbf{x}_i (\boldsymbol{\eta}_k)_h' B(t_{il}) + \sum_{m=1}^M(\chi_{im})_h \left((\boldsymbol{\phi}_{kn})_h + (\boldsymbol{\xi}_{kn})_h\mathbf{x}_i'\right)'B(t_{il}) \right]\right)\right],
\end{aligned}$$
we have that 
$$\left(\boldsymbol{\nu}_j\right)_h| \boldsymbol{\Theta}_{-\left(\boldsymbol{\nu}_j\right)_h}, \mathbf{Y}_1, \dots, \mathbf{Y}_N, \mathbf{X} \sim \mathcal{N}\left(\left(\mathbf{B}_{\boldsymbol{\nu}_j}\right)_h\left(\mathbf{b}_{\boldsymbol{\nu}_j}\right)_h, \left(\mathbf{B}_{\boldsymbol{\nu}_j}\right)_h\right).$$

Let $\left(\boldsymbol{\eta}_{jd}\right)_h$ denote the $d^{th}$ column of the matrix $(\boldsymbol{\eta}_j)_h$. Thus, letting
$$\left(\mathbf{B}_{\boldsymbol{\eta}_{jd}}\right)_h = \left( \left(\tau_{\boldsymbol{\eta}_{jd}}\right)_h\mathbf{P} + \frac{\beta_h}{(\sigma^2)_h} \sum_{i =1}^N \sum_{l=1}^{n_i}(Z_{ij})_h^2 x_{id}^2 B(t_{il})B'(t_{il}) \right)^{-1}$$
and 
$$\begin{aligned} \left(\mathbf{b}_{\boldsymbol{\eta}_{jd}}\right)_h = & \frac{\beta_h}{(\sigma^2)_h}\sum_{i=1}^N\sum_{l=1}^{n_i}(Z_{ij})_hx_{id}B(t_{il})\left[y_i(t_{il}) - \left(\sum_{r\ne d}(Z_{ij})_hx_{ir} (\boldsymbol{\eta}_{jr})_h'B(t_{il})\right)  \right.  \\ 
& - \left(\sum_{k \ne j} (Z_{ik})_h \mathbf{x}_i(\boldsymbol{\eta}_k)_h' B(t_{il}) \right) \\
& - \left.\left(\sum_{k=1}^K (Z_{ik})_h\left[(\boldsymbol{\nu}_k)_h' B(t_{il}) + \sum_{m=1}^M(\chi_{im})_h \left((\boldsymbol{\phi}_{kn})_h + (\boldsymbol{\xi}_{kn})_h\mathbf{x}_i'\right)'B(t_{il})\right] \right)\right],
\end{aligned}$$
we have that 
$$\left(\boldsymbol{\eta}_{jd}\right)_h| \boldsymbol{\Theta}_{-\left(\boldsymbol{\eta}_{jd}\right)_h}, \mathbf{Y}_1, \dots, \mathbf{Y}_N, \mathbf{X} \sim \mathcal{N}\left(\left(\mathbf{B}_{\boldsymbol{\eta}_{jd}}\right)_h\left(\mathbf{b}_{\boldsymbol{\eta}_{jd}}\right)_h, \left(\mathbf{B}_{\boldsymbol{\eta}_{jd}}\right)_h\right).$$

If we let 
$$\begin{aligned}
\left(\beta_{\sigma}\right)_h =\frac{\beta_h}{2}\sum_{i=1}^N\sum_{l=1}^{n_i}\left(y_i(t_{il}) -  \sum_{k=1}^K (Z_{ik})_h\Bigg(\left((\boldsymbol{\nu}_k)_h + (\boldsymbol{\eta}_k)_h \mathbf{x}_i'\right)'B(t_{il}) \right. \\+ \left. \left. \sum_{n=1}^M(\chi_{in})_h\left((\boldsymbol{\phi}_{kn})_h + (\boldsymbol{\xi}_{kn})_h\mathbf{x}_i'\right)'B(t_{il})\right)\right)^2,
\end{aligned}$$
then we have
$$(\sigma^2)_h| \boldsymbol{\Theta}_{-(\sigma^2)_h}, \mathbf{Y}_1, \dots, \mathbf{Y}_N, \mathbf{X}  \sim  IG\left(\alpha_0 + \frac{\beta_h\sum_{i=1}^N n_i}{2} , \beta_0 +\left(\beta_{\sigma}\right)_h\right).$$
Lastly, we can update the $\chi_{im}$ parameters, for $i = 1, \dots, N$ and $m = 1, \dots, M$, using a Gibbs update. If we let 
$$\begin{aligned}
\left(\mathbf{w}_{im}\right)_h = & \frac{\beta_h}{(\sigma^2)_h}\left[\sum_{l=1}^{n_i} \left(\sum_{k = 1}^K (Z_{ik})_h\left((\boldsymbol{\phi}_{km})_h + (\boldsymbol{\xi}_{km})_h\mathbf{x}_i'\right)'B(t_{il})\right)\right. \Bigg(y_i(t_{il})\\
& \left. - \sum_{k = 1}^K (Z_{ik})_h\left(\left((\boldsymbol{\nu}_k)_h + (\boldsymbol{\eta}_k)_h \mathbf{x}_i'\right)'B(t_{il})  + \sum_{n\ne m}(\chi_{in})_h\left((\boldsymbol{\phi}_{kn})_h + (\boldsymbol{\xi}_{kn})_h\mathbf{x}_i'\right)'B(t_{il})\right)\Bigg)\right]
\end{aligned}$$

and 
$$\left(\mathbf{W}_{im}\right)_h^{-1} = 1 + \frac{\beta_h}{\sigma^2} \sum_{l=1}^{n_i}\left(\sum_{k = 1}^K (Z_{ik})_h\left((\boldsymbol{\phi}_{km})_h + (\boldsymbol{\xi}_{km})_h\mathbf{x}_i'\right)'B(t_{il})\right)^2,$$
then we have that 
$$(\chi_{im})_h| \boldsymbol{\zeta}_{-(\chi_{im})_h}, \mathbf{Y}_1, \dots, \mathbf{Y}_N, \mathbf{X} \sim \mathcal{N}\left(\left(\mathbf{W}_{im}\right)_h \left(\mathbf{w}_{im}\right)_h, \left(\mathbf{W}_{im}\right)_h\right).$$

\end{document}